# Architectural Design of a RAM Arbiter

**Sourangsu Banerji**
*Department of Electronics & Communication Engineering,*
*RCC-Institute of Information Technology,*
*Under West Bengal University of Technology,*
**April, 2014.**

*Mini Project Report*

# Architectural Design of a RAM Arbiter

**Submitted By**

**Sourangsu Banerji**
Department of Electronics & Communication Engineering
RCC-Institute of Information Technology

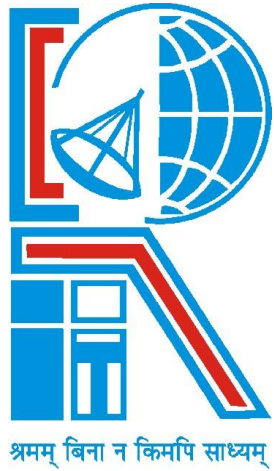

श्रमम् बिना न किमपि साध्यम्

**To**

**Department of Electronics & Communication Engineering,**
RCC-Institute of Information Technology,
(Affiliated to West Bengal University of Technology),
Canal South Road, Tangra, Kolkata- 700015,
West Bengal, India

**April, 2014.**

# ACKNOWLEDGEMENT

I take this opportunity to express my profound gratitude and deep regards to my guide Assistant Professor Abhishek Basu, for his exemplary guidance, monitoring and constant encouragement throughout the course of this report. The blessing, help and guidance given by him time to time shall carry me a long way in the journey of life on which I am about to embark.

I also take this opportunity to express a deep sense of gratitude to Dr. Tirtha Sankar Das, Head of Department (HOD) of Electronics & Communication Engineering department for his cordial support, valuable information and guidance, which helped me in completing this task through various stages.

I am obliged to faculty and technical staff members of Electronics & Communication Engineering department, for the valuable information provided by them in their respective fields. I am grateful for their cooperation during the period of my assignment.

Lastly, I would like to thank almighty, my parents, brother, sisters and friends for their constant encouragement without which this thesis would not be possible.

# DECLARATION

I certify that

1. The work contained in this report is original and has been done by me under the guidance of my supervisor(s).
2. The work has not been submitted to any other Institute for any degree or diploma.
3. I have followed the guidelines provided by the Institute in preparing the report.
4. I have conformed to the norms and guidelines given in the Ethical Code of Conduct of the Institute.
5. Whenever I have used materials (data, theoretical analysis, figures, and text) from other sources, I have given due credit to them by citing them in the text of the report and giving their details in the references. Further, we have taken permission from the copyright owners of the sources, whenever necessary.

Sourangsu Banerji.

# CERTIFICATE

This is to certify that Mr. Sourangsu Banerji, a student of Department of Electronics & Communication Engineering, RCC-Institute of Information Technology, has done a project work titled "**Architectural Design of a RAM Arbiter**" from January 15th 2014 to April 25, 2014 under my guidance and supervision.

Signature of the Supervisor
Date:

## ABSTRACT

Standard memory modules to store (and access) data are designed for use with a single system accessing it. More complicated memory modules would be accessed through a memory controller, which are also designed for one system. For multiple systems to access a single memory module there must be some facilitation that allows them to access the memory without overriding or corrupting the access from the others. This was done with the use of a memory arbiter, which controls the flow of traffic into the memory controller. The arbiter has a set of rules to abide to in order to choose which system gets through to the memory controller. In this project, a regular RAM module is designed for use with one system. Furthermore, a memory arbiter is also designed in Verilog that allows for more than one system to use a single RAM module in a controlled and synchronized manner. The arbiter uses a fixed priority scheme to avoid starvation of the system. In addition one of the major problems associated with such systems i.e. "The Address Clash Problem" has been nicely tackled and solved. The design is verified in simulation and validated on a Xilinx ML605 evaluation board with a Virtex-6 FPGA.

**Keyword:** RAM, RAM Arbiter, VHDL and FPGA.

# LIST OF FIGURES

# TABLE OF CONTENTS

**Conclusion**

**References**

# Chapter 1

## Introduction

Without a clock to govern its actions, an asynchronous system must rely on local coordination circuits instead. These circuits exchange completion signals to ensure that the actions at each stage begin only when the circuits have the data they need. The two most important coordination circuits are called the Rendezvous and the Arbiter.

A Rendezvous element indicates when the last of two or more signals has arrived at a particular stage. Asynchronous systems use these elements to wait until all the concurrent actions finish before starting the next action. For instance, an arithmetic division circuit must have both the dividend (say, 16) and the divisor (say, 2) before it can divide one by the other (to reach the answer 8).One form of Rendezvous circuit is called the Muller C-element, named after David Muller, a retired from a professor at the University of Illinois. A Muller C-element is a logic circuit with two inputs and one output. When both inputs of a Muller C-element are TRUE, its output becomes TRUE. When both inputs are FALSE, its output becomes FALSE. Otherwise the output remains unchanged. For the Muller C-element to act as a Rendezvous circuit, its inputs must not change again until its output responds. A chain of Muller C-elements can control the flow of data down an electronic bucket brigade.

An arbiter circuit performs another task essential for asynchronous computers. An Arbiter is like a traffic officer at an intersection who decides which car may pass through next. Given only one request, an Arbiter promptly permits the corresponding action, delaying any second request until the first action is completed. When an Arbiter gets two requests at once, it must decide which request to grant first. For example, when two processors request access to a shared memory at approximately the same time, the Arbiter puts the requests into a sequence, granting access to only one processor at a time. The Arbiter guarantees that there are never two actions under way at once, just as the traffic officer prevents accidents by ensuring that there are never two cars passing through the intersection on a collision course.

Although Arbiter circuits never grant more than one request at a time, there is no way to build an Arbiter that will always reach a decision within a fixed time limit. Present-day Arbiters reach decisions very quickly on average, usually within about a few hundred picoseconds. (A picosecond is a trillionth of a second.) When faced with close calls, however, the circuits may occasionally take twice as long, and in very rare cases the time needed to make a decision may be 10 times as long as normal.

The fundamental difficulty in making these decisions is nicely illustrated by the parable of Buridan's ass. Attributed to Buridan, a 14th-century French philosopher, this parable suggests that an ass placed exactly between two equal piles of hay might starve to death because it would be unable to choose which pile to eat. Similar minor dilemmas are familiar in everyday life. For example, two people approaching a doorway at the same time may pause before deciding who will go through first. They can go through in either order, and Buridan's ass can eat from either pile of hay. In both cases, all that is needed is a way to break the tie.

An Arbiter breaks ties. Like a flip-flop circuit, an Arbiter has two stable states corresponding to the two choices. One can think of these states as the Pacific Ocean and the Gulf of Mexico. Each request to an Arbiter pushes the circuit toward one stable state or the other, just as a hailstone that falls in the Rocky Mountains can roll downhill toward the Pacific or the Gulf. Between the two stable states, however, there must be a meta-stable line, which is equivalent to the Continental Divide. If a hailstone falls precisely on the divide, it may balance momentarily on that sharp mountain ridge before tipping toward the Pacific or the Gulf. Similarly, if two requests arrive at an Arbiter within a few picoseconds of each other, the circuit may pause in its meta-stable state before reaching one of its stable states to break the tie.

Novice Arbiter designers often seek to avoid even the occasional long delay by fashioning complicated circuits. A common error involves a circuit that notices the "hung" meta-stable state and pushes the Arbiter toward a particular decision. This is like training Buridan's ass to go left when it notices a hard choice. Such training, however, merely gives the ass three choices rather than two: go left, go right, or notice a hard choice and therefore go left. Even a trained ass will starve when unable to decide between the last two choices. Or, to use the geographic metaphor, you can move the Continental Divide with a shovel, but you cannot get rid of it. Although there is no way to eliminate meta-stability, simple, well-designed Arbiter circuits can ensure that virtually all delays are very brief. A typical contemporary Arbiter has a normal delay of 100 picoseconds and experiences a delay of 400 picoseconds less than once every 10 hours of operation. The probability of delays decreases exponentially with the length of the delay: an 800-picosecond pause occurs less than once every billion years of operation.

# Chapter 2

## Random Access Memory (RAM)

Random-access memory is a form of computer data storage. A random-access memory device allows data items to be read and written in roughly the same amount of time regardless of the order in which data items are accessed [1]. In contrast, with other direct-access data storage media such as hard disks, CD-RWs, DVD-RWs and the older drum memory, the time required to read and write data items varies significantly depend on their physical locations on the recording medium, due to mechanical limitations such as media rotation speeds and arm movement delays. Today random access memory (RAM) is widely used in computers and other electronics as a way to access and store data. This type of computer memory can be accessed randomly and without the need to access preceding or following data addresses. However, RAM is volatile memory and will only retain data as long as power is on. Once the system loses power, it loses any data stored in memory. RAM has evolved over time as engineers try to achieve better speed and efficiency [2].

### 2.1. Static Random Access Memory (SRAM)

Static Random Access Memory (SRAM) is a variation of RAM. SRAM is designed to fill two needs: provide a direct interface to CPUs at speeds unattainable by DRAMs and replace DRAMs in systems that require very low power consumption [3]. SRAM performs very well in low power applications due to the nature of the device. SRAM cells are comprised of six MOSFETs. Figure 1 below shows this.

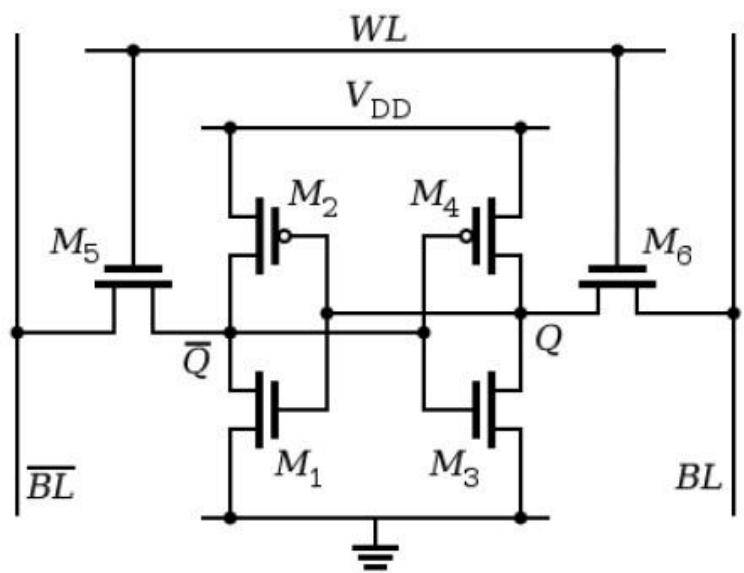

**Figure 1:** SRAM Six Transistors

Four transistors act as cross coupled inverters holding the bit information, while the remaining transistors control access to data during read/write operations. SRAM is preferred in portable equipment because of its low power capability and because it does not require a refresh cycle due to the absence of capacitors in its design. Although SRAM is still volatile, data will not leak away. This type of RAM is not used in more applications simply due to its price [5].

## 2.2. **Dynamic Random Access Memory (DRAM)**

The other major category of RAM is Dynamic Random Access Memory (DRAM). As with SRAM, DRAM fundamentally holds onto the information of individual bits, but unlike SRAM, it is designed with capacitors along with transistors, shown below in Figure 2.

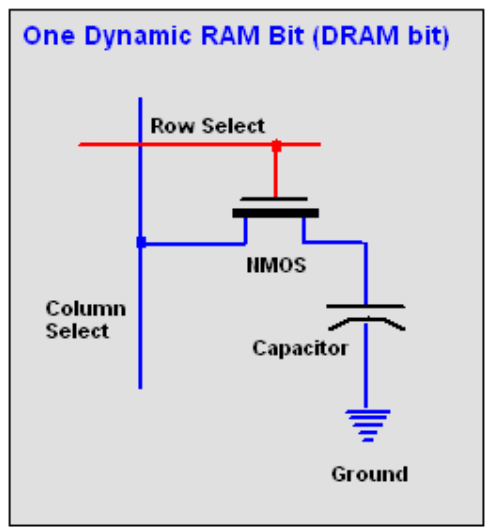

**Figure 2:** DRAM Schematic

The number of transistors is reduced to one in DRAM making it fundamentally simpler than SRAM. However since capacitors lose their charge over time, a refresh is needed to maintain stored data, which increases power usage due to the voltage of the capacitors The inability to maintain information without a refresh is why DRAM is considered dynamic as opposed to its "static" counterpart [7].

DRAM is used for its simplicity and its lower cost, however at the price of performance and efficiency. Engineers have been pushing the technology forward to improve its performance. When analyzing RAM it is important to look at bandwidth and latency. Bandwidth is the amount of data transferred per second, and latency is the time between sending an address to memory and receiving the data back on the data bus [8].These improvements are seen in various iterations of DRAM.

### 2.2.1. Synchronous DRAM (SDRAM)

SDRAM is synchronized with the system's clock. This interface waits for a clock signal before responding to inputs, resulting in data being available at every clock cycle. Asynchronous RAM attempts to respond to commands as soon as possible. To increase efficiency, memory is divided into several banks, enabling simultaneous processing of memory access commands. An address is comprised of bank, row, and column information [7].

### 2.2.2. Double Data Rate 1 (DDR1) SDRAM

To increase bandwidth, double data rate is introduced in DDR1 memory. Without having to increase clock speed, DDR1 transfers data on both the rising and falling edge of the clock. Additional power efficiency is achieved by reducing the supply voltage from 3.3V to 2.5V.

A 2n-prefetch architecture is introduced which allows 2 bits of data to be transferred to the queue in two separate pipelines. Without changing the clock, bandwidth is doubled with this interface [7].

### 2.2.3. DDR2 SDRAM

DDR2 makes further improvements upon earlier variations of SDRAM. Operation voltage lowered to 1.8V, decreasing total power consumption. Additionally, a 4n-prefetch buffer is added. Improving upon the previous 2 bits, 4 bits are now able to be transferred per clock cycle from the memory array to the data bus. DDR2 data rates are up to eight times faster than the original SDRAM [7].

### 2.2.4. DDR3 SDRAM

As with previous generations, DDR3 decreases power consumption and increases bandwidth. DDR3 uses a 1.5V power supply as opposed to the 1.8V power supply used in DDR2 and its bandwidth can be up to twice that of DDR2 [8]. DDR3 has eight banks, which allows more efficient bank access than in previous interfaces with four. Additionally, the pre-fetch buffer is increased to 8 bits wide, resulting in an 8n interface [7].

Two modes are used in DDR3 memory interface: burst chop (BC4) and burst length eight (BL8). BL8 can be seen in the timing diagram below in Figure 3. BL8 allows for addressing to occur once every eight data packets are sent, because consecutive memory addresses are used. BC4 allows bursts of four by treating data as though half of it is masked. Of the two options, BL8 is more widely used [8].

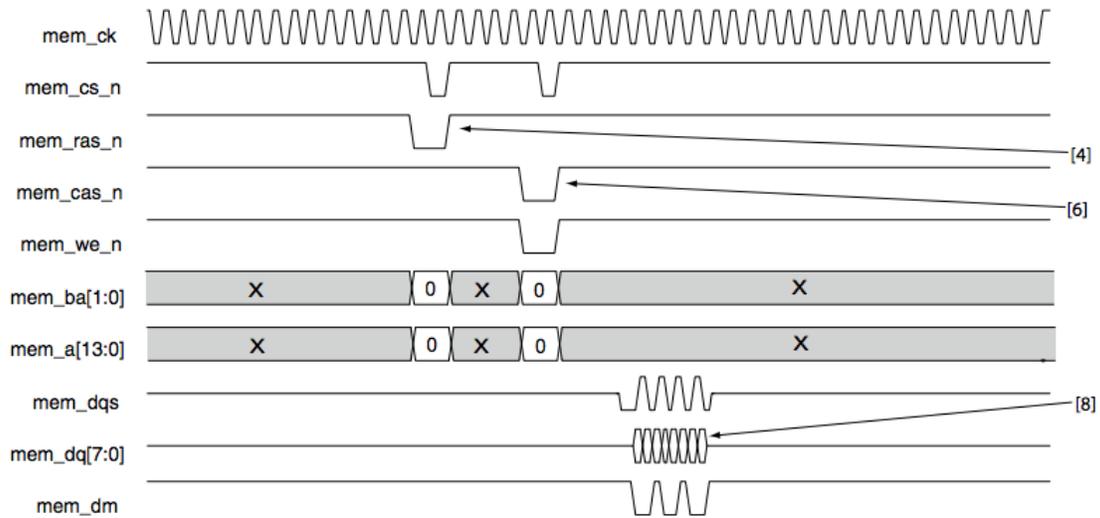

**Figure 3:** DDR3 Timing Diagram

## 2.3. Implementation of the RAM Module using Verilog

### 2.3.1. VHDL Code

```
---------------------------------------------------------------------------------
-- Entity for RAM
---------------------------------------------------------------------------------
library IEEE;
use IEEE.STD_LOGIC_1164.ALL;
use IEEE.STD_LOGIC_ARITH.ALL;
use IEEE.STD_LOGIC_UNSIGNED.ALL;
entity RAM is
generic
(
---------------------------------------------------------------------------------
-- Generics for scalability
---------------------------------------------------------------------------------
G_ADDR_WIDTH:    integer := 4;
G_DATA_WIDTH:    integer := 8
-- G_ADDR_WIDTH = Number of bits required to address the ram
-- G_DATA_WIDTH = Number of bits in a data
---------------------------------------------------------------------------------
);
port
(
---------------------------------------------------------------------------------
-- RAM Inputs
---------------------------------------------------------------------------------
CLOCK:        in  std_logic;
RST_N:        in  std_logic;
RD_EN:        in  std_logic;              --read enb--
WR_EN:        in  std_logic;              --write enb--
RD_ADDR:      in  std_logic_vector(G_ADDR_WIDTH-1 downto 0);--read addr---
```

```
WR_ADDR:        in  std_logic_vector(G_ADDR_WIDTH-1 downto 0);--write addr--
WR_DATA:        in  std_logic_vector(G_DATA_WIDTH-1 downto 0);--data input----
RD_DATA:        out std_logic_vector(G_DATA_WIDTH-1 downto 0) --data output--
);
end RAM;

-----------------------------------------------------------------------------------------------------------
-----------------------------------------------------------------------------------------------------------
-- Architecture for RAM
-----------------------------------------------------------------------------------------------------------
architecture RTL of RAM is
constant RAM_DEPTH:          integer := 2**G_ADDR_WIDTH;
type MEMORY_ARRAY is array(0 to RAM_DEPTH-1) of
std_logic_vector(G_DATA_WIDTH-1 downto 0);--ram type--
signal MEMORY:              MEMORY_ARRAY;
signal count:               integer:=0;
signal reset_done:          std_logic;
begin
memorypr:process(CLOCK)
begin
if (RST_N='0')then
reset_done<='1';
elsif(CLOCK'EVENT and CLOCK='1')then
if(count <(2**G_ADDR_WIDTH) and reset_done='1' )then
MEMORY(conv_integer(count))<=(others =>'0');
count<=count+1;
else
count<=0;
reset_done<='0';
end if;
if(reset_done='0')then
-----------------------------------------------------------------------------------------------------------
-----WRITING PROCESS IN RAM
-----------------------------------------------------------------------------------------------------------
if(WR_EN='1')then
MEMORY(conv_integer(WR_ADDR))<= WR_DATA;
end if;
-----------------------------------------------------------------------------------------------------------
-----READING PROCESS IN RAM
-----------------------------------------------------------------------------------------------------------
if(RD_EN='1')then
RD_DATA<=MEMORY(conv_integer(RD_ADDR));
end if;
end if;
end if;
end process;
end RTL;

-----------------------------------------------------------------------------------------------------------
```

### 2.3.2. Test Cases for RAM

1. RAM write operation
2. RAM read operation
3. RAM read & write operation

### 2.3.3. VHDL Testbench

```
LIBRARY ieee;
USE ieee.std_logic_1164.ALL;
USE ieee.std_logic_unsigned.all;
USE ieee.numeric_std.ALL;
ENTITY ram_test IS
END ram_test;
ARCHITECTURE behavior OF ram_test IS
-- Component Declaration for the Unit Under Test (UUT)
COMPONENT RAM
PORT(
CLOCK : IN  std_logic;
RST_N : IN  std_logic;
RD_EN : IN  std_logic;
WR_EN : IN  std_logic;
RD_ADDR : IN  std_logic_vector(3 downto 0);
WR_ADDR : IN  std_logic_vector(3 downto 0);
WR_DATA : IN  std_logic_vector(7 downto 0);
RD_DATA : OUT  std_logic_vector(7 downto 0)
);
END COMPONENT;
-----------------------------------------------------------------------------------------------
--Inputs
-----------------------------------------------------------------------------------------------
  signal CLOCK : std_logic := '0';
  signal RST_N : std_logic := '0';
  signal RD_EN : std_logic := '0';
  signal WR_EN : std_logic := '0';
  signal RD_ADDR : std_logic_vector(3 downto 0) := (others => '0');
  signal WR_ADDR : std_logic_vector(3 downto 0) := (others => '0');
  signal WR_DATA : std_logic_vector(7 downto 0) := (others => '0');
-----------------------------------------------------------------------------------------------
--Outputs
-----------------------------------------------------------------------------------------------
signal RD_DATA : std_logic_vector(7 downto 0);

BEGIN
-- Instantiate the Unit Under Test (UUT)
uut: RAM PORT MAP (
CLOCK => CLOCK,
RST_N => RST_N,
RD_EN => RD_EN,
WR_EN => WR_EN,
RD_ADDR => RD_ADDR,
```

```
WR_ADDR => WR_ADDR,
WR_DATA => WR_DATA,
RD_DATA => RD_DATA
);
CLOCK_process :process
begin
CLOCK <= '0';
wait for 50 ns;
CLOCK <= '1';
wait for 50 ns;
end process;
-- Stimulus process
stim_proc: process
begin
-- hold reset state for 100ms.
wait for 100 ns;
RST_N<='1';
wait for 100 ns;
---------------------------------------------------------------------------------------------
-- Test Case 1: RAM Write Operation --
---------------------------------------------------------------------------------------------
WR_EN<='1';
RD_EN<='0';
WR_ADDR<="1101";
WR_DATA<="11100111";
wait for 1700 ns;
---------------------------------------------------------------------------------------------
-- Test Case 2: RAM Read Operation --
---------------------------------------------------------------------------------------------
WR_EN <='0';
RD_EN <='1';
RD_ADDR<= "1101";
wait for 1700 ns;
---------------------------------------------------------------------------------------------
-- Test Case 3: RAM Read & Write Operation --
---------------------------------------------------------------------------------------------
WR_EN<='1';
RD_EN<='1';
RD_ADDR<= "1101";
WR_ADDR<="1011";
WR_DATA<="10111001";
wait for 1700 ns;
RD_ADDR<= "1011";
WR_ADDR <="1000";
WR_DATA <="10011111";
wait;
end process;

END;
---------------------------------------------------------------------------------------------
```

### 2.3.4. Waveforms

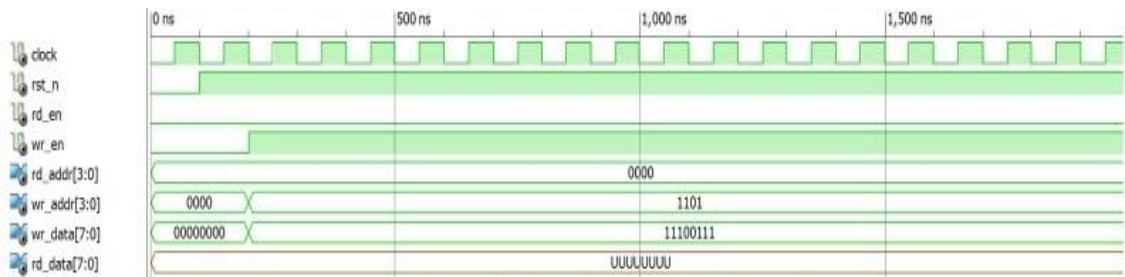

Figure 4: RAM Write Operation (Test Case 1)

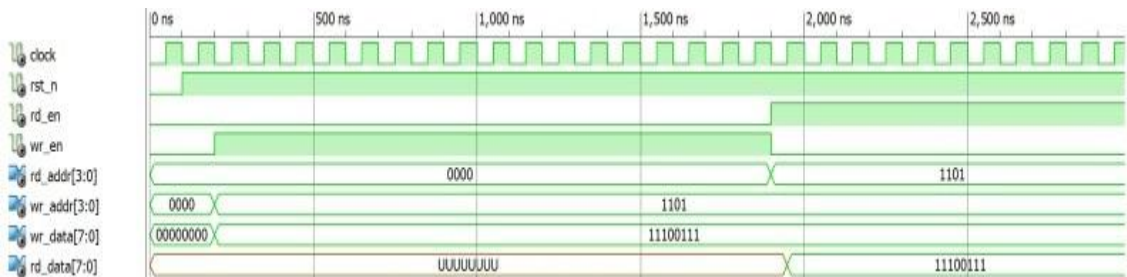

Figure 5: RAM Read Operation (Test Case 2)

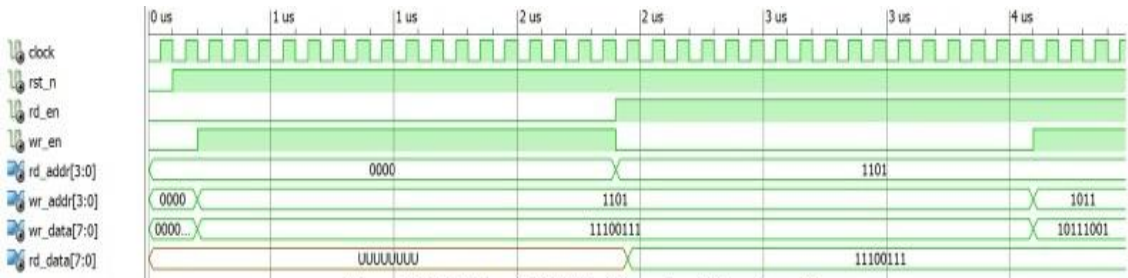

Figure 6: RAM Read & Write Operation (Test Case 3)

### 2.3.5. Analysis of the Test Cases

1. **RAM write operation**
   In this case, within the first RAM cycle, it is initialized and the write operation is performed at the address location "1101". The data written is "11100111". The RAM starts working when the RST_N pin is enabled.

2. **RAM read operation**
   Similar to the previous condition, when the RST_N pin is enabled within the first RAM cycle, data is read from the address location "1101".

3. **RAM read & write operation**
   In this case, the RAM is enabled when the reset pin is set HIGH. In the first RAM cycle, data "11100111" is written in the address "1101". In the second RAM cycle we seek to show that the RAM designed was able to perform both the operation of read and write in the same RAM cycle. Data is read from location "1101" and simultaneously data "10111001" was written in the address location "1011".

# Chapter 3

## Arbiter

Many systems exist in which a large number of requesters must access a common resource. In this case, the resource is shared memory. An arbiter is needed to control the flow of traffic between the requestors and shared memory.

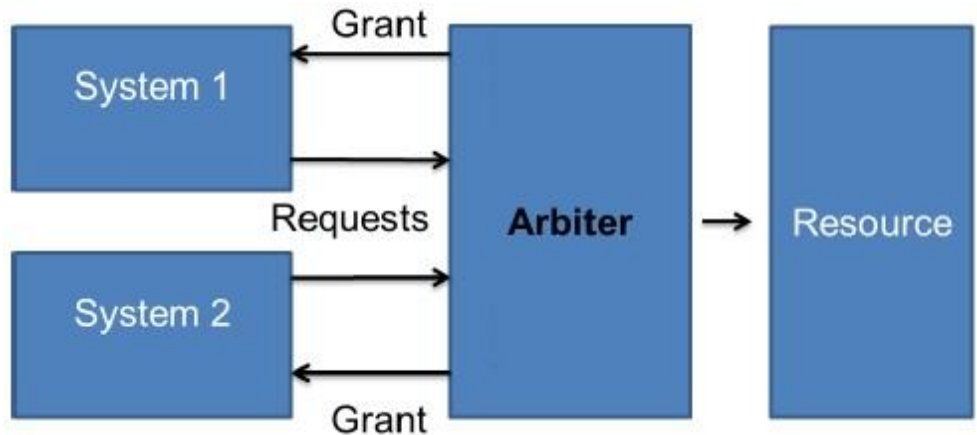

**Figure 7:** Example Arbiter Block Diagram

It determines how the resource is allocated amongst the requesters as shown in Figure 4 above. Arbiters have internal logic that determines how requesters get access based off of the applications needs. When designing an arbiter it is important to keep the interface, size, and speed in mind.

### 3.1. Arbitration Schemes

Many arbitration schemes already exist. These include round robin, first in first out, priority, and dynamic priority. The section below describes some of the schemes that we researched before the design of our own arbiter.

#### 3.1.1. Round Robin

In round robin, each system receives a specific amount of time to access memory and the systems cycle through in a pre-defined order [1]. A timing diagram of this technique can be seen below. The ready signal seen at the top of the timing diagram in Figure 5 is a depiction of the memory cycles, with each grant occurring at the positive edge of each cycle.

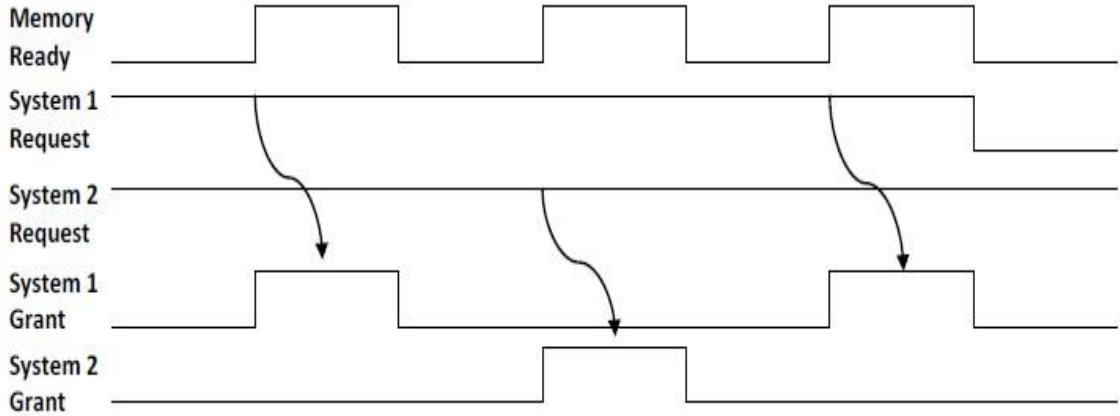

**Figure 8:** Round Robin Timing Diagram

In round robin systems are granted access entirely based off order rather than requests. This makes it an unfavorable option for arbiter designs stressing efficiency as systems may be granted access when not requesting, which results in arbiter idle time.

### 3.1.2. First In First Out

In first in first out whichever system asks for the memory first receives it. The arbiter keeps track of which system asserted its ready signal first and then gives it to that system. An example of FIFO timing can be seen below in Figure 6. System 1 requests access first and receives initial grant. System 2 then requests access to memory and is then shown to be given access after System 1 de-asserts its request.

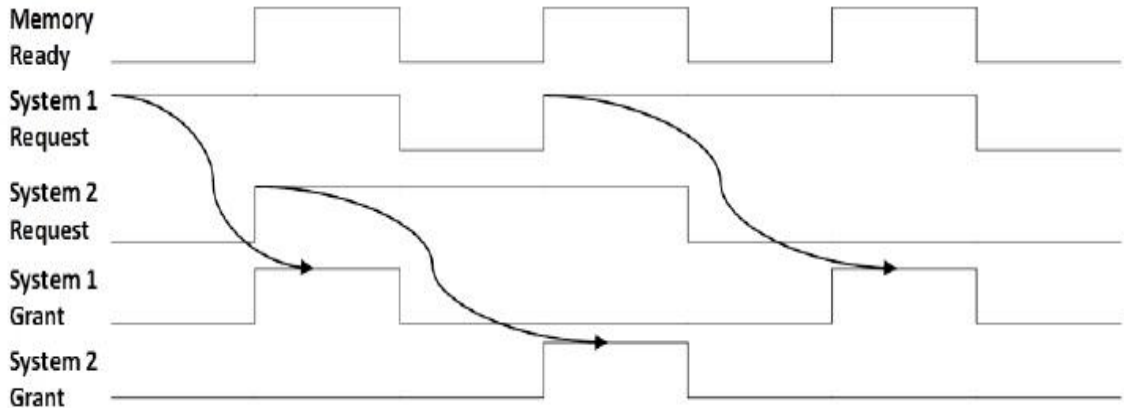

**Figure 9:** FIFO Timing Diagram

Unlike round robin, FIFO only provides access to those systems who are requesting. Resources are never wasted in this situation.

### 3.1.3. Priority

The priority method stores a specific priority value to each of the systems (low, medium, high etc.) and then grant access to whichever system has the highest priority. How the priority is assigned depends on the application. It might assign higher priorities to slower systems in order to prevent them from starving or assign high priorities to important systems [2]. In a priority system, care must be taken in order to avoid starvation of a system. An example of priority is shown below in Figure 7 where System 1 has high priority.

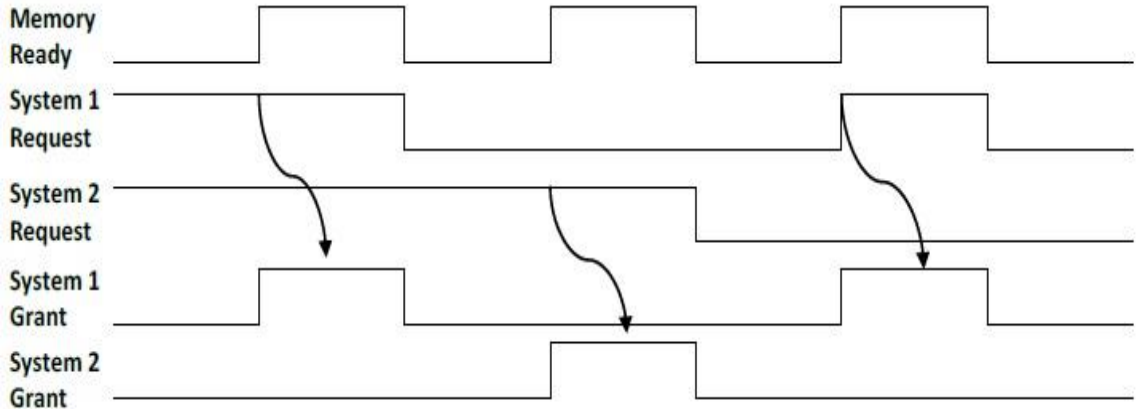

**Figure 10:** Priority Timing Diagram

Each time System 1 requests access to memory it is given it. System 2 will only receive resources if & when System 1 is no longer requesting.

A priority system is either fixed or dynamic. Fixed priorities are either hard coded within the arbiter design or assigned on the fly using features on-board. A dynamic priority system allows the priorities of different systems to change based on pre-determined factors [4]. For example, if one system was accessing memory more often than another system, the higher traffic system could gain higher priority. Another example would be to change priority based on how long a system goes without accessing memory. The scheme chosen depends on the requirements of the system.

## 3.2. Implementation of the Arbiter Module using Verilog

### 3.2.1. VHDL Code

```
Library IEEE;
use IEEE.STD_LOGIC_1164.ALL;
use IEEE.STD_LOGIC_ARITH.ALL;
use IEEE.STD_LOGIC_UNSIGNED.ALL;
---------------------------------------------------------------------------------------------------------------------
-- Entity for ARBITER
---------------------------------------------------------------------------------------------------------------------
entity ARBITER_NEW is
generic
(
---------------------------------------------------------------------------------------------------------------------
-- Generics for scalability
---------------------------------------------------------------------------------------------------------------------
```

```
G_ADDR_WIDTH:           integer := 4;
G_DATA_WIDTH:           integer := 8;
G_REGISTERED_DATA:      integer :=0
-- G_ADDR_WIDTH = Number of bits required to address the ram
-- G_DATA_WIDTH = Number of bits in a data
-- G_REGISTERED_DATA =1 for registered data in output 0 for nonregistered
------------------------------------------------------------------------------------------------
);
port
(
------------------------------------------------------------------------------------------------
-- General Inputs And Output
------------------------------------------------------------------------------------------------
RST_N:          in  std_logic;
CLOCK:          in  std_logic;
RST_DONE:       out std_logic;
------------------------------------------------------------------------------------------------
-- Inputs from --------client1--------------
------------------------------------------------------------------------------------------------
RD_EN_C1:       in  std_logic;                      --read enb--
WR_EN_C1:       in  std_logic;                      --write enb--
RDADDR_C1:      in  std_logic_vector(G_ADDR_WIDTH-1 downto 0);--read addr---
WRADDR_C1:      in  std_logic_vector(G_ADDR_WIDTH-1 downto 0);--write addr--
WRDATA_C1:      in  std_logic_vector(G_DATA_WIDTH-1 downto 0);--data in----
------------------------------------------------------------------------------------------------
-- Inputs from --------client2--------------
------------------------------------------------------------------------------------------------
DATAIN_C2:              in  std_logic_vector(G_DATA_WIDTH-1 downto 0);--input data--
REQUEST_C2:         in  std_logic;                  --request to access memory--
RD_NOT_WRITE_C2:    in  std_logic;                  --if '0' then write or read--
ADDR_C2:            in  std_logic_vector(G_ADDR_WIDTH-1 downto 0);--addr for rd or wr--
------------------------------------------------------------------------------------------------
--output from --------client1--------------
------------------------------------------------------------------------------------------------
RDDATA_C1:          out std_logic_vector(G_DATA_WIDTH-1 downto 0);--data out--
------------------------------------------------------------------------------------------------
--output from --------client2--------------
------------------------------------------------------------------------------------------------
DATAOUT_C2:         out std_logic_vector(G_DATA_WIDTH-1 downto 0);--out data--
ACK_C2:             out std_logic;                  --acknowledgement--
------------------------------------------------------------------------------------------------
-- Others Input And Output
------------------------------------------------------------------------------------------------
RD_EN:          out std_logic;
WR_EN:          out std_logic;
WR_ADDR:        out std_logic_vector(G_ADDR_WIDTH-1  downto 0);
RD_ADDR:        out std_logic_vector(G_ADDR_WIDTH-1  downto 0);
WR_DATA:        out std_logic_vector(G_DATA_WIDTH-1  downto 0);
RD_DATA:        in  std_logic_vector(G_DATA_WIDTH-1  downto 0));
end ARBITER_NEW;
------------------------------------------------------------------------------------------------
```

---------------------------------------------------------------------------------------------------
-- Architecture for ARBITER
---------------------------------------------------------------------------------------------------
architecture RTL of ARBITER_NEW is
--------------Temporary registers------------------
signal TEMP_RD_DATA:          std_logic_vector(G_DATA_WIDTH-1  downto 0);
signal TEMP_RD_DATA1:          std_logic_vector(G_DATA_WIDTH-1  downto 0);
signal TEMP_RD_DATA2:          std_logic_vector(G_DATA_WIDTH-1  downto 0);
signal TEMP_RD_EN:          std_logic;
signal TEMP_WR_EN:          std_logic;
signal TEMP_WR_ADDR:          std_logic_vector(G_ADDR_WIDTH-1  downto 0);
signal TEMP_RD_ADDR:          std_logic_vector(G_ADDR_WIDTH-1  downto 0);
signal TEMP_WR_DATA:          std_logic_vector(G_DATA_WIDTH-1  downto 0);
-------------Client type and state for FSM-----------
type client is (reset,idle,client1_read,client2_read,client1_write,client2_write);
signal pr_client_read:     client;                         --present client read-
signal pr_client_write:    client;                         --present client write-
signal nx_client_read:     client;                         --next client read--
signal nx_client_write:    client;                          --next client write--
-------------Acknowledgement reg for client2---------
signal TEMP_ACK:          std_logic:='0';
signal TEMP_ACK1:          std_logic;
signal TEMP_ACK2:          std_logic;
signal TEMP_WR:          std_logic:='0';
signal TEMP_WR1:          std_logic;
------------Generic consideration------------------------------
signal REGISTERED_DATA:    integer range 0 to 1 :=0;
------------Reset done generation Counter & register-----------
signal RESET_DONE_REG:     std_logic;
signal COUNT:          integer range 0 to 2**G_ADDR_WIDTH-1:=0;
------------Address Clash check register---------------------
signal ADDR_CLASHI:          std_logic:='0';
signal ADDR_CLASH:          std_logic:='0';
begin
---------------------------------------------------------------------------------------------------
--FSM for ARBITER
---------------------------------------------------------------------------------------------------
---------------------------------------------------------------------------------------------------
-------Sequential section & reset condition- ----
---------------------------------------------------------------------------------------------------
p1:process(RST_N,CLOCK)
begin
if (RST_N='0') then
pr_client_read  <= reset;
pr_client_write <= reset;
elsif (CLOCK'EVENT and CLOCK='1' ) then
pr_client_read  <= nx_client_read;
pr_client_write <= nx_client_write;
end if;
end process;

```
---------------------------------------------------------------------------------------------------
--------Generate for registered data-------
---------------------------------------------------------------------------------------------------
g1:  if (G_REGISTERED_DATA=1) generate
REGISTERED_DATA<=G_REGISTERED_DATA;
end generate g1;
---------------------------------------------------------------------------------------------------
---------Combinational section & client state------
---------------------------------------------------------------------------------------------------
p2:process(pr_client_read,pr_client_write,clock)
begin
if(RST_N='1' and clock='1')then
if(nx_client_read=reset and nx_client_write=reset)then
if(count<(2**G_ADDR_WIDTH))then
RESET_DONE_REG <= '0';
count<=count+1;
else
nx_client_read  <= idle;
nx_client_write <= idle;
RESET_DONE_REG <= '1';
count<=0;
end if;
end if;
elsif(RST_N='0') then
nx_client_read <= reset;
nx_client_write<=reset;
end if;
if(pr_client_read=idle)then              ----when arbiter idle--
if(RD_EN_C1='0')then
if(REQUEST_C2='0')then
nx_client_read<= idle;
elsif(RD_NOT_WRITE_C2='1')then
nx_client_read<= client2_read;
elsif(RD_NOT_WRITE_C2='0')then
nx_client_write<= client2_write;
end if;
else
nx_client_read <=client1_read;
end if;
end if;
if(pr_client_write=idle)then
if( WR_EN_C1='0')then
if(REQUEST_C2='0')then
nx_client_write<= idle;
elsif(RD_NOT_WRITE_C2='0')then
nx_client_write<= client2_write;
elsif(RD_NOT_WRITE_C2='1')then
nx_client_read<= client2_read;
end if;
else
nx_client_write <=client1_write;
```

```
end if;
end if;--------------------------------------------------------
if(pr_client_read=client1_read)then                -----when arbiter allow client 1---
if(RD_EN_C1='1')then
nx_client_read <=client1_read;
else
if(REQUEST_C2='0')then
nx_client_read<= idle;
elsif(RD_NOT_WRITE_C2='1')then
nx_client_read<= client2_read;
elsif(RD_NOT_WRITE_C2='0')then
nx_client_read<= idle;
end if;
end if;
end if;
if(pr_client_write=client1_write)then
if(WR_EN_C1='1')then
nx_client_write <=client1_write;
else
if(REQUEST_C2='0')then
nx_client_write<= idle;
elsif(RD_NOT_WRITE_C2='0')then
nx_client_write<= client2_write;
elsif(RD_NOT_WRITE_C2='1')then
nx_client_write<= idle;
end if;
end if;
end if;--------------------------------------------------------------------
if(pr_client_read=client2_read)then                ------when arbiter allow client 2-----
if(RD_EN_C1='0')then
if(REQUEST_C2='1')then
if( RD_NOT_WRITE_C2='1')then
nx_client_read<= client2_read;
else
nx_client_read<=idle;
nx_client_write<= client2_write;
end if;
else
nx_client_read<=idle;
end if;
else
nx_client_read <=client1_read;
end if;
end if;
if(pr_client_write=client2_write)then
if(WR_EN_C1='0')then
if(REQUEST_C2='1')then
if( RD_NOT_WRITE_C2='0')then
nx_client_write<= client2_write;
else
nx_client_write<=idle;
```

```
nx_client_read<= client2_read;
end if;
else
nx_client_write<=idle;
end if;
else
nx_client_write <=client1_write;
end if;
end if;
```
-------------------------------------------------------------------------------------------------------------------
```
end process;
```
-------------------------------------------------------------------------------------------------------------------
--Assigning Temp Registers according to the client----------------
-------------------------------------------------------------------------------------------------------------------
```
pram:process(CLOCK)
begin
```
-------------------------------------------------------------------------------------------------------------------
----------------Read & Write operation ------------
-------------------------------------------------------------------------------------------------------------------
```
if(RST_N = '0')then
TEMP_RD_DATA  <= (others =>'0');
TEMP_RD_DATA1 <= (others =>'0');
TEMP_RD_DATA2 <= (others =>'0');
elsif(CLOCK'EVENT and CLOCK='1')then
if(nx_client_read = idle)then
TEMP_RD_EN<='0';
TEMP_RD_ADDR<=(others =>'0');
elsif (nx_client_read=client1_read)then
TEMP_RD_EN   <= RD_EN_C1;
TEMP_RD_ADDR <= RDADDR_C1;
elsif(nx_client_read=client2_read)then
if(TEMP_ACK='0')then
TEMP_RD_EN  <= '1';
TEMP_RD_ADDR<= ADDR_C2;
TEMP_ACK    <= '1';
end if;
end if;
if(nx_client_write = idle)then
TEMP_WR_EN   <= '0';
TEMP_WR_DATA <= (others =>'0');
TEMP_WR_ADDR <= (others =>'0');
elsif (nx_client_write=client1_write)then
TEMP_WR_EN   <= WR_EN_C1;
TEMP_WR_DATA <= WRDATA_C1;
TEMP_WR_ADDR <= WRADDR_C1;
elsif(nx_client_write=client2_write)then
if(TEMP_WR='0')then
TEMP_WR_EN   <= '1';
TEMP_WR_ADDR <= ADDR_C2;
TEMP_WR_DATA <= DATAIN_C2;
TEMP_WR     <= '1';
```

end if;
end if;
------------------------------------------------------------------------------------------------------------
-----If Addr Clash occurs -------
------------------------------------------------------------------------------------------------------------
if (TEMP_RD_EN='1' and TEMP_WR_EN ='1') then
if(TEMP_WR_ADDR = TEMP_RD_ADDR )then
ADDR_CLASH <='1';
TEMP_RD_DATA<=TEMP_WR_DATA;
else
ADDR_CLASH <='0';
end if;
else
ADDR_CLASH <='0';
end if;
------------------------------------------------------------------------------------------------------------
if(TEMP_WR1='1')then ------For ACK generation during client2_Write------
TEMP_WR<='0';
end if;
TEMP_ACK1<=TEMP_ACK; ------For ACK generation during client2_Read------
if(TEMP_ACK1='1')then
TEMP_ACK1<='0';
TEMP_ACK<='0';
end if;          ---------------------------------------------------
ADDR_CLASHI<=ADDR_CLASH;---One clock cycle delay in addr clash for Registered data----
TEMP_RD_DATA1<=TEMP_RD_DATA;---One clock cycle delay in output for Registered data with addr clash -----
TEMP_RD_DATA2<=RD_DATA; ---One clock cycle delay in output for Registered data without addr clash -----
end if;
end process;
------------------------------------------------------------------------------------------------------------
--------Data in output from temp registers---------
------------------------------------------------------------------------------------------------------------
RD_EN<= TEMP_RD_EN;
WR_EN<= TEMP_WR_EN;
WR_DATA<=TEMP_WR_DATA;
WR_ADDR<=TEMP_WR_ADDR;
RD_ADDR<=TEMP_RD_ADDR;
------------------------------------------------------------------------------------------------------------
TEMP_WR1<=TEMP_WR;-----For ACK generation during client2_write--------
ACK_C2<='1' when   (TEMP_ACK1='1' or TEMP_WR1='1') else '0';--output ACK generation during client2_write and read---
RST_DONE<=RESET_DONE_REG;       ----------Indication for reset compleate----
------------------------------------------------------------------------------------------------------------
-----------------Data out for client 2 ----------------
------------------------------------------------------------------------------------------------------------
DATAOUT_C2<=RD_DATA when (ADDR_CLASH='0') else
          TEMP_RD_DATA;

---------------------------------------------------------------------------------------------------------------------
-----------------------Data out for client 1----------------
---------------------------------------------------------------------------------------------------------------------
RDDATA_C1<=RD_DATA when (REGISTERED_DATA =0 and ADDR_CLASH='0' ) else
TEMP_RD_DATA when (REGISTERED_DATA =0 and ADDR_CLASH='1' ) else
TEMP_RD_DATA2 when (REGISTERED_DATA =1 and ADDR_CLASHI='0' ) else
TEMP_RD_DATA1 when (REGISTERED_DATA =1 and ADDR_CLASHI='1' );
end RTL;

---------------------------------------------------------------------------------------------------------------------

### 3.2.2. Test Cases for Arbiter

1. Only Client1 wants to write.

2. Only Client1 wants to read.

3. Only Client2 wants to write.

4. Only Client2 wants to read.

5. Client1 wants to read and write in different RAM location at same time.

6. Client1 wants to read and write in different RAM location at different time.

7. Client1 wants to read and write in same RAM location at same time.

8. Client1 wants to read and write in same RAM location at different time.

9. Client2 wants to read and write in different RAM location at same time.

10. Client2 wants to read and write in different RAM location at different time.

11. Client2 wants to read and write in same RAM location at same time.

12. Client2 wants to read and write in same RAM location at different time.

13. Client1 wants to write and Client2 wants to read in different RAM location at same time.

14. Client1 wants to write and client2 wants to read in different RAM location at different time.

15. Client1 wants to write and Client2 wants to read in same RAM location at same time.

16. Client1 wants to write and Client2 wants to read in same RAM location at different time.

17. Client1 wants to read and Client2 wants to write in same RAM location at same time.

18. Client1 wants to read and Client2 wants to write in same RAM location at different time.

19. Client1 wants to read and Client2 wants to write in different RAM location at same time.

20. Client1 wants to read and Client2 wants to write in different RAM location at different time.

21. Client1 wants to read and Client2 wants to read in same RAM location at different time.

22. Client1 wants to read and Client2 wants to read in same RAM location at same time.

23. Client1 wants to write and Client2 wants to write in different RAM location at different time.

24. Client1 wants to write and Client2 wants to write in different RAM location at same time.

25. Client1 wants to read and write in the same RAM location and Client2 also wants to read in the RAM location where Client1 has written at different time.

26. Client1 wants to read and write in the same RAM location and Client2 also wants to read in the RAM location where Client1 has written at same time.

27. Client1 wants to read and write in the same RAM location and Client2 also wants to write in the RAM location where Client1 has written at different time.

28. Client1 wants to read and write in the same RAM location and Client2 also wants to write in the RAM location where Client1 has written at same time.

29. Client2 wants to read and write in the same RAM location and Client1 also wants to write in the RAM location where Client2 has written at different time.

30. Client2 wants to read and write in the same RAM location and Client1 also wants to write in the RAM location where Client2 has written at same time.

31. Client2 wants to read and write in the same RAM location and Client1 also wants to read in the RAM location where Client2 has written at same time.

32. Client2 wants to read and write in the same RAM location and Client1 also wants to read in the RAM location where Client2 has written at different time.

33. If any client resets (RST_N=0) the system at any time.

34. If any client gives the inputs until RST_DONE is high.

### 3.3.3. VHDL Testbench

```
LIBRARY ieee;
USE ieee.std_logic_1164.ALL;
ENTITY ARBITER_TEST IS
END ARBITER_TEST;
ARCHITECTURE behavior OF ARBITER_TEST IS
-- Component Declaration for the Unit Under Test (UUT)
COMPONENT ARBITER_NEW
PORT(
RST_N : IN  std_logic;
CLOCK : IN  std_logic;
RST_DONE : OUT  std_logic;
RD_EN_C1 : IN  std_logic;
WR_EN_C1 : IN  std_logic;
RDADDR_C1 : IN  std_logic_vector(3 downto 0);
WRADDR_C1 : IN  std_logic_vector(3 downto 0);
WRDATA_C1 : IN  std_logic_vector(7 downto 0);
DATAIN_C2 : IN  std_logic_vector(7 downto 0);
REQUEST_C2 : IN  std_logic;
RD_NOT_WRITE_C2 : IN  std_logic;
ADDR_C2 : IN  std_logic_vector(3 downto 0);
RDDATA_C1 : OUT  std_logic_vector(7 downto 0);
DATAOUT_C2 : OUT  std_logic_vector(7 downto 0);
ACK_C2 : OUT  std_logic;
RD_EN : OUT  std_logic;
WR_EN : OUT  std_logic;
WR_ADDR : OUT  std_logic_vector(3 downto 0);
RD_ADDR : OUT  std_logic_vector(3 downto 0);
WR_DATA : OUT  std_logic_vector(7 downto 0);
RD_DATA : IN  std_logic_vector(7 downto 0)
);
END COMPONENT;
--Inputs
signal RST_N : std_logic := '0';
signal CLOCK : std_logic := '0';
signal RD_EN_C1 : std_logic := '0';
signal WR_EN_C1 : std_logic := '0';
signal RDADDR_C1 : std_logic_vector(3 downto 0) := (others => '0');
signal WRADDR_C1 : std_logic_vector(3 downto 0) := (others => '0');
signal WRDATA_C1 : std_logic_vector(7 downto 0) := (others => '0');
signal DATAIN_C2 : std_logic_vector(7 downto 0) := (others => '0');
signal REQUEST_C2 : std_logic := '0';
signal RD_NOT_WRITE_C2 : std_logic := '0';
signal ADDR_C2 : std_logic_vector(3 downto 0) := (others => '0');
signal RD_DATA : std_logic_vector(7 downto 0) := (others => '0');
--Outputs
signal RST_DONE : std_logic;
signal RDDATA_C1 : std_logic_vector(7 downto 0);
signal DATAOUT_C2 : std_logic_vector(7 downto 0);
signal ACK_C2 : std_logic;
```

```
signal RD_EN : std_logic;
signal WR_EN : std_logic;
signal WR_ADDR : std_logic_vector(3 downto 0);
signal RD_ADDR : std_logic_vector(3 downto 0);
signal WR_DATA : std_logic_vector(7 downto 0);
-- Clock period definitions
constant CLOCK_period : time := 100 ns;
BEGIN
-- Instantiate the Unit Under Test (UUT)
uut: ARBITER_NEW PORT MAP (
RST_N => RST_N,
CLOCK => CLOCK,
RST_DONE => RST_DONE,
RD_EN_C1 => RD_EN_C1,
WR_EN_C1 => WR_EN_C1,
RDADDR_C1 => RDADDR_C1,
WRADDR_C1 => WRADDR_C1,
WRDATA_C1 => WRDATA_C1,
DATAIN_C2 => DATAIN_C2,
REQUEST_C2 => REQUEST_C2,
RD_NOT_WRITE_C2 => RD_NOT_WRITE_C2,
ADDR_C2 => ADDR_C2,
RDDATA_C1 => RDDATA_C1,
DATAOUT_C2 => DATAOUT_C2,
ACK_C2 => ACK_C2,
RD_EN => RD_EN,
WR_EN => WR_EN,
WR_ADDR => WR_ADDR,
RD_ADDR => RD_ADDR,
WR_DATA => WR_DATA,
RD_DATA => RD_DATA
);
-- Clock process definitions
CLOCK_process :process
begin
CLOCK <= '0';
wait for CLOCK_period/2;
CLOCK <= '1';
wait for CLOCK_period/2;
end process;
-- Stimulus process
stim_proc: process
begin
-- hold reset state for 100 ns.
wait for 100 ns;
-------------------------------------------------------------------------------------------
----Test Case-1:  Only Client1 wants to write
-------------------------------------------------------------------------------------------
RST_N<='1';
wait for 500 ns;
WR_EN_C1<= '1';
```

```
WRADDR_C1 <="1010";
WRDATA_C1 <="10100011";
```
----------------------------------------------------------------------------------------------------
----------------------------------------------------------------------------------------------------
----Test Case-2:  Only Client1 wants to read
----------------------------------------------------------------------------------------------------
```
RST_N<='1';
wait for 500 ns;
WR_EN_C1<= '1';
WRADDR_C1 <="1010";
WRDATA_C1 <="10100011";
wait for 1700 ns;
WR_EN_C1<= '0';
RD_EN_C1 <= '1';
RDADDR_C1 <="1010";
```
----------------------------------------------------------------------------------------------------
----------------------------------------------------------------------------------------------------
----Test Case-3:  Only Client2 wants to write
----------------------------------------------------------------------------------------------------
```
RST_N<='1';
WR_EN_C1<= '0';
REQUEST_C2<= '1';
RD_NOT_WRITE_C2<= '0';
ADDR_C2 <="1110";
DATAIN_C2 <="11100011";
```
----------------------------------------------------------------------------------------------------
----------------------------------------------------------------------------------------------------
----Test Case-4:  Only Client2 wants to read
----------------------------------------------------------------------------------------------------
```
RST_N<='1';
WR_EN_C1<= '0';
REQUEST_C2<= '1';
RD_NOT_WRITE_C2<= '0';
ADDR_C2 <="1110";
DATAIN_C2 <="11100011";
wait for 1700 ns;
WR_EN_C1<= '0';
RD_NOT_WRITE_C2<= '1';
ADDR_C2 <="1110";
```
----------------------------------------------------------------------------------------------------
----------------------------------------------------------------------------------------------------
----Test Case-5:  Client1 wants to read and write in different RAM location at same time
----------------------------------------------------------------------------------------------------
```
RST_N<='1';
wait for 500 ns;
WR_EN_C1<= '1';
WRADDR_C1 <="1010";
WRDATA_C1 <="10100011";
wait for 1700 ns;
RD_EN_C1<= '1';
RDADDR_C1 <="1010";
```

WRADDR_C1 <="1110";
WRDATA_C1 <="10111011";
--------------------------------------------------------------------------------------------------------------
--------------------------------------------------------------------------------------------------------------
----Test Case-6:  Client1 wants to read and write in different RAM location at different time
--------------------------------------------------------------------------------------------------------------
RST_N<='1';
wait for 500 ns;
WR_EN_C1<= '1';
WRADDR_C1 <="1010";
WRDATA_C1 <="10100011";
wait for 1700 ns;
RD_EN_C1<= '1';
RDADDR_C1 <="1010";
wait for 500 ns;
WRADDR_C1 <="1110";
WRDATA_C1 <="10111011";
--------------------------------------------------------------------------------------------------------------
--------------------------------------------------------------------------------------------------------------
----Test Case-7:  Client1 wants to read and write in same RAM location at same time
--------------------------------------------------------------------------------------------------------------
RST_N<='1';
wait for 500 ns;
WR_EN_C1<= '1';
WRADDR_C1 <="1010";
WRDATA_C1 <="10100011";
wait for 1700 ns;
RD_EN_C1<= '1';
RDADDR_C1 <="1010";
WRADDR_C1 <="1010";
WRDATA_C1 <="10111011";
--------------------------------------------------------------------------------------------------------------
--------------------------------------------------------------------------------------------------------------
----Test Case-8:  Client1 wants to read and write in same RAM location at different time
--------------------------------------------------------------------------------------------------------------
RST_N<='1';
wait for 500 ns;
WR_EN_C1<= '1';
WRADDR_C1 <="1010";
WRDATA_C1 <="10100011";
wait for 1700 ns;
RD_EN_C1<= '1';
RDADDR_C1 <="1010";
wait for 500 ns;
WRADDR_C1 <="1010";
WRDATA_C1 <="10111011";
--------------------------------------------------------------------------------------------------------------
--------------------------------------------------------------------------------------------------------------
----Test Case-9:  Client2 wants to read and write in different RAM location at same time.
--------------------------------------------------------------------------------------------------------------
RST_N<='1';

```
WR_EN_C1<= '0';
RD_EN_C1<= '0';
REQUEST_C2<= '1';
RD_NOT_WRITE_C2<= '0';
ADDR_C2 <="1010";
DATAIN_C2 <="11100011";
wait for 1700 ns;
RD_NOT_WRITE_C2<= '0';
ADDR_C2 <="1001";
DATAIN_C2 <="00100011";
RD_NOT_WRITE_C2<= '1';
ADDR_C2 <="1010";
```
--------------------------------------------------------------------------------------------------------------
--------------------------------------------------------------------------------------------------------------
----Test Case-10:  Client2 wants to read and write in different RAM location at different time.
--------------------------------------------------------------------------------------------------------------
```
RST_N<='1';
WR_EN_C1<= '0';
RD_EN_C1<= '0';
REQUEST_C2<= '1';
RD_NOT_WRITE_C2<= '0';
ADDR_C2 <="1010";
DATAIN_C2 <="11100011";
wait for 1700 ns;
RD_NOT_WRITE_C2<= '0';
ADDR_C2 <="1001";
DATAIN_C2 <="00100011";
wait for 500 ns;
RD_NOT_WRITE_C2<= '1';
ADDR_C2 <="1010";
```
--------------------------------------------------------------------------------------------------------------
--------------------------------------------------------------------------------------------------------------
----Test Case-11:  Client2 wants to read and write in same RAM location at same time.
--------------------------------------------------------------------------------------------------------------
```
RST_N<='1';
WR_EN_C1<= '0';
RD_EN_C1<= '0';
REQUEST_C2<= '1';
RD_NOT_WRITE_C2<= '0';
ADDR_C2 <="1010";
DATAIN_C2 <="11100011";
wait for 1700 ns;
RD_NOT_WRITE_C2<= '1';
ADDR_C2 <="1010";
RD_NOT_WRITE_C2<= '0';
ADDR_C2 <="1010";
DATAIN_C2 <="00100011";
```
--------------------------------------------------------------------------------------------------------------

-------------------------------------------------------------------------------
----Test Case-12:  Client2 wants to read and write in same RAM location at different time.
-------------------------------------------------------------------------------
RST_N<='1';
WR_EN_C1<= '0';
RD_EN_C1<= '0';
REQUEST_C2<= '1';
RD_NOT_WRITE_C2<= '0';
ADDR_C2 <="1010";
DATAIN_C2 <="11100011";
wait for 1700 ns;
RD_NOT_WRITE_C2<= '1';
ADDR_C2 <="1010";
wait for 500 ns;
RD_NOT_WRITE_C2<= '0';
ADDR_C2 <="1010";
DATAIN_C2 <="00100011";
-------------------------------------------------------------------------------
-------------------------------------------------------------------------------
----Test Case-13:  Client1 wants to write and client2 wants to read in different RAM location at same time
-------------------------------------------------------------------------------
RST_N<='1';
WR_EN_C1<= '0';
REQUEST_C2<= '1';
RD_NOT_WRITE_C2<= '0';
ADDR_C2 <="1110";
DATAIN_C2 <="11100011";
wait for 1700 ns;
WR_EN_C1<= '1';
RD_EN_C1<= '0';
WRADDR_C1 <="1001";
WRDATA_C1 <="10111011";
REQUEST_C2<= '1';
RD_NOT_WRITE_C2<= '1';
ADDR_C2 <="1110";
-------------------------------------------------------------------------------
-------------------------------------------------------------------------------
----Test Case-14:  Client1 wants to write and client2 wants to read in different RAM location at different time
-------------------------------------------------------------------------------
RST_N<='1';
WR_EN_C1<= '0';
REQUEST_C2<= '1';
RD_NOT_WRITE_C2<= '0';
ADDR_C2 <="1110";
DATAIN_C2 <="11100011";
wait for 1700 ns;
WR_EN_C1<= '1';
RD_EN_C1<= '0';
WRADDR_C1 <="1001";

```
WRDATA_C1 <="10111011";
wait for 500 ns;
REQUEST_C2<= '1';
RD_NOT_WRITE_C2<= '1';
ADDR_C2 <="1110";
```
-------------------------------------------------------------------------------------------------------
-------------------------------------------------------------------------------------------------------
----Test Case-15: Client1 wants to write and client2 wants to read in same RAM location at same time
-------------------------------------------------------------------------------------------------------
```
RST_N<='1';
WR_EN_C1<= '0';
REQUEST_C2<= '1';
RD_NOT_WRITE_C2<= '0';
ADDR_C2 <="1110";
DATAIN_C2 <="11100011";
wait for 1700 ns;
WR_EN_C1<= '1';
RD_EN_C1<= '0';
WRADDR_C1 <="1110";
WRDATA_C1 <="10111011";
REQUEST_C2<= '1';
RD_NOT_WRITE_C2<= '1';
ADDR_C2 <="1110";
```
-------------------------------------------------------------------------------------------------------
-------------------------------------------------------------------------------------------------------
----Test Case-16: Client1 wants to write and client2 wants to read in same RAM location at different time
-------------------------------------------------------------------------------------------------------
```
RST_N<='1';
WR_EN_C1<= '0';
REQUEST_C2<= '1';
RD_NOT_WRITE_C2<= '0';
ADDR_C2 <="1110";
DATAIN_C2 <="11100011";
wait for 1700 ns;
WR_EN_C1<= '1';
RD_EN_C1<= '0';
WRADDR_C1 <="1110";
WRDATA_C1 <="10111011";
wait for 500 ns;
REQUEST_C2<= '1';
RD_NOT_WRITE_C2<= '1';
ADDR_C2 <="1110";
```
-------------------------------------------------------------------------------------------------------
-------------------------------------------------------------------------------------------------------
----Test Case-17: Client1 wants to read and Client2 wants to write in same RAM location at same time
-------------------------------------------------------------------------------------------------------
```
RST_N <='1';
WR_EN_C1<= '1';
```

```
RD_EN_C1<= '0';
WRADDR_C1<="1010";
WRDATA_C1<="10101111";
wait for 1700 ns;
RD_EN_C1<='1';
WR_EN_C1<='0';
RDADDR_C1<="1010";
REQUEST_C2<='1';
RD_NOT_WRITE_C2<= '0';
ADDR_C2<="1010";
DATAIN_C2<="10111011";
```
-------------------------------------------------------------------------------------------------------
-------------------------------------------------------------------------------------------------------
----Test Case-18:  Client1 wants to read and Client2 wants to write in same RAM location at different time
-------------------------------------------------------------------------------------------------------
```
RST_N <='1';
WR_EN_C1<= '1';
RD_EN_C1<= '0';
WRADDR_C1<="1010";
WRDATA_C1<="10101111";
wait for 1700 ns;
RD_EN_C1<='1';
WR_EN_C1<='0';
RDADDR_C1<="1010";
wait for 500 ns;
REQUEST_C2<='1';
RD_NOT_WRITE_C2<= '0';
ADDR_C2<="1010";
DATAIN_C2<="10111011";
```
-------------------------------------------------------------------------------------------------------
-------------------------------------------------------------------------------------------------------
----Test Case-19:  Client1 wants to read and Client2 wants to write in different RAM location at same time
-------------------------------------------------------------------------------------------------------
```
RST_N <='1';
WR_EN_C1<= '1';
RD_EN_C1<= '0';
WRADDR_C1<="1000";
WRDATA_C1<="10101111";
wait for 1700 ns;
RD_EN_C1<='1';
WR_EN_C1<='0';
RDADDR_C1<="1000";
REQUEST_C2<='1';
RD_NOT_WRITE_C2<= '0';
ADDR_C2<="1010";
DATAIN_C2<="10100011";
```
-------------------------------------------------------------------------------------------------------

-------------------------------------------------------------------------------------------------------
----Test Case-20:  Client1 wants to read and Client2 wants to write in different RAM location at
different time
-------------------------------------------------------------------------------------------------------
RST_N <='1';
WR_EN_C1<= '1';
RD_EN_C1<= '0';
WRADDR_C1<="1000";
WRDATA_C1<="10101111";
wait for 1700 ns;
RD_EN_C1<='1';
WR_EN_C1<='0';
RDADDR_C1<="1000";
wait for 500 ns;
REQUEST_C2<='1';
RD_NOT_WRITE_C2<= '0';
ADDR_C2<="1010";
DATAIN_C2<="10100011";
-------------------------------------------------------------------------------------------------------
-------------------------------------------------------------------------------------------------------
----Test Case-21:  Client1 wants to read and Client2 wants to read in same RAM location at
different time.
-------------------------------------------------------------------------------------------------------
RST_N <='1';
WR_EN_C1<= '1';
RD_EN_C1<= '0';
WRADDR_C1<="1010";
WRDATA_C1<="10101111";
wait for 1700 ns;
RD_EN_C1<='1';
WR_EN_C1<='0';
RDADDR_C1<="1010";
wait for 300 ns;
REQUEST_C2<='1';
RD_NOT_WRITE_C2<='1';
ADDR_C2<="1010";
wait for 200 ns;
RD_EN_C1<='0';
-------------------------------------------------------------------------------------------------------
-------------------------------------------------------------------------------------------------------
----Test Case-22:  Client1 wants to read and Client2 wants to read in same RAM location at same
time.
-------------------------------------------------------------------------------------------------------
RST_N <='1';
WR_EN_C1<= '1';
RD_EN_C1<= '0';
WRADDR_C1<="1010";
WRDATA_C1<="10101111";
wait for 1700 ns;
RD_EN_C1<='1';
WR_EN_C1<='0';

RDADDR_C1<="1010";
REQUEST_C2<='1';
RD_NOT_WRITE_C2<='1';
ADDR_C2<="1010";
-------------------------------------------------------------------------------------------------------------
-------------------------------------------------------------------------------------------------------------
----Test Case-23:  Client1 wants to read and Client2 wants to read in different RAM location at
different time.
-------------------------------------------------------------------------------------------------------------
RST_N <='1';
WR_EN_C1<= '1';
RD_EN_C1<= '0';
WRADDR_C1<="1001";
WRDATA_C1<="10101111";
wait for 1700 ns;
RD_EN_C1<='1';
WR_EN_C1<='0';
RDADDR_C1<="1001";
wait for 300 ns;
REQUEST_C2<='1';
RD_NOT_WRITE_C2<='1';
ADDR_C2<="1010";
wait for 200 ns;
RD_EN_C1<='0';
-------------------------------------------------------------------------------------------------------------
-------------------------------------------------------------------------------------------------------------
----Test Case-24:  Client1 wants to read and Client2 wants to read in different RAM location at
same time.
-------------------------------------------------------------------------------------------------------------
RST_N <='1';
WR_EN_C1<= '1';
RD_EN_C1<= '0';
WRADDR_C1<="1001";
WRDATA_C1<="10101111";
wait for 1700 ns;
RD_EN_C1<='1';
WR_EN_C1<='0';
RDADDR_C1<="1001";
REQUEST_C2<='1';
RD_NOT_WRITE_C2<='1';
ADDR_C2<="1010";
-------------------------------------------------------------------------------------------------------------
-------------------------------------------------------------------------------------------------------------
----Test Case-25:  Client1 wants to read and write in the same RAM location and Client2 also
wants to read in the RAM location where Client1 has written at different time.
-------------------------------------------------------------------------------------------------------------
RST_N <='1';
WR_EN_C1<= '1';
RD_EN_C1<= '0';
WRADDR_C1<="1001";
WRDATA_C1<="10101111";

wait for 1700 ns;
RD_EN_C1<='1';
RDADDR_C1<="1001";
WRADDR_C1<="1001";
WRDATA_C1<="10100011";
wait for 300 ns;
REQUEST_C2<='1';
RD_NOT_WRITE_C2<='1';
ADDR_C2<="1001";
wait for 200 ns;
RD_EN_C1<='0';
-------------------------------------------------------------------------------------------------------------
-------------------------------------------------------------------------------------------------------------
----Test Case-26:  Client1 wants to read and write in the same RAM location and Client2 also wants to read in the RAM location where Client1 has written at same time.
-------------------------------------------------------------------------------------------------------------
RST_N <='1';
WR_EN_C1<= '1';
RD_EN_C1<= '0';
WRADDR_C1<="1001";
WRDATA_C1<="10101111";
wait for 1700 ns;
RD_EN_C1<='1';
RDADDR_C1<="1001";
WRADDR_C1<="1001";
WRDATA_C1<="10100011";
REQUEST_C2<='1';
RD_NOT_WRITE_C2<='1';
ADDR_C2<="1001";
-------------------------------------------------------------------------------------------------------------
-------------------------------------------------------------------------------------------------------------
----Test Case-27:  Client1 wants to read and write in the same RAM location and Client2 also wants to write in the RAM location where Client1 has written at different time.
-------------------------------------------------------------------------------------------------------------
RST_N <='1';
WR_EN_C1<= '1';
WRADDR_C1<="1001";
WRDATA_C1<="10101111";
wait for 1700 ns;
RD_EN_C1<='1';
RDADDR_C1<="1001";
WRADDR_C1<="1001";
WRDATA_C1<="10100011";
wait for 300 ns;
REQUEST_C2<='1';
RD_NOT_WRITE_C2<='0';
ADDR_C2 <="1001";
DATAIN_C2 <="11100011";
wait for 200 ns;
WR_EN_C1<='0';
-------------------------------------------------------------------------------------------------------------

---------------------------------------------------------------------------------------------------------
----Test Case-28: Client1 wants to read and write in the same RAM location and Client2 also wants to write in the RAM location where Client1 has written at same time.
---------------------------------------------------------------------------------------------------------
RST_N <='1';
WR_EN_C1<= '1';
RD_EN_C1<= '0';
WRADDR_C1<="1001";
WRDATA_C1<="10101111";
wait for 1700 ns;
RD_EN_C1<='1';
RDADDR_C1<="1001";
WRADDR_C1<="1001";
WRDATA_C1<="10100011";
REQUEST_C2<='1';
RD_NOT_WRITE_C2<='0';
ADDR_C2 <="1001";
DATAIN_C2 <="11100011";
---------------------------------------------------------------------------------------------------------

---------------------------------------------------------------------------------------------------------
----Test Case-29: Client2 wants to read and write in the same RAM location and Client1 also wants to write in the RAM location where Client2 has written at different time.
---------------------------------------------------------------------------------------------------------
RST_N <='1';
WR_EN_C1<= '0';
RD_EN_C1<= '0';
REQUEST_C2<='1';
RD_NOT_WRITE_C2<='0';
ADDR_C2 <="1001";
DATAIN_C2 <="11100011";
wait for 1700 ns;
RD_NOT_WRITE_C2<='1';
ADDR_C2 <="1001";
wait for 300 ns;
WRADDR_C1<="1001";
WRDATA_C1<="10101111";
wait for 200 ns;
WR_EN_C1<= '1';
---------------------------------------------------------------------------------------------------------

---------------------------------------------------------------------------------------------------------
----Test Case-30: Client2 wants to read and write in the same RAM location and Client1 also wants to write in the RAM location where Client2 has written at same time.
---------------------------------------------------------------------------------------------------------
RST_N <='1';
WR_EN_C1<= '0';
RD_EN_C1<= '0';
REQUEST_C2<='1';
RD_NOT_WRITE_C2<='0';
ADDR_C2 <="1001";
DATAIN_C2 <="11100011";
wait for 1700 ns;

```
WR_EN_C1<= '1';
RD_NOT_WRITE_C2<='1';
ADDR_C2 <="1001";
WRADDR_C1<="1001";
WRDATA_C1<="10101111";
```
------------------------------------------------------------------------------------------
------------------------------------------------------------------------------------------
----Test Case-31:  Client2 wants to read and write in the same RAM location and Client1 also wants to read in RAM location where Client2 has written at same time.
------------------------------------------------------------------------------------------
```
RST_N <='1';
WR_EN_C1<= '0';
RD_EN_C1<= '0';
REQUEST_C2<='1';
RD_NOT_WRITE_C2<='0';
ADDR_C2 <="1001";
DATAIN_C2 <="11100011";
wait for 1700 ns;
RD_EN_C1<= '1';
RD_NOT_WRITE_C2<='1';
ADDR_C2 <="1001";
RDADDR_C1 <="1001";
```
------------------------------------------------------------------------------------------
------------------------------------------------------------------------------------------
----Test Case-32:  Client2 wants to read and write in the same RAM location and Client1 also wants to read in the RAM location where Client2 has written at different time.
------------------------------------------------------------------------------------------
```
RST_N <='1';
WR_EN_C1<= '0';
RD_EN_C1<= '0';
REQUEST_C2<='1';
RD_NOT_WRITE_C2<='0';
ADDR_C2 <="1001";
DATAIN_C2 <="11100011";
wait for 1700 ns;
RD_NOT_WRITE_C2<='1';
ADDR_C2 <="1001";
wait for 300 ns;
RD_EN_C1<= '1';
RDADDR_C1 <="1001";
```
------------------------------------------------------------------------------------------
------------------------------------------------------------------------------------------
----Test Case-33:  If any client resets (RST_N=0) the system at any time.
------------------------------------------------------------------------------------------
```
RST_N<='1';
WR_EN_C1<= '1';
RD_EN_C1<= '0';
WRADDR_C1<="1010";
WRDATA_C1<="10101111";
wait for 1700 ns;
RST_N<='0';
```

```
RD_EN_C1<= '1';
WR_EN_C1<='0';
RDADDR_C1<="1010";
REQUEST_C2<='1';
RD_NOT_WRITE_C2<='0';
ADDR_C2<="0110";
DATAIN_C2<="10111011";
wait for 500 ns;
RST_N<='1';
RDADDR_C1<="1010";
wait for 300 ns;
RDADDR_C1<="0110";
```
--------------------------------------------------------------------------------------------------------------
--------------------------------------------------------------------------------------------------------------
----Test Case-34:  If any client gives the inputs until RST_DONE is high.
--------------------------------------------------------------------------------------------------------------
```
RST_N<='1';
wait for 200 ns;
WR_EN_C1<= '1';
RD_EN_C1<= '0';
WRADDR_C1<="1010";
WRDATA_C1<="10101111";
RST_N<='0';
wait for 1700 ns;
RD_EN_C1<= '1';
WR_EN_C1<='0';
RDADDR_C1<="1010";
REQUEST_C2<='1';
RD_NOT_WRITE_C2<='0';
ADDR_C2<="0110";
DATAIN_C2<="10111011";
wait for 800 ns;
RST_N<='1';
RDADDR_C1<="1010";
```
--------------------------------------------------------------------------------------------------------------
```
wait;
end process;
END;
```

--------------------------------------------------------------------------------------------------------------

### 3.3.4. Waveforms

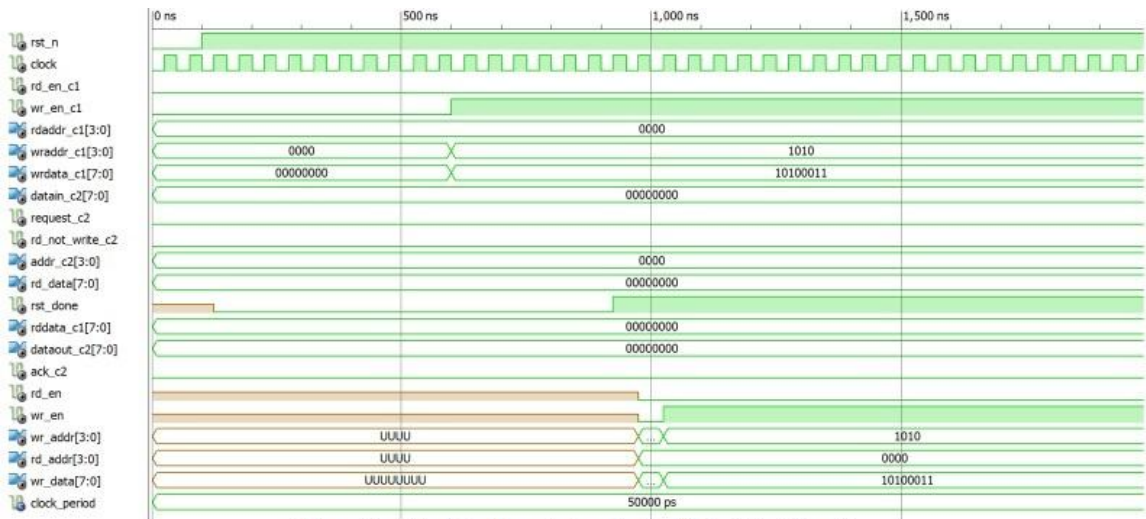

Figure 11: Only Client1 wants to write for Arbiter (Test Case 1)

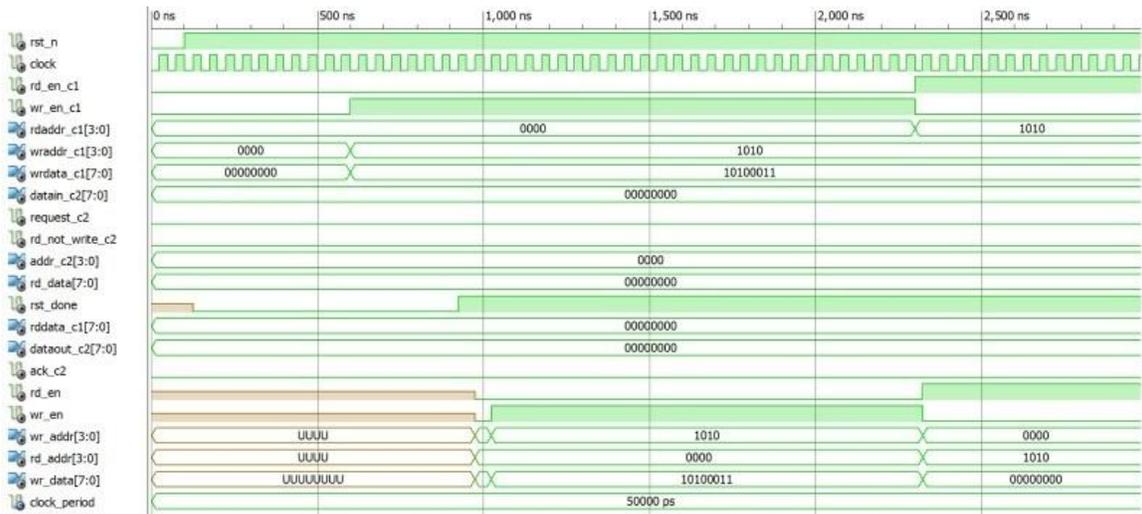

Figure 12: Only Client1 wants to read for Arbiter (Test Case 2)

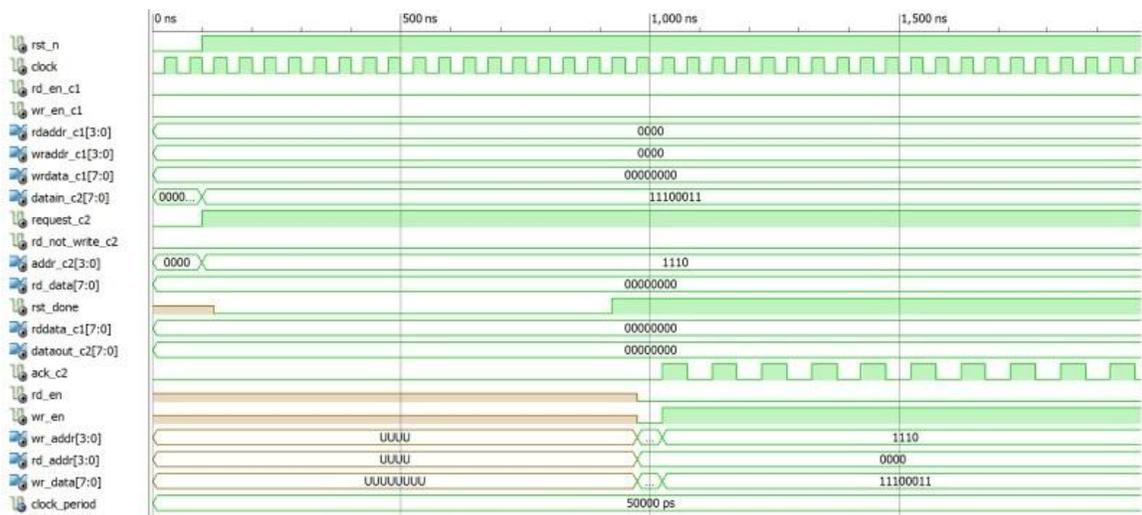

Figure 13: Only Client2 wants to write for Arbiter (Test Case 3)

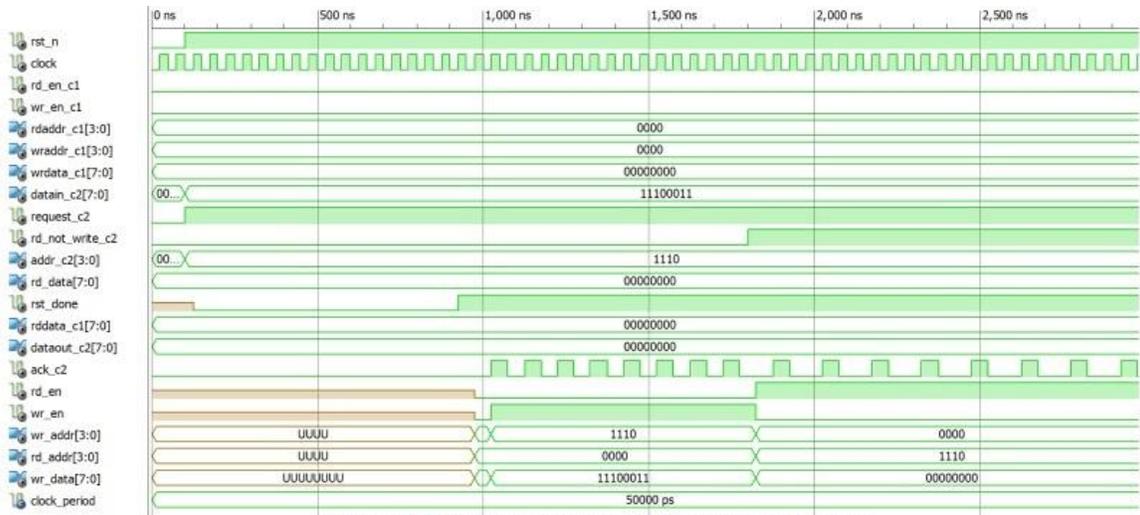

Figure 14: Only Client2 wants to read for Arbiter (Test Case 4)

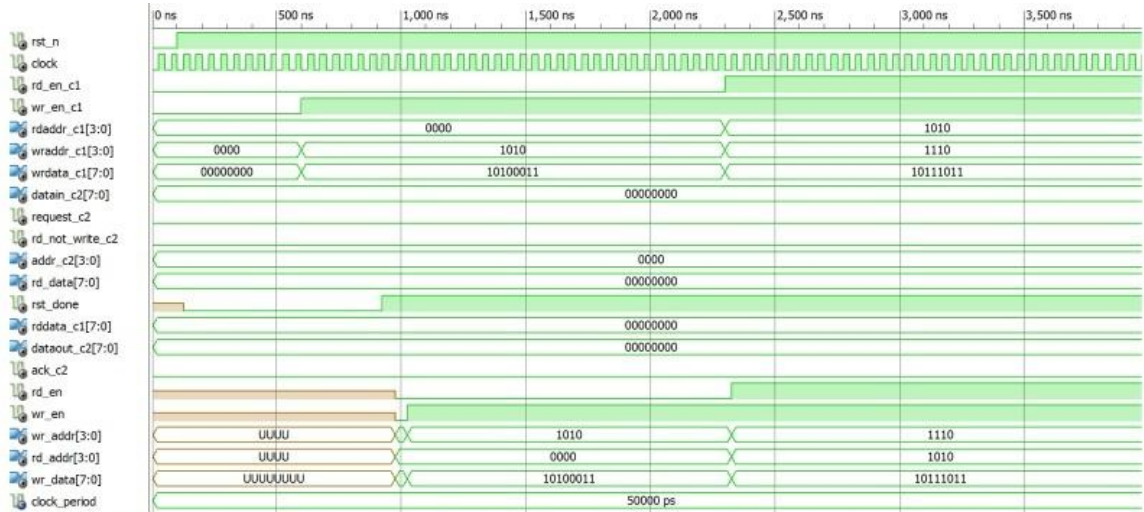

Figure 15: Client1 wants to read and write in different RAM location at same time for Arbiter (Test Case 5)

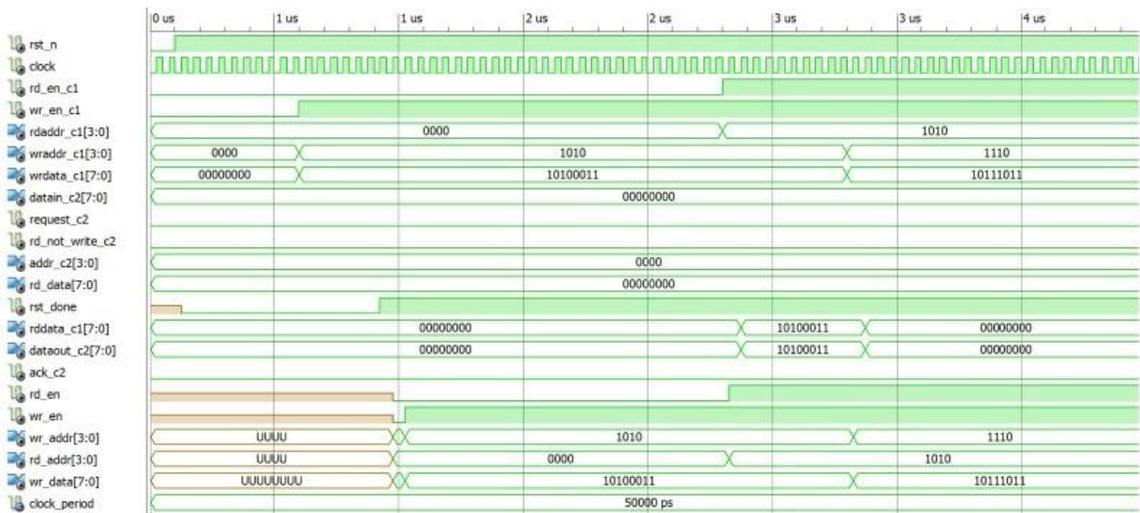

Figure 16: Client1 wants to read and write in different RAM location at different time for Arbiter (Test Case 6)

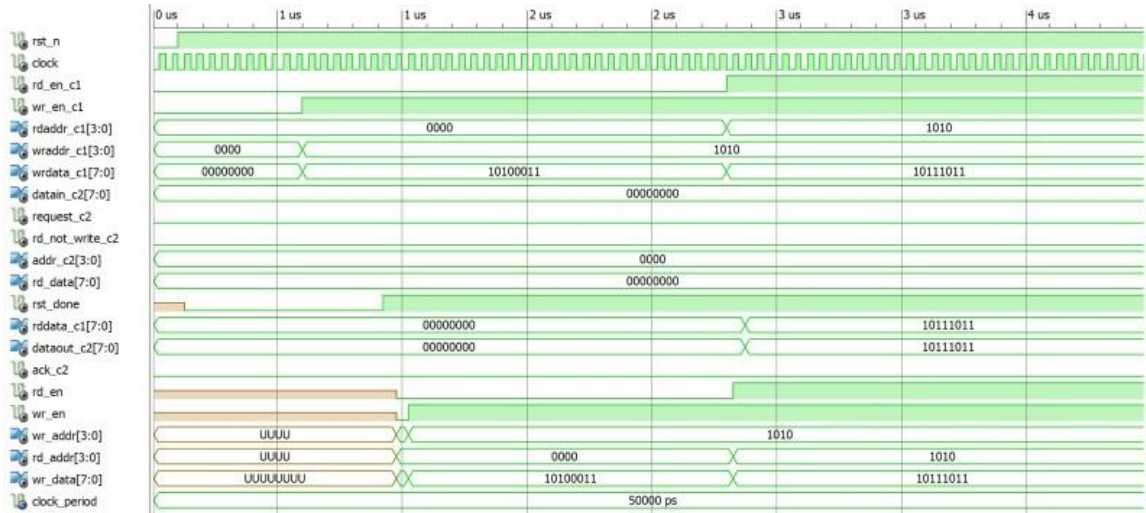

Figure 17: Client1 wants to read and write in same RAM location at same time for Arbiter (Test Case 7)

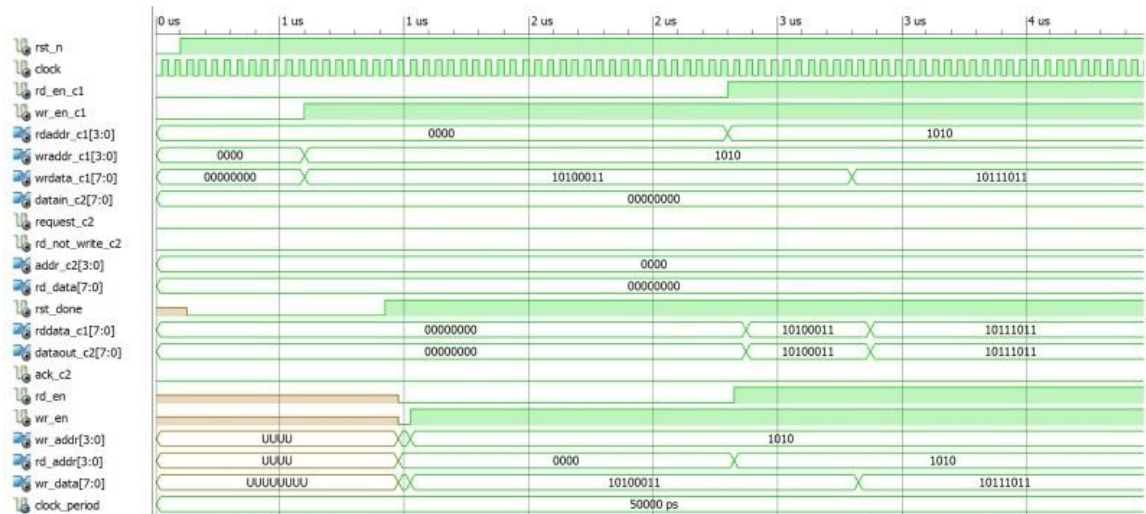

Figure 18: Client1 wants to read and write in same RAM location at different time for Arbiter (Test Case 8)

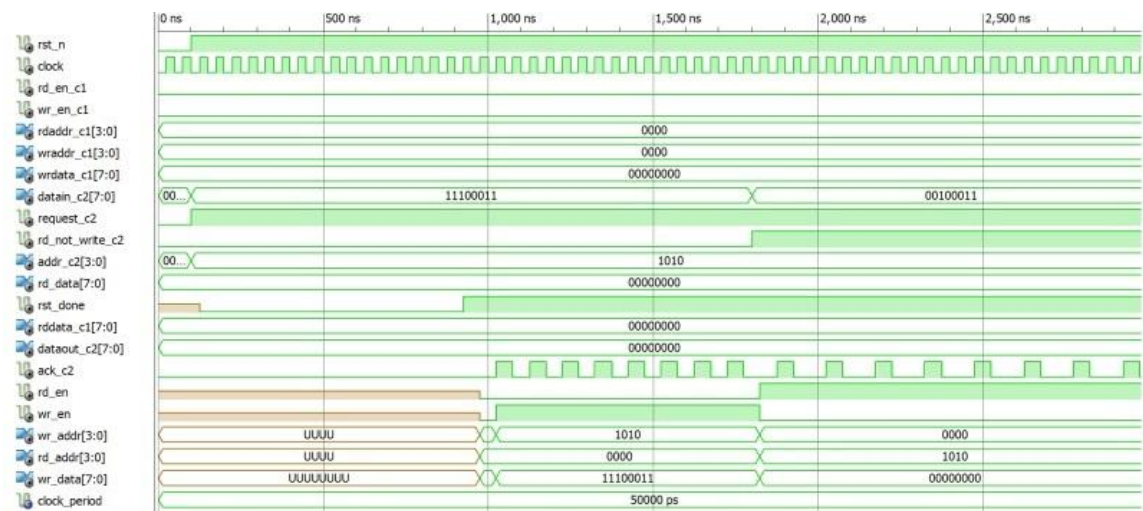

Figure 19: Client2 wants to read and write in different RAM location at same time at Arbiter (Test Case 9)

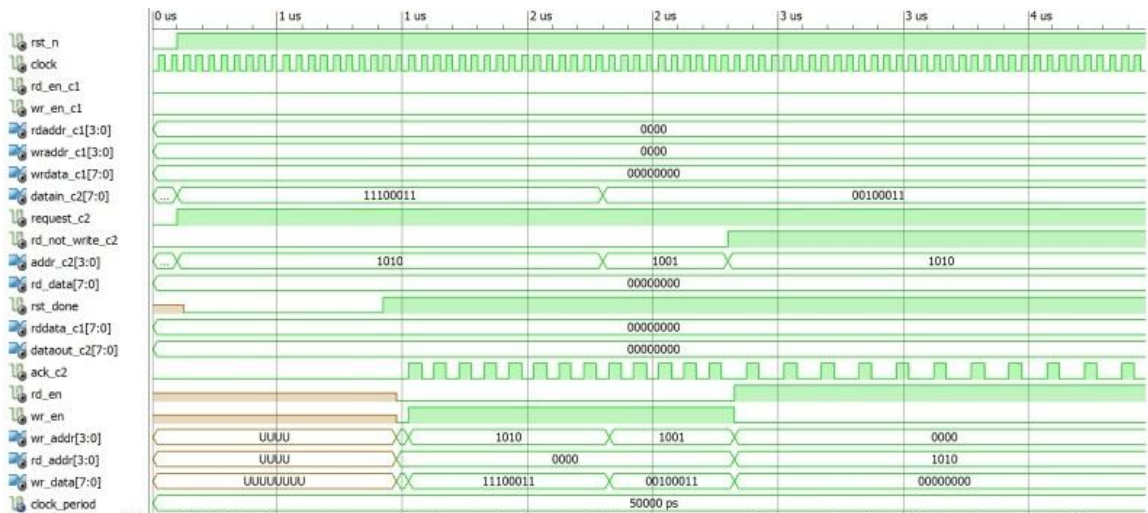

Figure 20: Client2 wants to read and write in different RAM location at different time for Arbiter (Test Case 10)

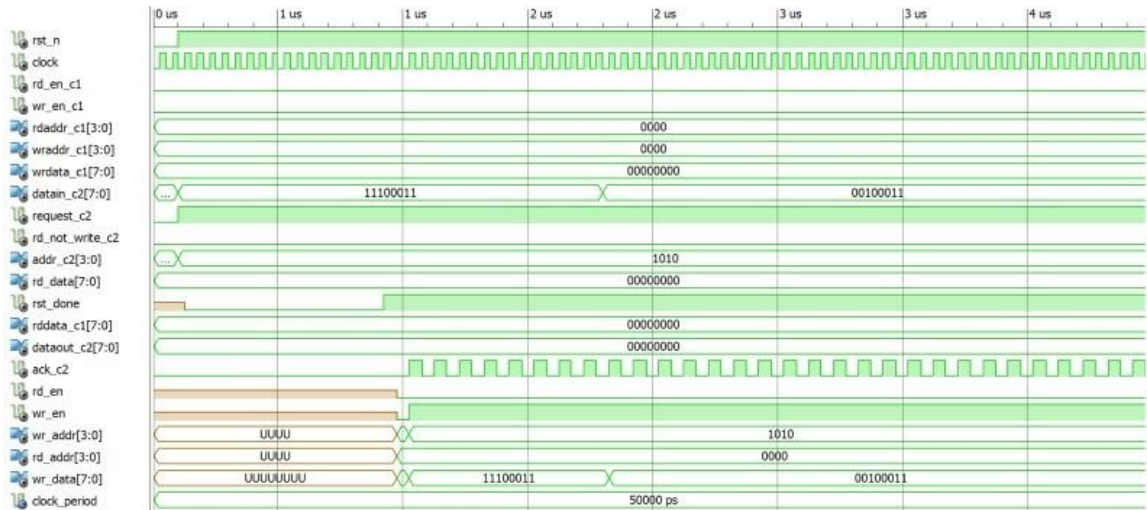

Figure 21: Client2 wants to read and write in same RAM location at same time for Arbiter (Test Case 11)

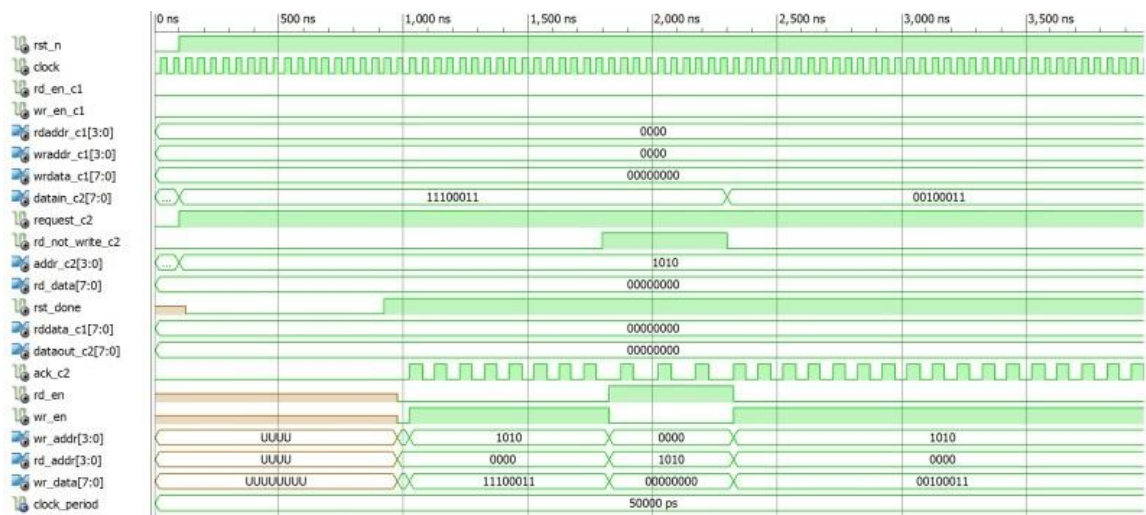

Figure 22: Client2 wants to read and write in same RAM location at different time for Arbiter (Test Case 12)

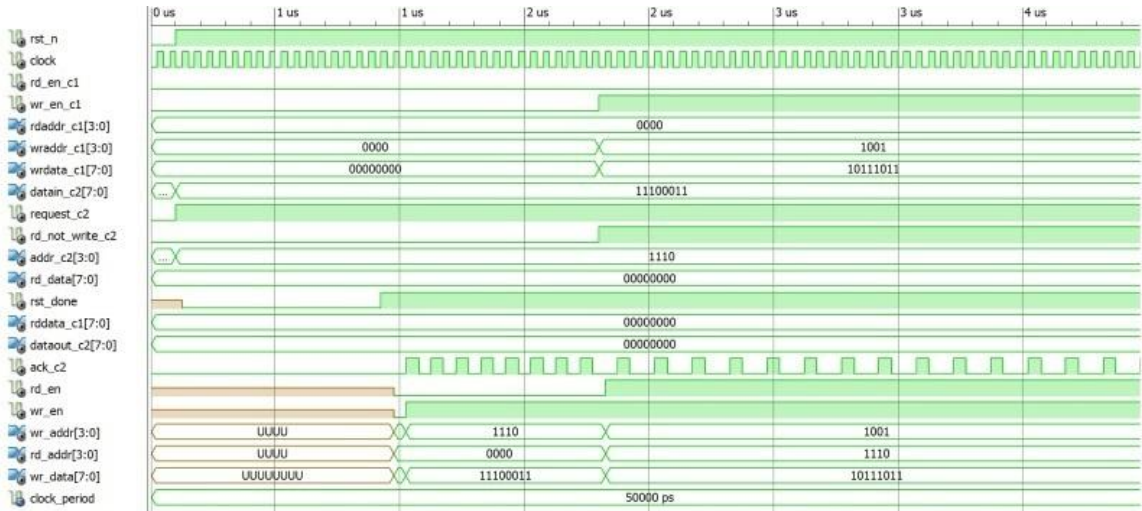

Figure 23: Client1 wants to write and Client2 wants to read in different RAM location at same time for Arbiter (Test Case 13)

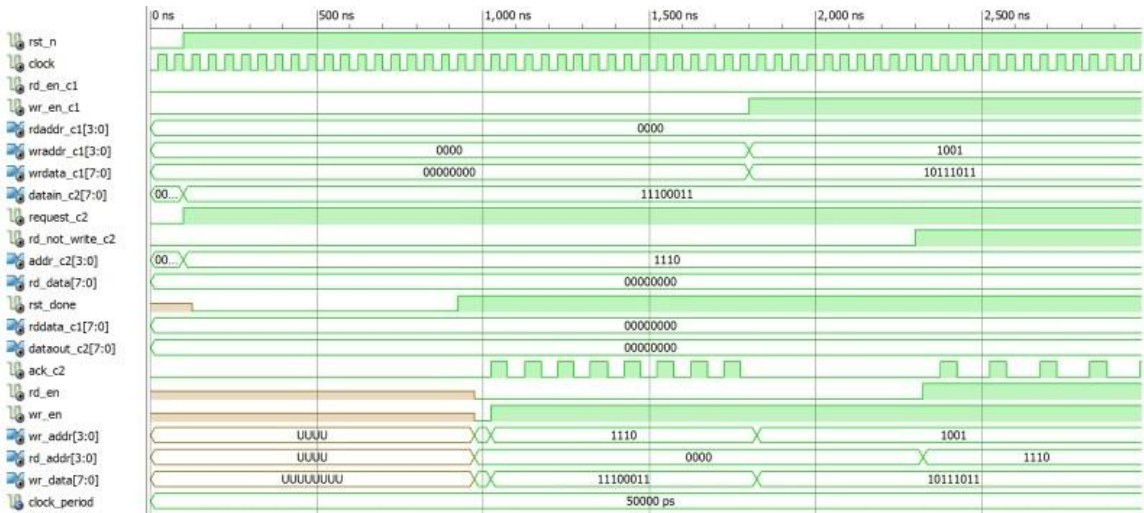

Figure 24: Client1 wants to write and client2 wants to read in different RAM location at different time for Arbiter (Test Case 14)

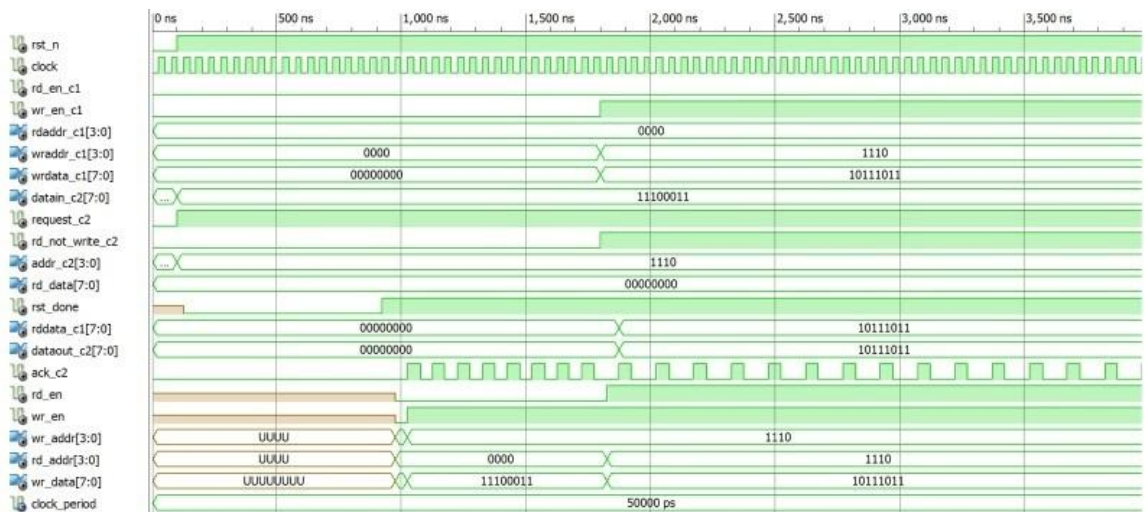

Figure 25:Client1 wants to write and Client2 wants to read in same RAM location at same time for Arbiter (Test Case 15)

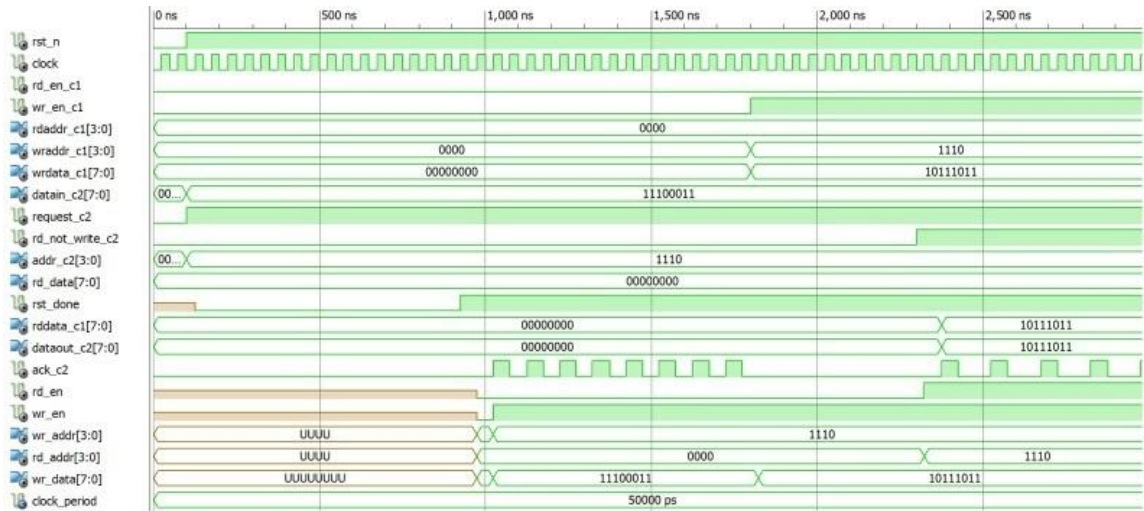

Figure 26: Client1 wants to write and Client2 wants to read in same RAM location at different time for Arbiter (Test Case 16)

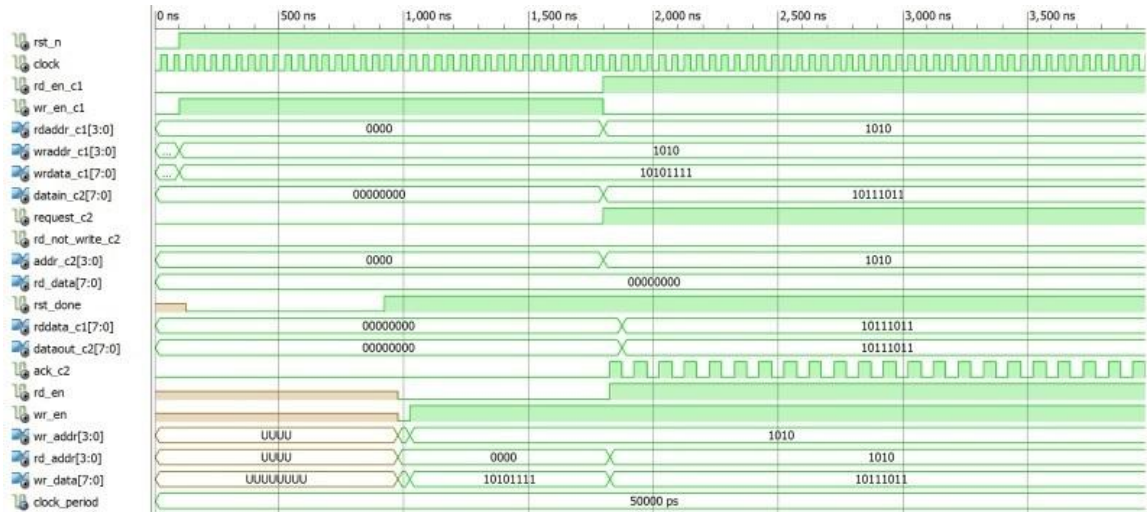

Figure 27: Client1 wants to write and Client2 wants to read in same RAM location at same time for Arbiter (Test Case 17)

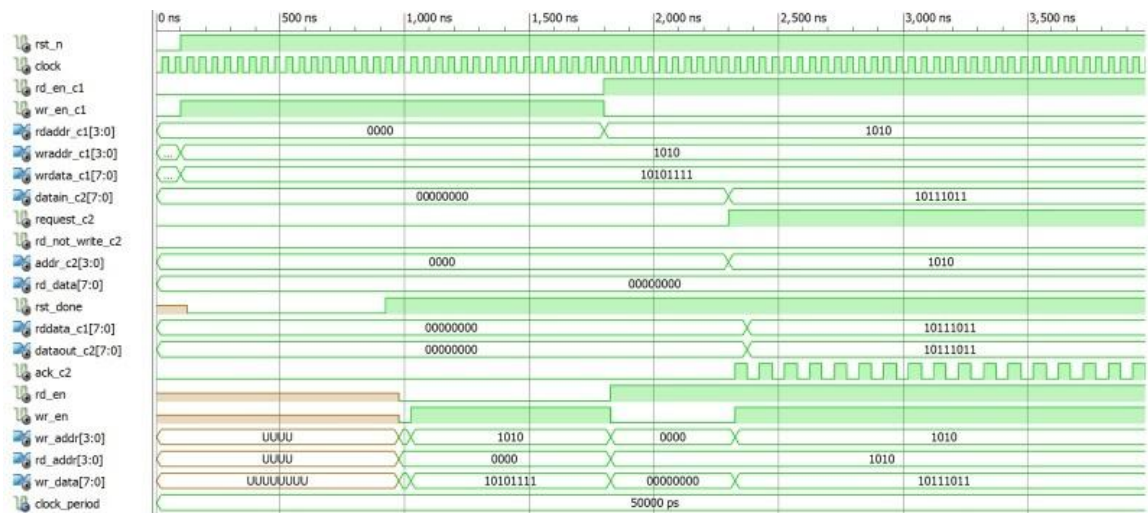

Figure 28: Client1 wants to read and Client2 wants to write in same RAM location at different time for Arbiter(Test Case 18)

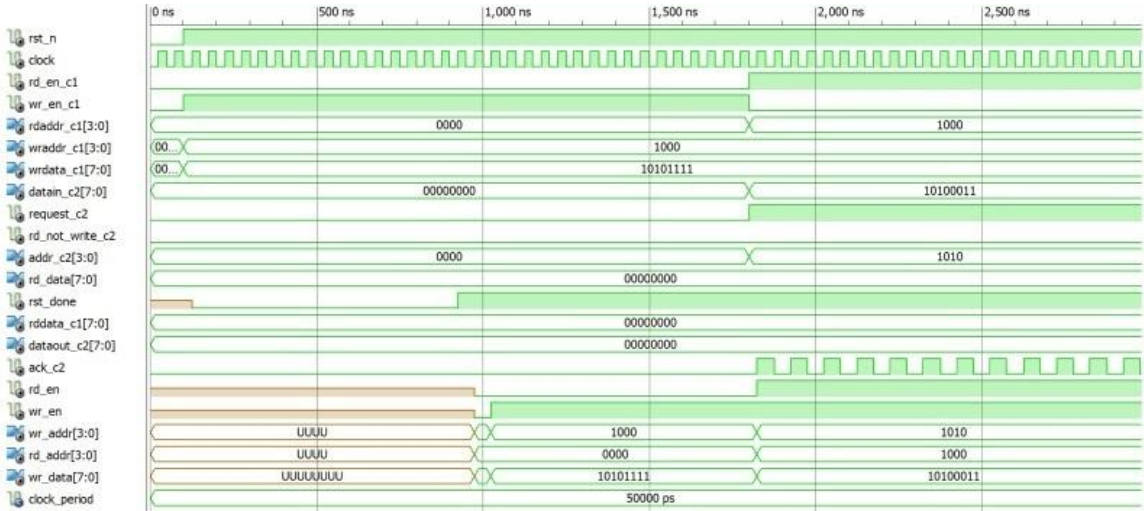

Figure 29: Client1 wants to read and Client2 wants to write in different RAM location at same time for Arbiter (Test Case 19)

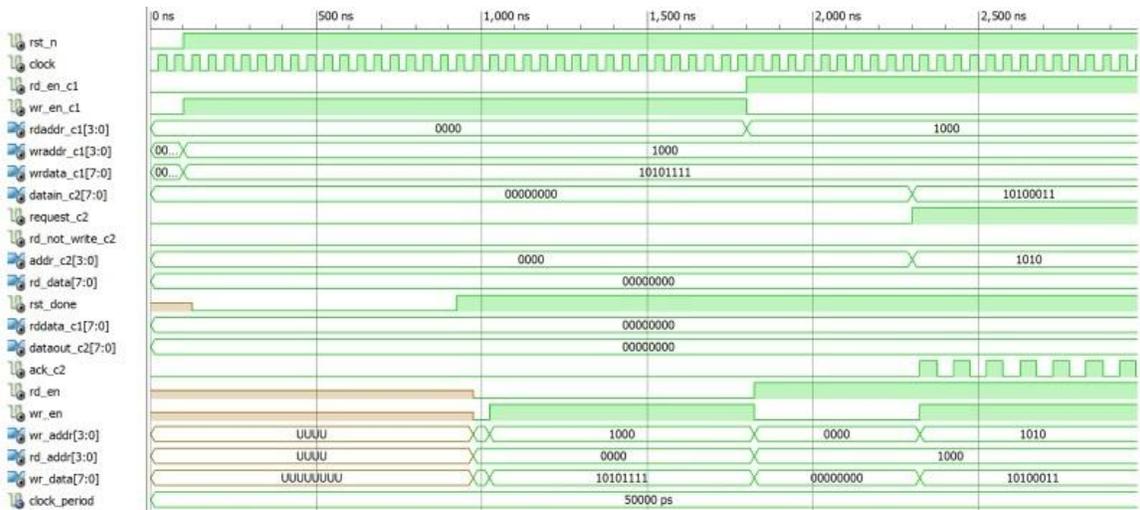

Figure 30: Client1 wants to read and Client2 wants to write in different RAM location at different time for Arbiter(Test Case 20)

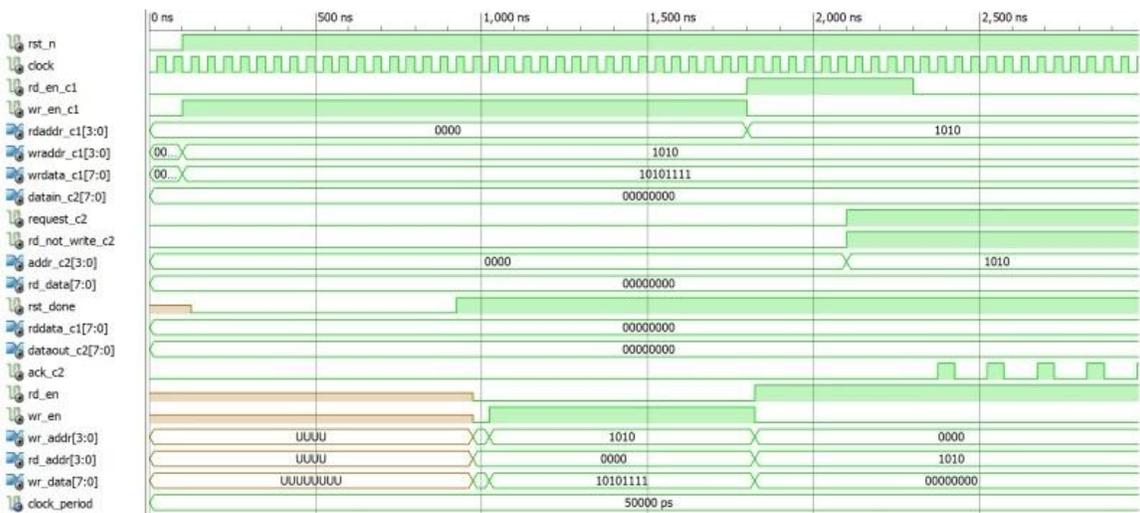

Figure 31: Client1 wants to read and Client2 wants to read in same RAM location at different time for Arbiter (Test Case 21)

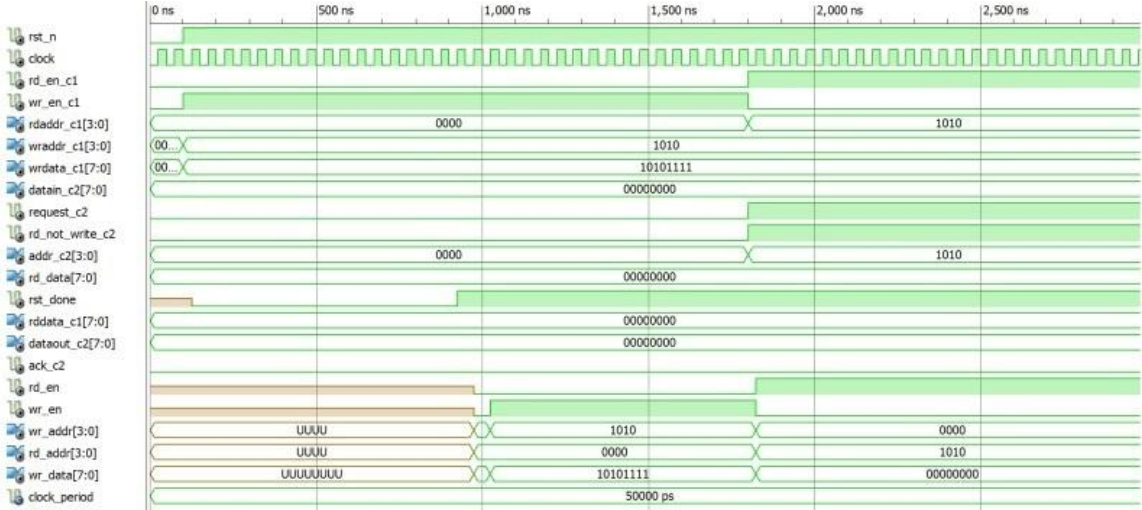

Figure 32: Client1 wants to read and Client2 wants to read in same RAM location at same time for Arbiter (Test Case 22)

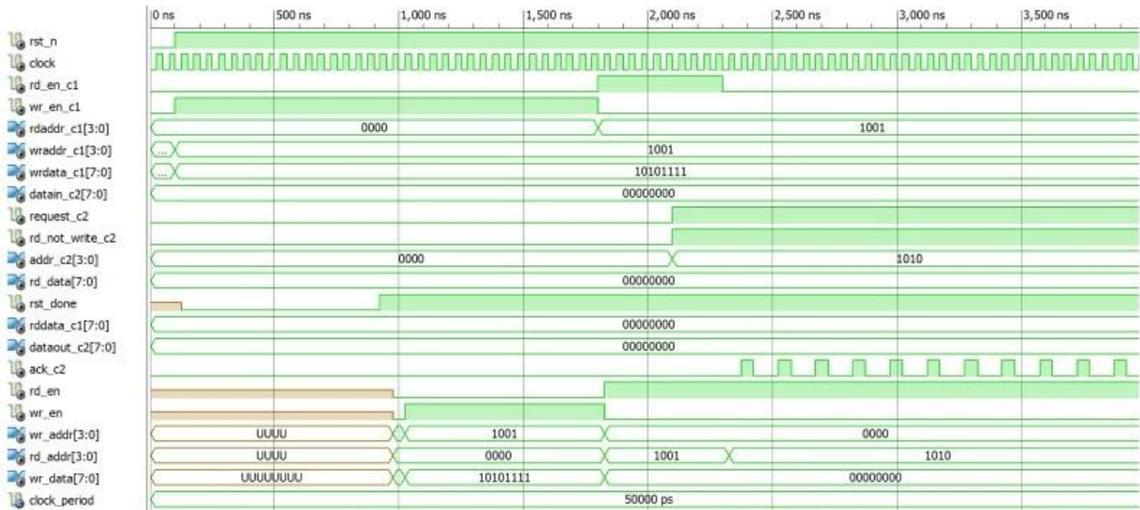

Figure 33: Client1 wants to write and Client2 wants to write in different RAM location at different time for Arbiter (Test Case 23)

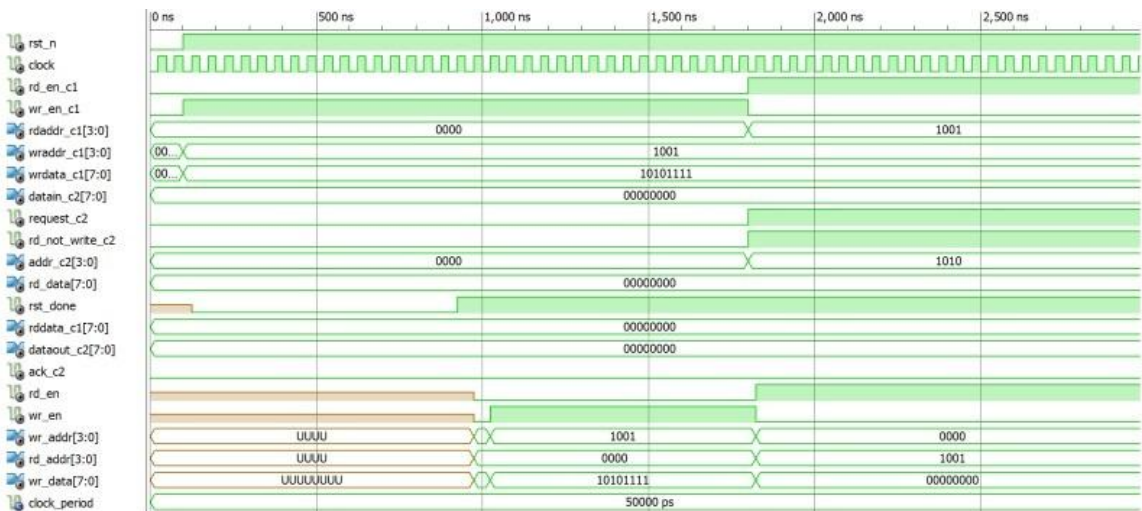

Figure 34: Client1 wants to write and Client2 wants to write in different RAM location at same time for Arbiter (Test Case 24)

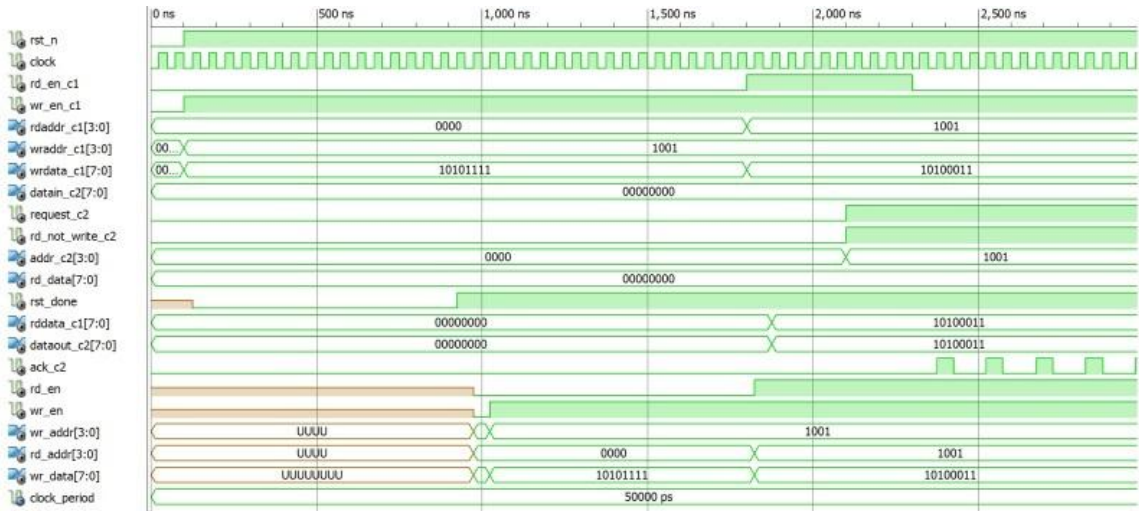

Figure 35: Client1 wants to read and write in the same RAM location and Client2 also wants to read in the RAM location where Client1 has written at different time for Arbiter (Test Case 25)

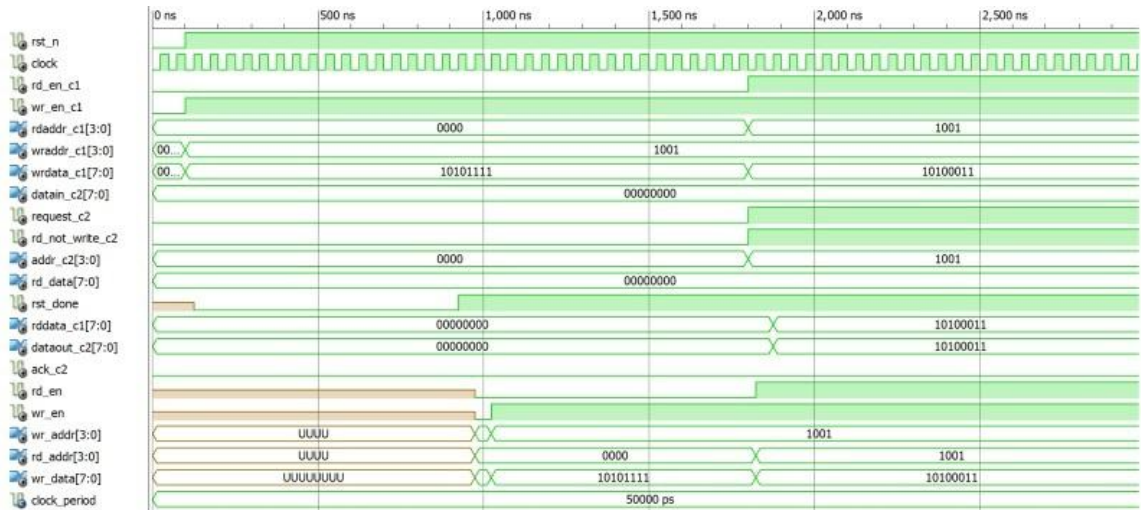

Figure 36: Client1 wants to read and write in the same RAM location and Client2 also wants to read in the RAM location where Client1 has written at same time for Arbiter (Test Case 26)

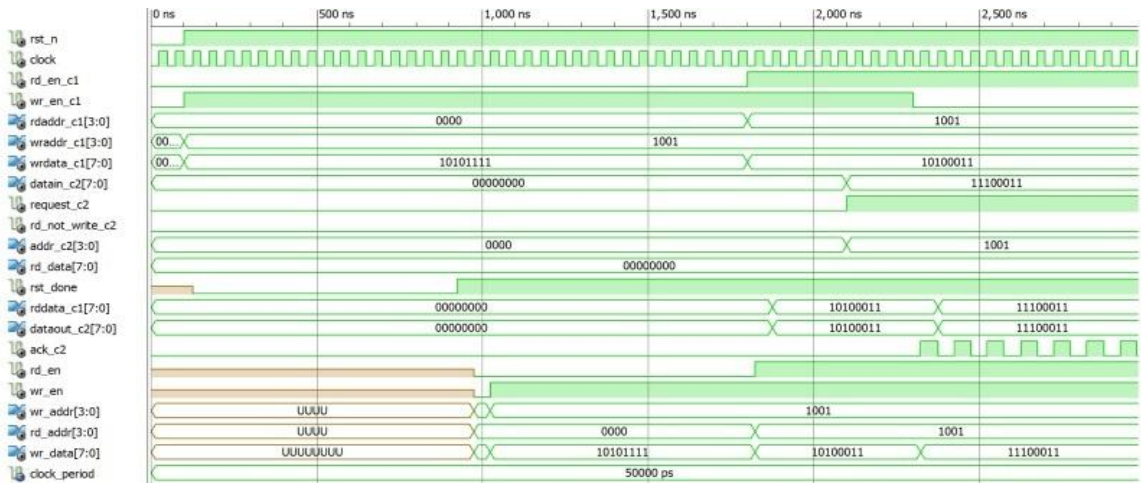

Figure 37: Client1 wants to read and write in the same RAM location and Client2 also wants to write in the RAM location where Client1 has written at different time for Arbiter (Test Case 27)

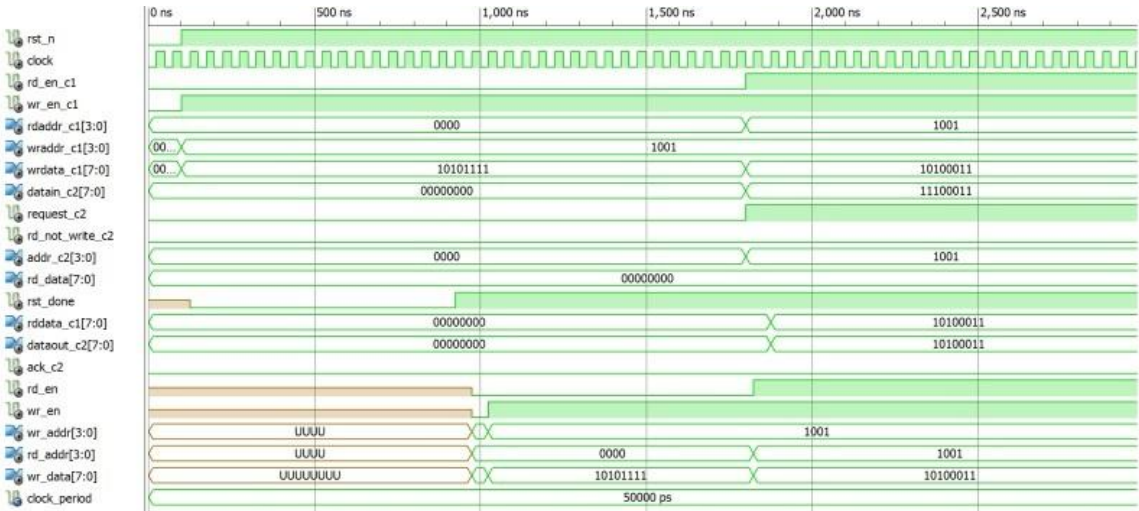

Figure 38: Client1 wants to read and write in the same RAM location and Client2 also wants to write in the RAM location where Client1 has written at same time for Arbiter (Test Case 28)

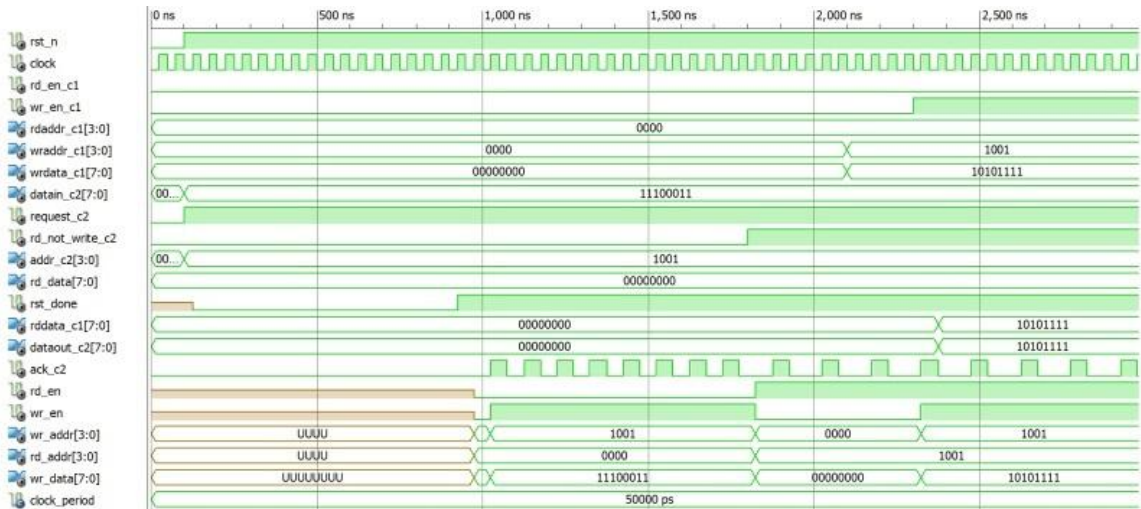

Figure 39: Client2 wants to read and write in the same RAM location and Client1 also wants to write in the RAM location where Client2 has written at different time for Arbiter (Test Case 29)

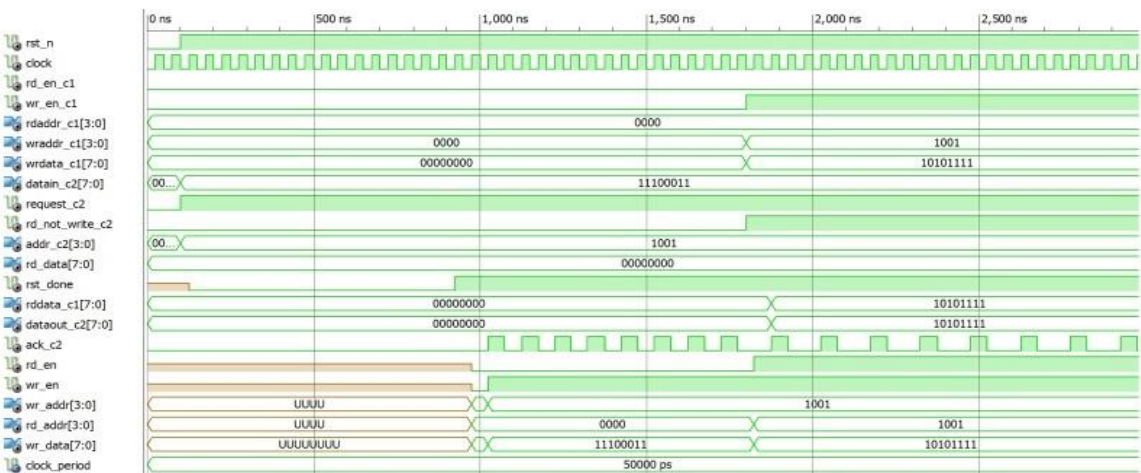

Figure 40: Client2 wants to read and write in the same RAM location and Client1 also wants to write in the RAM location where Client2 has written at same time for Arbiter (Test Case 30)

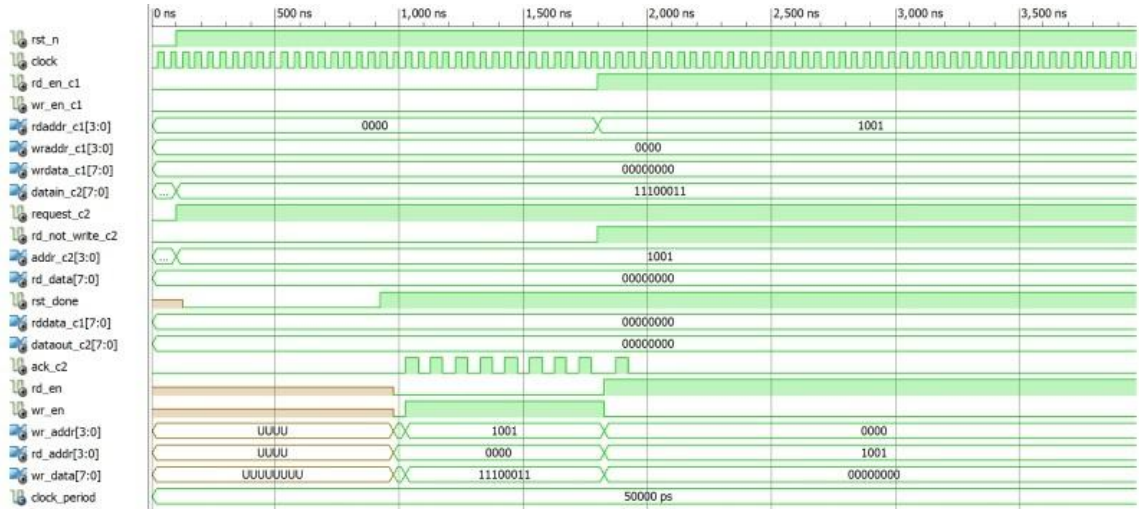

Figure 41: Client2 wants to read and write in the same RAM location and Client1 also wants to read in the RAM location where Client2 has written at same time for Arbiter (Test Case 31)

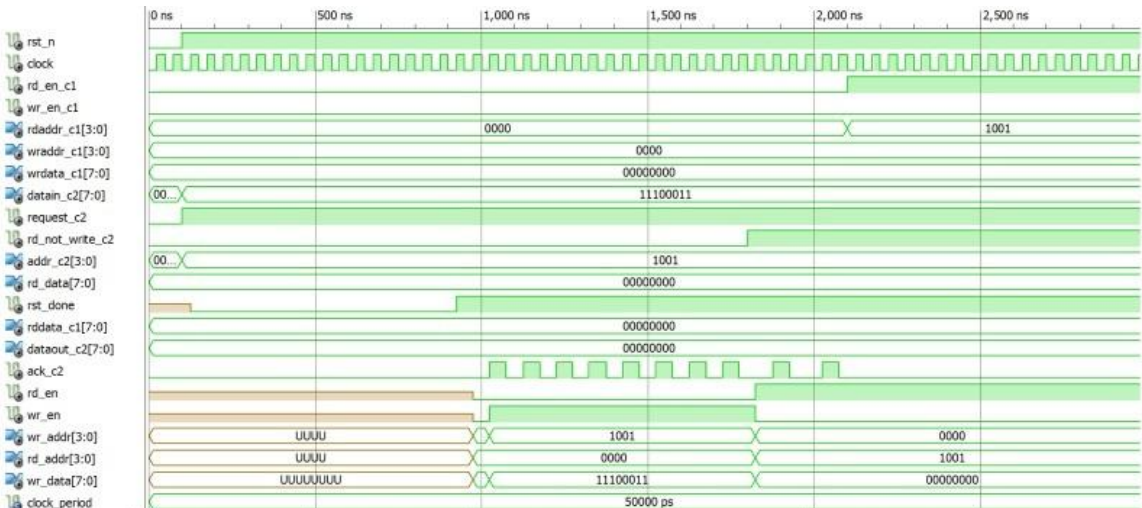

Figure 42: Client2 wants to read and write in the same RAM location and Client1 also wants to read in the RAM location where Client2 has written at different time for Arbiter (Test Case 32)

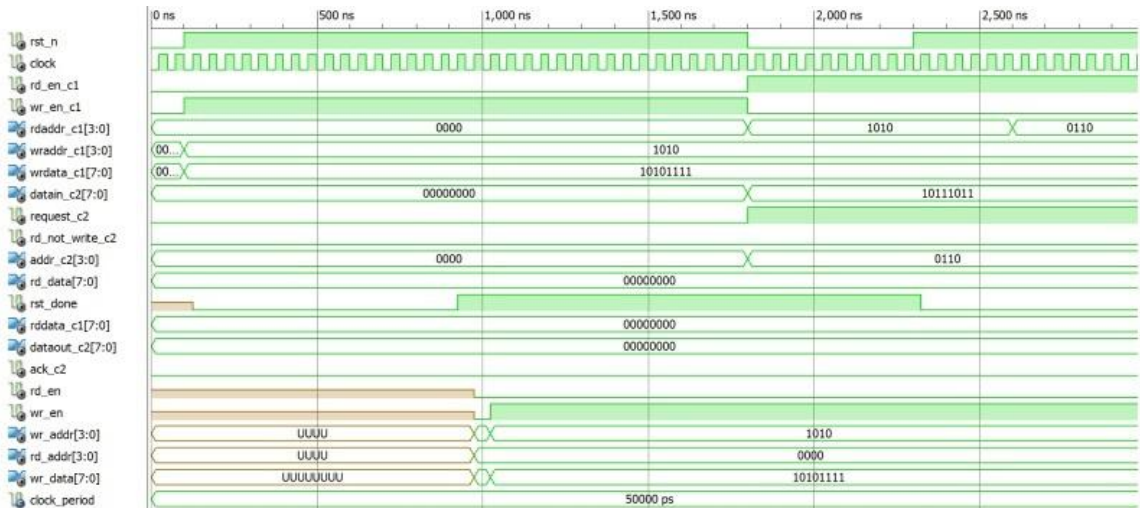

Figure 43: If any client resets (RST_N=0) the system at any time for Arbiter (Test Case 33)

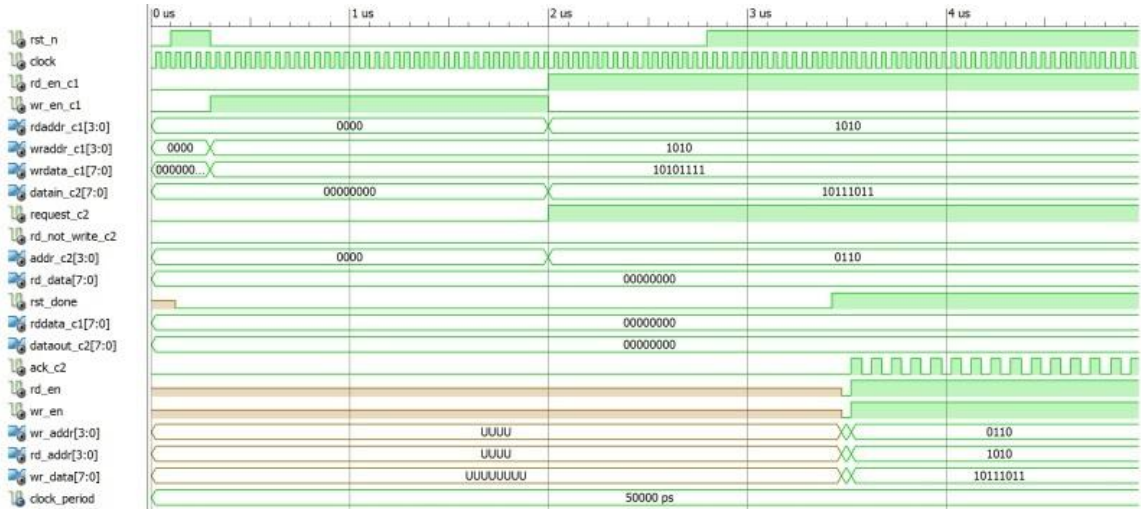

Figure 44: If any client gives the inputs until RST_DONE is high for Arbiter (Test Case 34)

## 3.3.5. Analysis of the Test Cases

**1. Only Client1 wants to write for Arbiter.**

In this case, when the RST_N pin is enabled, the arbiter starts its operation. Within the first RAM_DEPTH cycle, Client1 enables its WR_EN_C1 pin and writes the data "1010011" at the address location "1010".From the point, this operation starts to take place within 500 ns the RST_DONE pin goes high, the arbiter places the data it received, on the input pins of the RAM i.e. the WR_ADDR receives the address location and the WR_DATA receives the data. A typical RAM_DEPTH cycle duration is of 1700 ns.

**2. Only Client1 wants to read for Arbiter.**

The Arbiter starts working as soon as the RST_N pin is enabled. The clock period is set to 50ns. Within the first RAM_DEPTH cycle, a data "10100011" is written at the location "1010". The WR_EN_C1 pin is enabled to facilitate the operation. When the RST_DONE pin goes high, the Arbiter conveys this to the input pins of the RAM. During the next RAM_DEPTH cycle, Client1 wishes to read at the RAM location "1010" by enabling the RD_EN_C1 pin. Immediately, the information is sent to the RAM. As the RAM is yet to be connected to this block, hence the output RDDATA_C1 pin gives "00000000".

**3. Only Client2 wants to write for Arbiter.**

Here in this case, the Arbiter starts working as soon as the RST_N pin is set high. The clock period is set to 50 ns. As the arbitration scheme used is a priority based one, so Client2 needs to issue a signal on the REQUEST_C2 pin in order to access the RAM. Here the priority is given to Client1.Within the first RAM_DEPTH cycle, REQUEST_C2 pin is enabled to allow Client2 to access the RAM. Client2 uses two states of a single pin RD_NOT_WRITE to perform both of its read and write operation. This pin only works if REQUEST_C2 pin is enabled. Upon enabling REQUEST_C2, RD_NOT_WRITE is set to low to allow for the write operation. Data "11100011" is written at the location "1110". The Arbiter passes the address location and data on to RAM as soon as the RST_DONE pin goes high. Contrary to the previous case, here the RST_DONE goes high after a relative long time as since in this priority based scheme, Client1 enjoys a higher priority compared to Client2; hence it takes a relatively long time to set up the system for Client2. The arbiter acknowledges the write operation by setting up periodic pulses at the ACK_C2 output pin.

**4. Only Client2 wants to read for Arbiter.**

In this case, within the first RAM_DEPTH cycle, data "11100011" is written in the address location "1110" by enabling the REQUEST_C2 pin and setting RD_NOT_WRITE pin to an active low state. The ACK_C2 acknowledges the write request with normal periodic pulses. However in the second RAM_DEPTH cycle, Client2 activates the high signal on RD_NOT_WRITE pin and issue a signal to the arbiter to read from location "1110". The read operation is acknowledged by the Arbiter by issuing clock pulses at the ACK_C2 pin whose clock period is two times that of the former one. As the RAM is yet to be connected to this block, hence the output RDDATA_C1 pin gives "00000000".

**5. Client1 wants to read and write in different RAM location at same time for Arbiter.**

Within the first RAM_DEPTH cycle data "10100011" is written at the address location "1010" by Client1 by enabling the WR_EN_C1 pin. In the successive RAM_DEPTH cycle, Client1 wishes to perform the simultaneous operation of reading and writing at different RAM location by enabling both the RD_EN_C1 and WR_EN_C1 pins. This is handled successfully by the Arbiter. Reading was done at the address location "1010" and writing of the data "10111011" was done at the location "1110".

**6. Client1 wants to read and write in different RAM location at different time for Arbiter.**

Within the first RAM_DEPTH cycle data "10100011" is written at the address location "1010" by Client1 by enabling the WR_EN_C1 pin. In the successive RAM_DEPTH cycle, Client1 wishes to perform the operation of reading and writing at different RAM location by enabling both the RD_EN_C1 and WR_EN_C1 pins. However in this case the client chooses to read and write at different time within the same RAM_DEPTH cycle. As shown there is a time delay of 500 ns between the successive operations. Reading was done at the address location "1010" and writing of the data "10111011" was done at the location "1110".

**7. Client1 wants to read and write in same RAM location at same time for Arbiter.**

In this case, within the first RAM_DEPTH cycle data "10100011" is written at the address location "1010" by Client1, enabling the WR_EN_C1 pin. In the successive RAM_DEPTH cycle, Client1 wishes to perform the operation of reading and writing at the same RAM location by enabling both the RD_EN_C1 and WR_EN_C1 pins. Reading was done at the address location "1010" and writing of the data "10111011" was done at the same location. In such cases we deploy the strategy of making the client read the updated value compared to the old one. One who has gone through the code would know that we made use of a temporary register in the Arbiter itself which holds the data for the current RAM_DEPTH cycle. Since such is the case, one could see that we get the output "101110011" on the output pin RDDATA_C1, which is nothing but the temporary data disguised to the client who perceives it to be coming from the RAM.

**8. Client1 wants to read and write in same RAM location at different time for Arbiter.**

Within the first RAM_DEPTH cycle data "10100011" is written at the address location "1010" by Client1, enabling the WR_EN_C1 pin. In the successive RAM_DEPTH cycle, Client1 wishes to perform the operation of reading and writing at the same RAM location by enabling both the RD_EN_C1 and WR_EN_C1 pins. Reading was done at the address location "1010" and writing of the data "10111011" was done at the same location. In such a case we deploy the strategy of making the client read the updated value compared to the old one. However in this case the client chooses to read and write at different time within the same RAM_DEPTH cycle. As shown there is a time delay of 500ns between the successive operations. We made use of a temporary register in the Arbiter itself which holds the data for the current clock cycle. Since such is the case, one could see that we get the output "101110011" on the output pin RDDATA_C1, which is nothing but the temporary data disguised to the client who perceives it to be coming from the RAM.

**9. Client2 wants to read and write in different RAM location at same time for Arbiter.**

Client2 initially performs the write operation in the first RAM_DEPTH cycle at the address location "1010". The data is "11100011". This is achieved by setting the REQUEST_C2 and the RD_NOT_WRITE pin to active high and low respectively. During the next RAM_DEPTH cycle, Client2 performs the read and write operation at the locations "1010" and "1001" respectively. But if we look into the plot closely, one would find no trace of the data being written at the address location "1001". This is because in the third RAM_DEPTH cycle, since Client2 uses a single pin for read & write operation so it can only perform a single operation at a given time contrary to Client1. In the third RAM_DEPTH cycle, RD_NOT_WRITE pin is set high, and the read operation is performed. We could see that data "00100011" is given at the pin DATAIN_C2 but the data is waiting to get recognized at the arbiter, which waits for the RD_NOT_WRITE signal to go low. However in this case since the RAM does not read & write at the same RAM location hence, there is no address clash problem.

## 10. Client2 wants to read and write in different RAM location at different time for Arbiter.

Client2 initially performs the write operation in the first RAM_DEPTH cycle at the address location "1010". The data is "11100011". This is achieved by setting the REQUEST_C2 and the RD_NOT_WRITE pin to active high and low respectively. During the next RAM_DEPTH cycle, Client2 performs the read and write operation at the locations "1010" and "1001" respectively at a different time. Contrary to the previous case, here we could see that Client2 could perform both the operations very smoothly without hiccups. During the second RAM_DEPTH cycle, write operation is performed at the location "1001", which is immediately sent to the input pins of the RAM. In the successive clock cycle, the reading operation is performed at the address location "1010". The interval between the two operations is 500 ns. The ACK_C2 signal gives us the correct signals verifying our claims.

## 11. Client2 wants to read and write in same RAM location at same time for Arbiter.

This case is similar in working to Test Case 9; the only difference is that the RAM locations are the same for both read and write operation. Since Client2 can perform only a single operation at any given time, hence one of the operations seems to be omitted as seen from the waveforms. However in the successive RAM_DEPTH cycles the arbiter repeatedly sends the data to the input pins of the RAM connected to it.

## 12. Client2 wants to read and write in same RAM location at different time for Arbiter.

This case is similar in working to Test Case 10; the only difference is that the RAM locations are the same for both read and write operation. There is a time delay of 500 ns between the read and write operation. However in the successive RAM_DEPTH cycles the arbiter repeatedly sends the data to the input pins of the RAM connected to it.

## 13. Client1 wants to write and Client2 wants to read in different RAM location at same time for Arbiter.

In this case Client1 performs the write operation in the second RAM_DEPTH cycle. In the first RAM_DEPTH cycle, the Arbiter is enabled by the RST_N pin. Along with which Client2 sends a REQUEST_C2 signal for sharing the RAM. We observe that when Client1 is not using the read as well as write function, the arbiter grants access to Client2 to read or write from the RAM. Similar is the case for the write operation also. In the first RAM_DEPTH cycle, Client2 write data "11100011" at the address location "1110'. During the second RAM_DEPTH cycle, both the client performs their respective operations smoothly. The instant data is given to the system the arbiter sends it to the input pins of the RAM. However due to absence of any RAM output is not achieved at the output pins of the Arbiter.

### 14. Client1 wants to write and client2 wants to read in different RAM location at different time for Arbiter.

Client1 performs the write operation in the second RAM_DEPTH cycle. In the first RAM_DEPTH cycle, the Arbiter is enabled by the RST_N pin. Along with which Client2 sends a REQUEST_C2 signal for sharing the RAM. Client2 write data "11100011" at the address location "1110'. During the second RAM_DEPTH cycle, both the client performs their respective operations at different times i.e. there is a time delay of 500 ns. Client1 performs the write operation at the address location "1001". The instant data is given to the system the arbiter sends it to the input pins of the RAM. After about 500 ns, Client2 performs the read operation at the address location "1110". However due to absence of any RAM output is not achieved at the output pins of the Arbiter.

### 15. Client1 wants to write and Client2 wants to read in same RAM location at same time for Arbiter.

In the first RAM_DEPTH cycle, Client2 performs the write operation by enabling the requisite pins. During the second RAM_DEPTH cycle, since Client1 is not using the read function, the arbiter grants the read operation to Client2 without any hassle and both the clients perform the operation smoothly. The RD_ADDR and the WR_DATA pins supply the data to the RAM. One thing to notice here is that output is seen at the output pins RDDATA_C1 and DATAOUT_C2. This is because since the both the clients performs read and write at the same address location, hence when client2 reads from the location where client1 has written it gets the updated value. This updated value comes from the Arbiter's temporary register rather than the RAM itself.

### 16. Client1 wants to write and Client2 wants to read in same RAM location at different time for Arbiter

In the first RAM_DEPTH cycle, Client2 performs the write operation by enabling the requisite pins. During the second RAM_DEPTH cycle, since Client1 is not using the read function, the arbiter grants the read operation to Client2 without any hassle and both the clients perform the operation smoothly as they operate at different times. The RD_ADDR and the WR_DATA pins supply the data to the RAM. One thing to notice here is that output is seen at the output pins RDDATA_C1 and DATAOUT_C2. This is because since the both the clients performs read and write at the same address location, hence when client2 reads from the location where client1 has written it gets the updated value. This updated value comes from the Arbiter's temporary register rather than the RAM itself. Our claims are testified by the appropriate pulses we get from the ACK_C2 pin output. In contrast to the previous test case, the only thing that is different here is that the operations are performed at a time interval of 500 ns.

### 17. Client1 wants to read and Client2 wants to write in same RAM location at same time for Arbiter.

This case is similar to test case 15 which had been discussed earlier. The only difference is that here Client1 performs a write operation in the first RAM_DEPTH cycle. Since there is no conflict of interest among the clients, the system allows both the clients to access the system simultaneously and sends the data immediately to the RAM.

### 18. Client1 wants to read and Client2 wants to write in same RAM location at different time for Arbiter.

This case is similar to test cased 16.

### 19. Client1 wants to read and Client2 wants to write in different RAM location at same time for Arbiter.

Client1 during the first RAM_DEPTH cycle performs the write operation at the address location "1000". The data written is "10101111". During the second RAM_DEPTH cycle, since the address location for both the clients are different hence, the arbiter grants them access to the RAM simultaneously. Client1 reads from address location "1000" and Client2 writes to the address location "1010". The RAM receives the data almost instantaneously.

### 20. Client1 wants to read and Client2 wants to write in different RAM location at different time for Arbiter.

This case is very similar in operation the previous test case. The only difference is that during the second RAM_DEPTH cycle the operations of both the clients differ by 500 ns.

### 21. Client1 wants to read and Client2 wants to read in same RAM location at different time for Arbiter.

During the first RAM_DEPTH cycle, Client1 perform write operation at the address location "1010". Subsequently in the next RAM_DEPTH cycle, both Client1 and Client2 try to access the RAM to read the same RAM address location. Since it a priority based arbitration system, Client1 gets access to the RAM compared to Client2. It is only when Client1 does not require the read operation any longer that the system grants access to Client2 which reads from the same location. There is a time delay of 500 ns after which Client1 sets its RD_EN_C1 to active low signal. Since both access the same RAM location at different time, this does not affect the clients on the whole.

### 22. Client1 wants to read and Client2 wants to read in same RAM location at same time for Arbiter.

The present case is similar to the previous test case differing only in the fact that here the scheme of priority based arbitration is clearly visible. During the second RAM_DEPTH cycle, both the clients try to access the same RAM location at the same time but only Client1 gets access to the RAM. Since there is no output at ACK_C2 pin it bears testimony to our claims.

### 23. Client1 wants to write and Client2 wants to write in different RAM location at different time for Arbiter.

It is similar in behavior to test case 21. The only difference is that the following case deals with the "write" operation compared to the "read" operation in the former one. The RAM is seen to receive the values form the arbiter almost instantaneously.

### 24. Client1 wants to write and Client2 wants to write in different RAM location at same time for Arbiter.

It is similar in behavior to test case 22. The only difference is that the following case deals with the "write" operation compared to the "read" operation in the former one. The RAM is seen to receive the values form the arbiter almost instantaneously. One thing to note is that the case of whether both the clients access the same or different RAM location is insignificant here since it is a priority based arbitration scheme.

### 25. Client1 wants to read and write in the same RAM location and Client2 also wants to read in the RAM location where Client1 has written at different time for Arbiter.

In this test case Client1 performs the write operation at the address location "1001". In the second RAM_DEPTH cycle, Client1 as it enjoys the higher priority among the two, can simultaneously perform the read and write operation. It writes the data "10100011" at the previous address location. However as previously also discussed, there is an output at the RDDATA_C1 pin as this apparent output seems to come from the temporary register located in the Arbiter itself which provides this output when client1 performs the read operation. Since Client2 access the system at a different time (i.e. after 500 ns), the arbiter grants permission to Client2 to read from the same location as Client1 is not using the read operation. This can be observed by the setting low of the RD_EN_C1 signal near 2500 ns mark in the waveform graph.

### 26. Client1 wants to read and write in the same RAM location and Client2 also wants to read in the RAM location where Client1 has written at same time for Arbiter.

Client1 performs the write operation at the address location "1001". In the second RAM_DEPTH cycle, Client1 as it enjoys the higher priority among the two, can simultaneously perform the read and write operation. It writes the data "10100011" at the previous address location. However as previously also discussed, there is an output at the RDDATA_C1 pin as this apparent output seems to come from the temporary register located in the Arbiter itself which provides this output when client1 performs the read operation. Since Client2 access the system at the same time, hence Client2 is unable to access the RAM at all.

**27. Client1 wants to read and write in the same RAM location and Client2 also wants to write in the RAM location where Client1 has written at different time for Arbiter.**

In this test case Client1 performs the write operation at the address location "1001". In the second RAM_DEPTH cycle, Client1 as it enjoys the higher priority among the two, can simultaneously perform the read and write operation. It writes the data "10100011" at the previous address location. However as previously also discussed, there is an output at the RDDATA_C1 pin as this apparent output seems to come from the temporary register located in the Arbiter itself which provides this output when client1 performs the read operation. Since Client2 access the system at a different time (i.e. after 500 ns), the arbiter grants permission to Client2 to write at same location as Client1 is not using the write operation. One thing to observe here is that when Client2 starts to write the new data at the same address location, the output changes at the output pin instantaneously since Client1 continues to read. This clearly shows that instead of the data coming from the RAM itself, the output comes from the temporary register located in the Arbiter.

**28. Client1 wants to read and write in the same RAM location and Client2 also wants to write in the RAM location where Client1 has written at same time for Arbiter.**

Client1 performs the write operation at the address location "1001". In the second RAM_DEPTH cycle, Client1 as it enjoys the higher priority among the two, can simultaneously perform the read and write operation. It writes the data "10100011" at the previous address location. However as previously also discussed, there is an output at the RDDATA_C1 pin as this apparent output seems to come from the temporary register located in the Arbiter itself which provides this output when client1 performs the read operation. Since Client2 tries access the system at a same time, it has to wait until Client1 has finished its write operation.

**29. Client2 wants to read and write in the same RAM location and Client1 also wants to write in the RAM location where Client2 has written at different time for Arbiter.**

This test case is similar to the previous test case. The thing to observe here is that during the second RAM_DEPTH cycle, compared to the former case, here Client1 performs the write operation at the time when Client2 performs the read operation at the same time. There would be no basic difference between the previous waveform graph and the current one since the read and write operation is being performed by both the clients and not one, hence it does not really affect us whether the operations are performed instantaneously or at different time.

**30. Client2 wants to read and write in the same RAM location and Client1 also wants to write in the RAM location where Client2 has written at same time for Arbiter.**

During the first RAM_DEPTH cycle, Client2 performs the write operation at the address location "1001". The data written is "11100011". However in the second RAM_DEPTH cycle, since Client2 has a lower priority in the system, it can only read or write at any given time. The Arbiter only allows Client2 to read from the address location "1001". This can verified from the pulses at the output pins of ACK_C2.Moreover, since Client2 reads, the arbiter at a later time allows Client1 to write to the RAM at the same address location. Almost instantaneously, the data given is sent to the input pins of the RAM and is also reflected at the output pins of DATAOUT_C2 and RDDATA_C1, the reason for which had already been discussed. We observe that the write operation of Client2 during the second RAM_DEPTH cycle is omitted.

**31. Client2 wants to read and write in the same RAM location and Client1 also wants to read in the RAM location where Client2 has written at same time for Arbiter.**

In this case, Client2 performs a write operation during the first RAM_DEPTH cycle. In the next cycle, one would observe that around 1700-1800 ns that Client2 performs the read operation on the address location "1010".This is more clear from the type of the output pulses at the output pin of ACK_C2. But almost within 25-50 ns, we see that the Arbiter grants the read operation to Client1 who enjoys a higher priority. The ACK_C2 signals stops giving any output. Since both Client1 and Client2 try to perform the read operation at the same time, it may be somewhat difficult to comprehend the change with the naked eye.

**32. Client2 wants to read and write in the same RAM location and Client1 also wants to read in the RAM location where Client2 has written at different time for Arbiter.**

Client2 performs a write operation during the first RAM_DEPTH cycle. In the next cycle, one would observe that around 1700-1800 ns that Client2 performs the read operation on the address location "1010".This is more clear from the type of the output pulses at the output pin of ACK_C2. But almost within 200-300 ns, we see that the Arbiter grants the read operation to Client1 who enjoys a higher priority. The ACK_C2 signals stops giving any output. Since both Client1 and Client2 try to perform the read operation at different time, the change is over a larger period of time as compared to the previous case and one could see the change quite easily.

**33. If any client resets (RST_N=0) the system at any time for Arbiter.**

This test case can only be analyzed properly only when the RAM will be connected to the systems otherwise it is similar to test case 1.

## 34. If any client gives the inputs until RST_DONE is high for Arbiter.

The analysis of this test case can only be done when the RAM is connected to the system.

# Chapter 4

## RAM Arbiter

This chapter describes the methods and codes used for completion of the project. A RAM arbiter will prevent collision of data's being accessed and maintain requests in order to ensure memory is holding accurate data.

In addition to being useful in their own right, RAM arbiters form the fundamental building block for controllers that match multiple requesters with multiple resources. Whenever a resource, such as a memory, buses or a multiplier is shared by many agents, an arbiter is required to assign access to the resource to one agent at a time.

### 4.1. Top Level Block Diagram

The figure shown below is a top level block diagram of our proposed design of the RAM Arbiter.

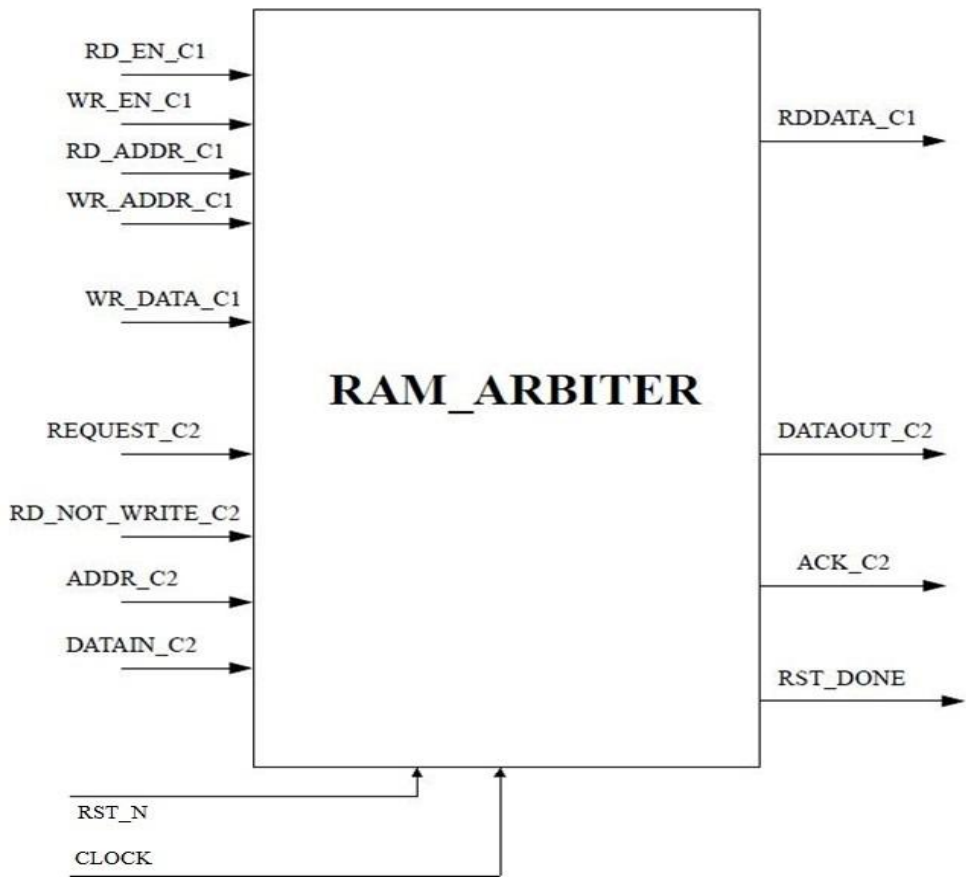

**Figure 45:** Top Level Implementation of our RAM Arbiter

## 4.2. Port List and their Functionality

| PORT-NAME | DIRECTION | FUNCTIONALITY |
| --- | --- | --- |
| RD_EN_C1 | IN | Active high signal for read operation from RAM by client1 |
| WR_EN_C1 | IN | Active high signal for write operation into RAM by client1 |
| RD_ADDR_C1 | IN | RAM address(bus signal) from where client1 wants to read |
| WR_ADDR_C1 | IN | RAM address(bus signal) where client1 wants to write |
| WR_DATA_C1 | IN | The data(bus signal),client1 wants to write in WR_ADDR |
| REQUEST_C2 | IN | Active high signal by client2 to access RAM |
| RD_NOT_WRITE_C2 | IN | Active high for reading, Active low for writing by client2 |
| ADDR_C2 | IN | RAM address(bus signal) either for read or write by client2 |
| DATAIN_C2 | IN | The data (bus signal) if client2 wants to write in ADDR_C2 |
| RD_DATA_C1 | OUT | Output data(bus signal) got by client1 from RAM |
| DATA_OUT_C2 | OUT | Output data(bus signal) got by client2 from RAM |
| ACK_C2 | OUT | Active high signal to indicate that request by client2 is done |
| CLOCK | IN | Positive edge input for valid operation (global signal) |
| RST_N | IN | Active low signal to clear all registers (global signal) |
| RST_DONE | OUT | Active low signal to indicate that RAM is being initialized |

This is the top level block diagram which can interface with two clients (client1 and client2) with their respective input and output ports. But client1 has the higher priority than client2. So client1 can have the access of the RAM any time it wants. But if client2 wants to access the RAM then it must send a request first and according to the following condition Client2 can get the access of the RAM:

1) If client1 is only writing then Client2 can get the access of the RAM for reading only
2) If client1 is only reading then client2 can get the access of the RAM for writing only
3) If client1 is reading and also writing then client2 cannot get any access of the RAM

4) If client1 is not doing anything then Client2 can get the access of the RAM either for reading or writing.

## 4.3. Internal Architecture

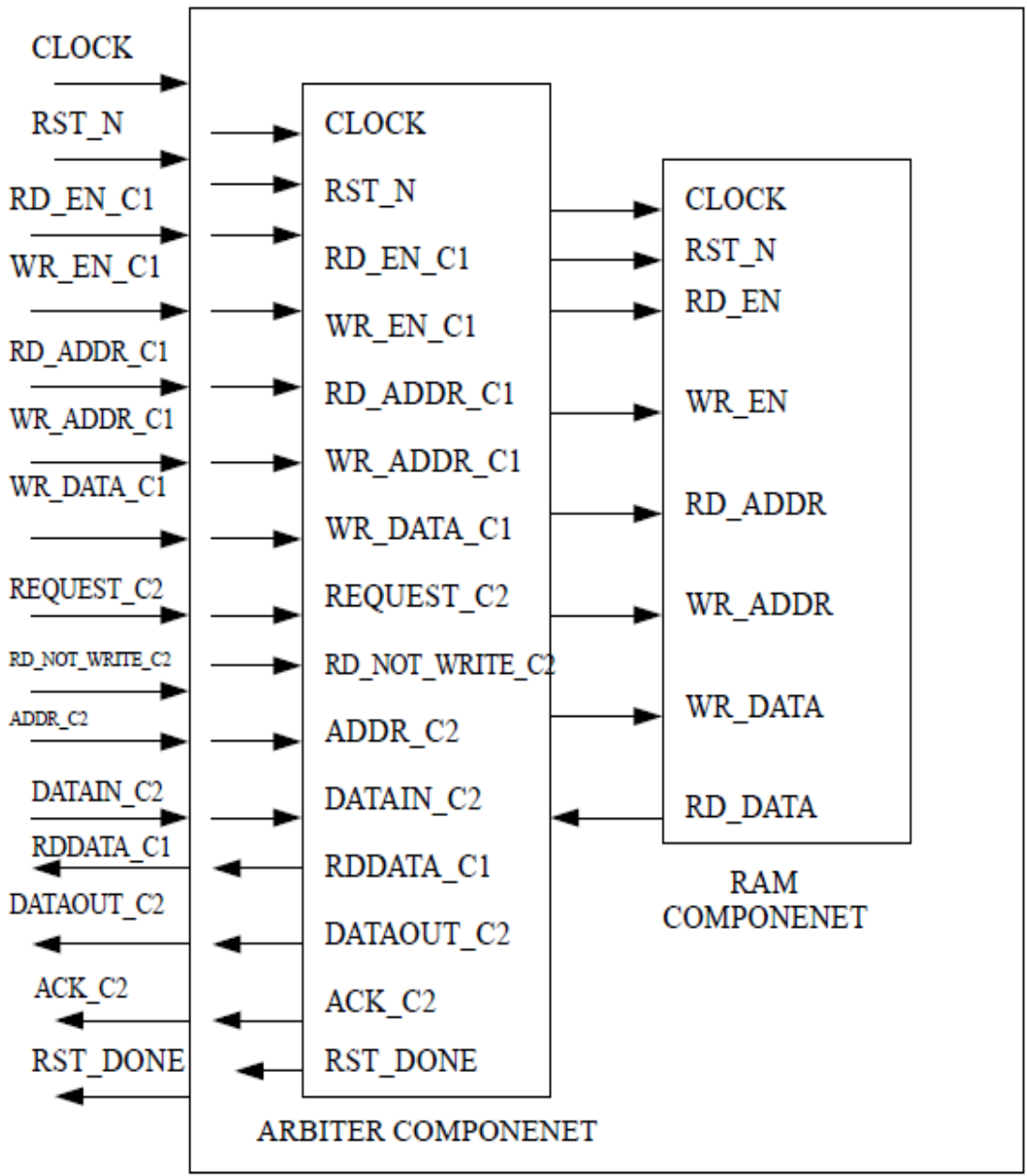

**Figure 46:** Internal Architecture

### 4.3.1. Functionality

This is the main entity block where two components (ARBITER component and RAM component) are port mapped to get our required output.

## 4.4. Arbiter Component

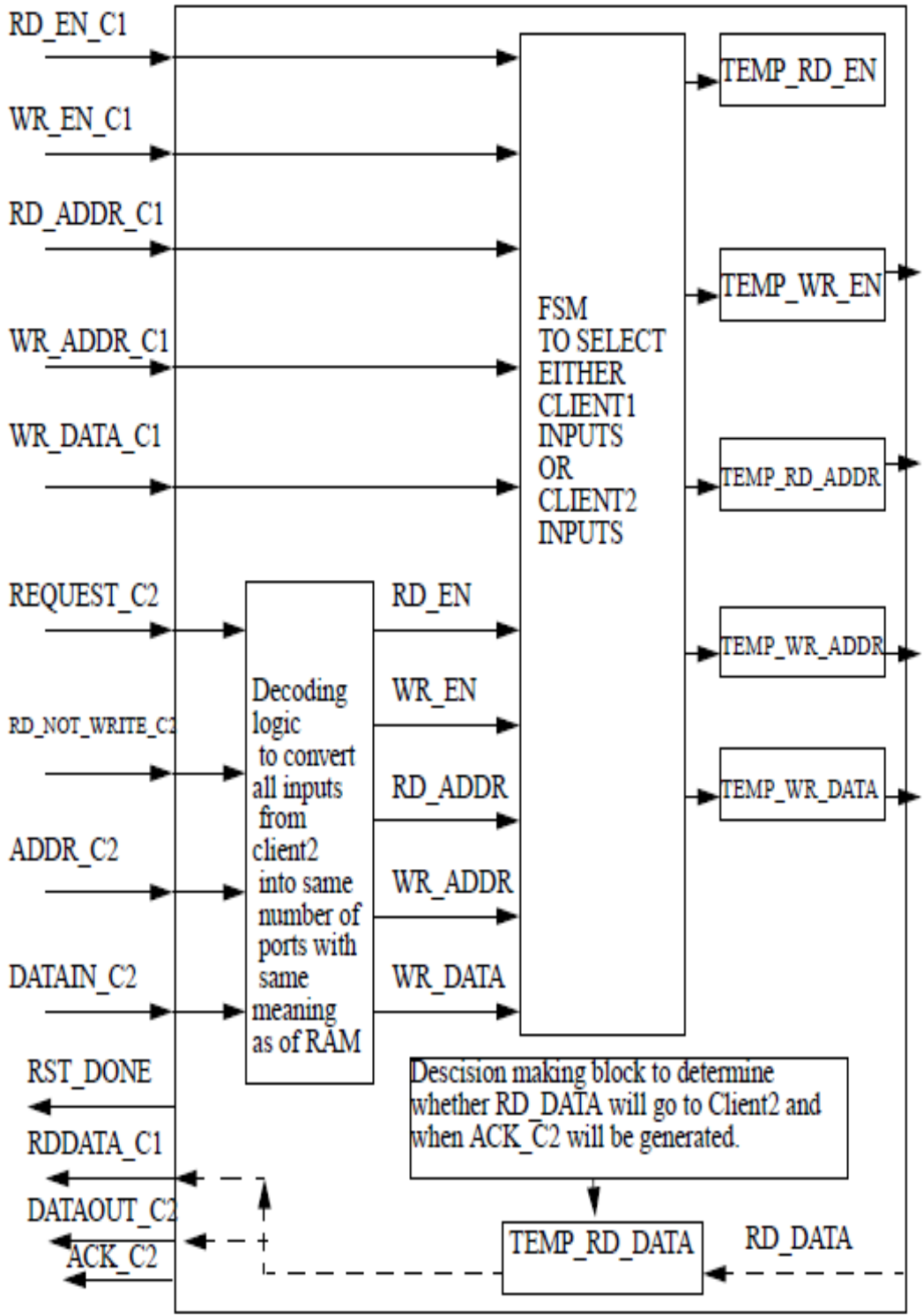

**Figure 47:** Arbiter Compoment

## 4.5. Address Clash Consideration and its Solution

During Address clash condition either client1 or client2 should read the updated value not the old value. In this condition the data from client1 or client2 is separately stored in a register called TEMP_RD_DATA and a signal ADDR_CLASH is generated.

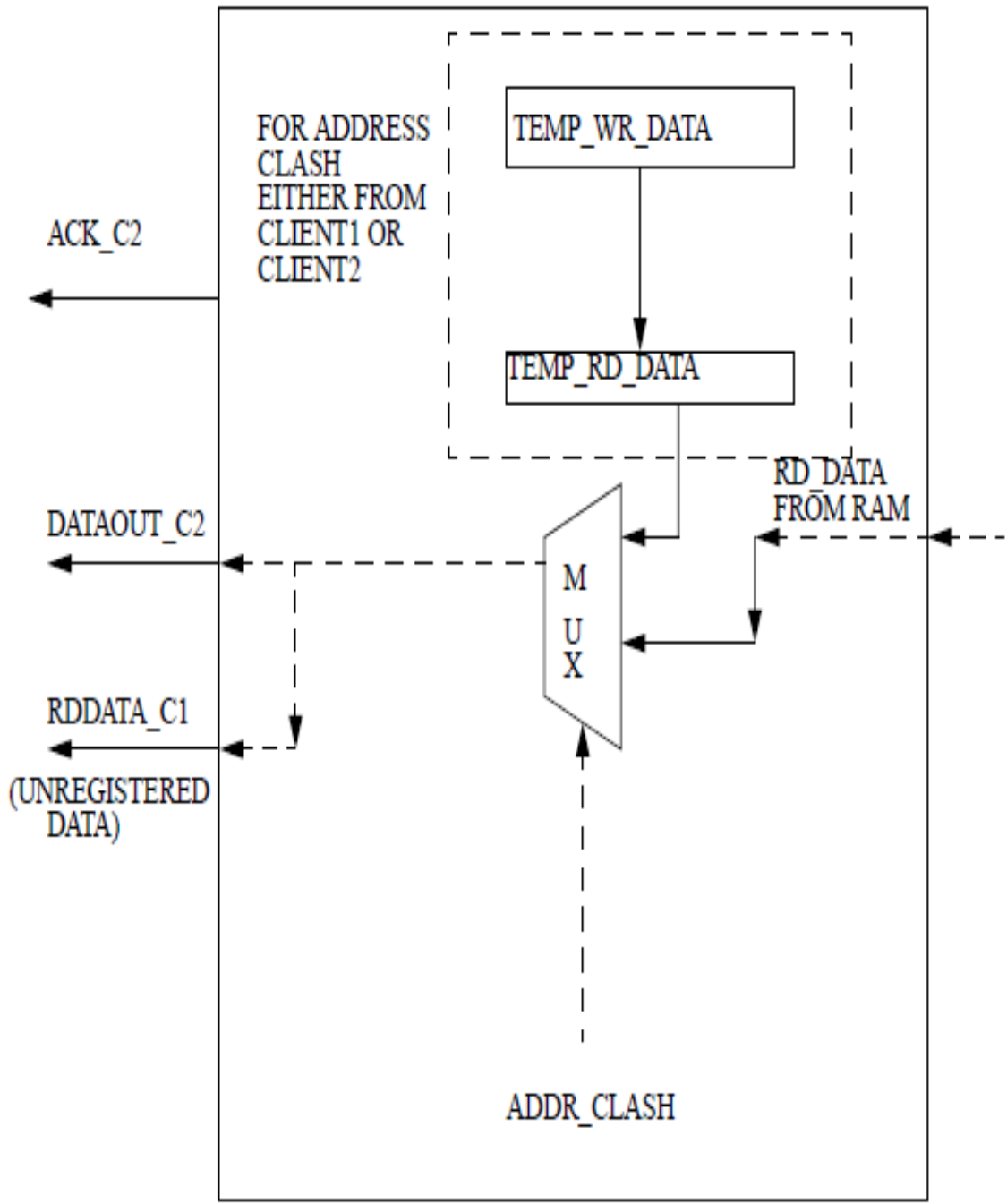

**Figure 48:** Address Clash Solution for Unregistered data

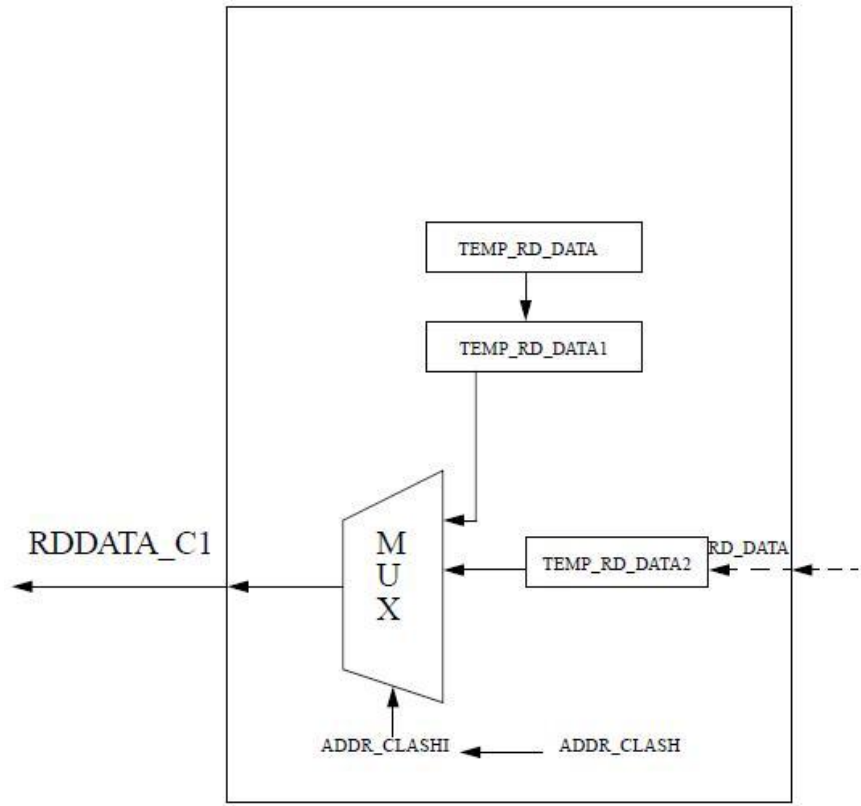

**Figure 49**: Address Clash Solution for Registered data for Client1

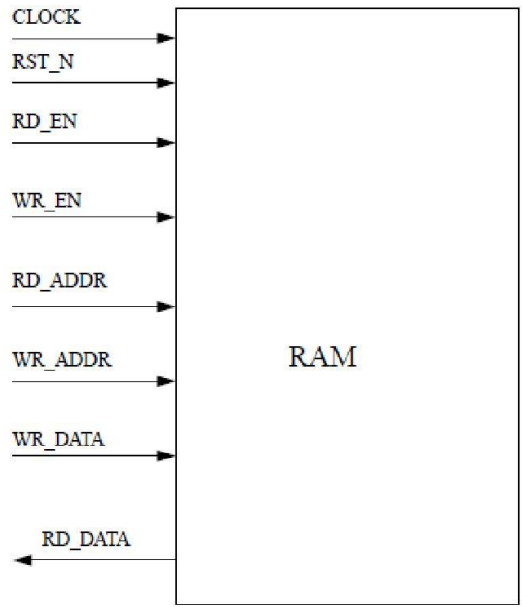

**Figure 50**: Input Ports of RAM

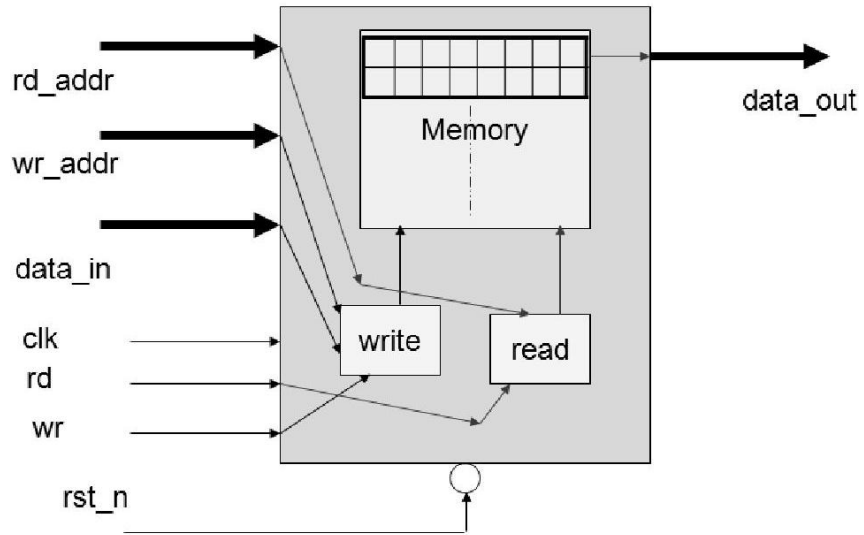

**Figure 51**: Read & Write Operation of RAM

## 4.6. Port Map in between RAM and Arbiter

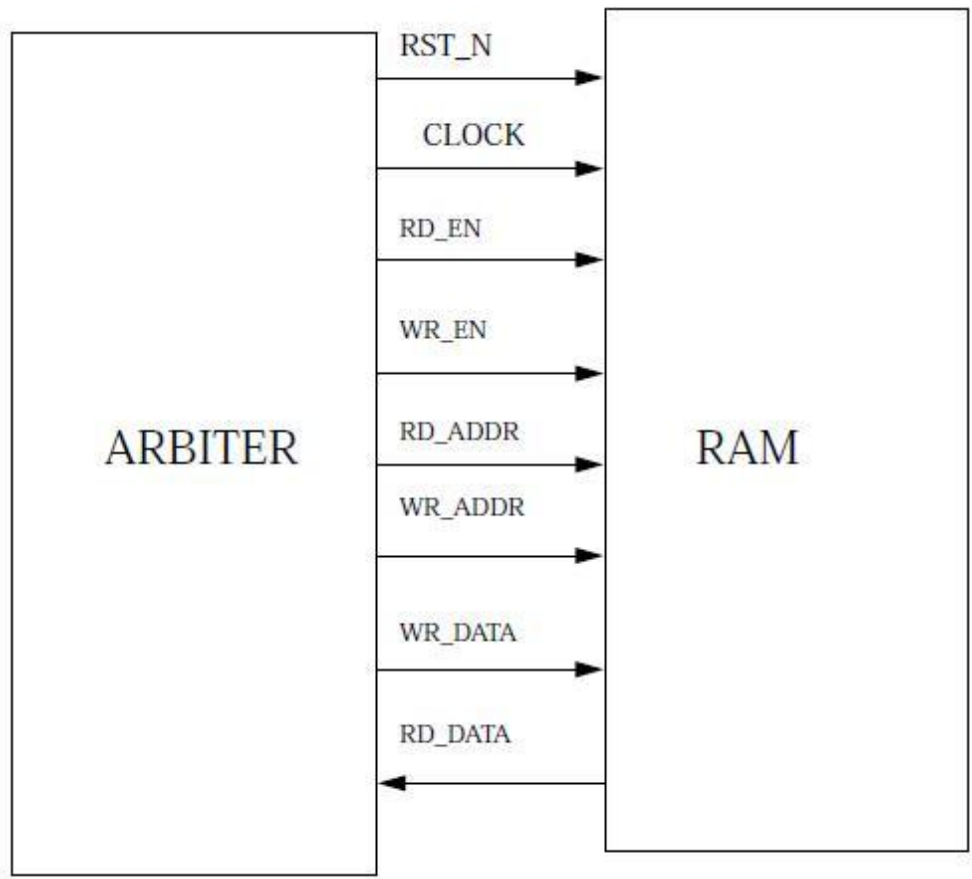

**Figure 52**: Port Map between RAM & Arbiter

## 4.7. Port Map in between Entity and Arbiter

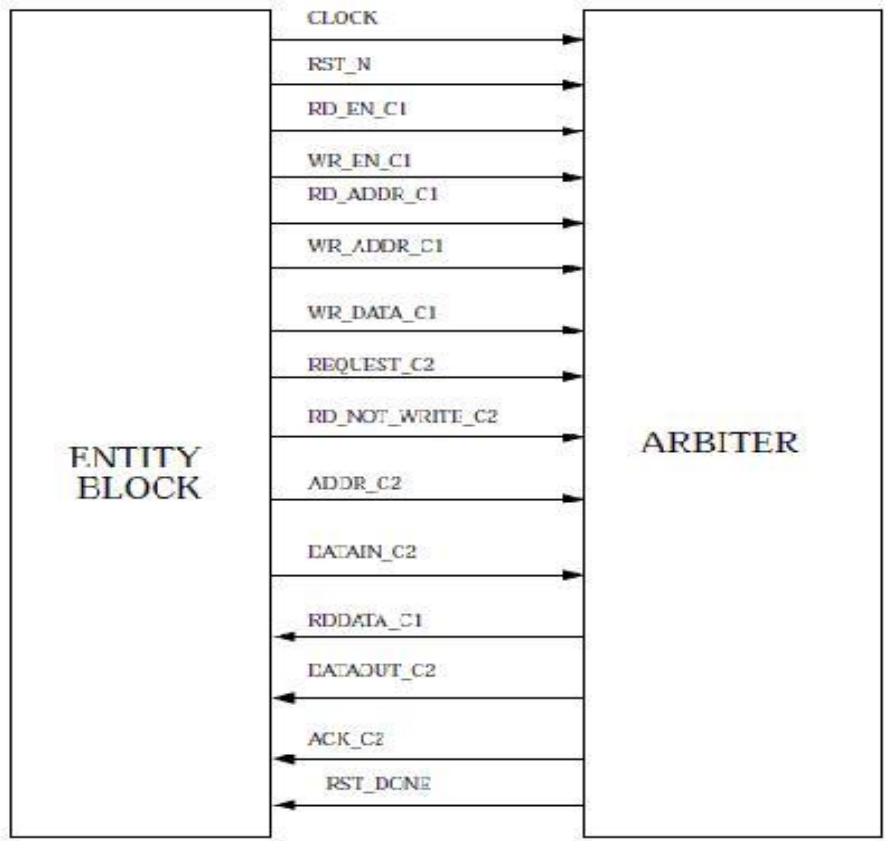

**Figure 53**: Port Map between Entity & Arbiter

## 4.8. FSM Design Block

This FSM determines which client gets access to the specified RAM depending on the following conditions:-

1) If Client1 is writing into RAM but not reading, then Client2 can read but can't write.

2) If Client1 is reading from RAM but not writing, then Client2 can write but can't read.

3) If Client2 is reading and Client1 wants to read then Client1 gets the access of RAM.

4) If Client2 is writing and Client1 wants to write then Client1 gets the access of RAM.

Thus, there is no clash in between read and write operation of both the clients. So, we have taken two separate states per client.

1)  For client1 there are client1_read and client1_write state.

2)  For client2 there are client2_read and client2_write state.

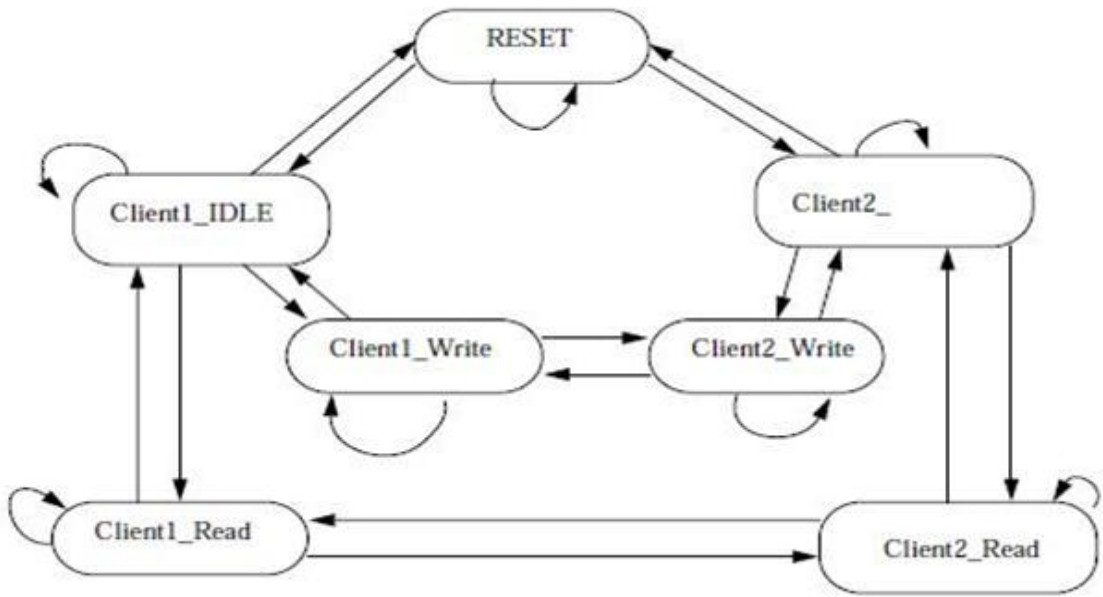

On applying RST_N=0 FSM will be in RESET state and after RST_N=1 it will take RAM_DEPTH cycles to reach in IDLE state and any input can be accepted after reaching IDLE state.

Transition of states depending upon different input conditions from the clients are given below:

1. IDLE----->Client1_Read==> RD_EN_C1=1.

2. Client1_Read----->IDLE==> RD_EN_C1=0.

3. IDLE---->Client1_Write==> WR_EN_C1=1.

4. Client1_Write---->IDLE==> WR_EN_C1=0.

5. IDLE--->Client2_Read==> RD_EN_C1=0 and REQUEST_C2=1 and RD_NOT_WRITE_C2=1.

6. Client2_Read--->IDLE==>REQUEST_C2=0 or REQUEST_C2=1 and RD_NOT_WRITE_C2=0.

7. IDLE--->Client2_Write==>WR_EN_C1=0 and REQUEST_C2=1 and RD_NOT_WRITE_C2=0.

8. Client2_Write-->IDLE==>REQUEST_C2=0 or REQUEST_C2=1 and RD_NOT_WRITE_C2=1.

9. Client2_Read---->Client1_Read===>RD_EN_C1=1.

10. Client1_Read----->Client2_Read===>RD_EN_C1=0 and REQUEST_C2=1 and RD_NOT_WRITE_C2=1.

11. Client2_Write----->Client1_Write==>WR_EN_C1=1.

12. Client1_Write----->Client2_Write==>WR_EN_C1=0 and REQUEST_C2=1 and RD_NOT_WRITE_C2=0.

13. Client1_Read--->Client1_Read==>RD_EN_C1=1.

14. Client1_Write--->Client1_Write==>WR_EN_C1=1.

15. Client2_Read--->Client2_Read===>REQUEST_C2=1 and RD_NOT_WRITE_C2=1 and RD_EN_C1=0.

16. Client2_Write---->Client2_Write===>REQUEST_C2=1 and RD_NOT_WRITE_C2=0 and WR_EN_C1=0.

17. IDLE---->IDLE===>RD_EN_C1=0 and WR_EN_C1=0 and REQUEST_C2=0.

18. RESET--->RESET===>if RST_N=0 holds or RST_N=0 for once and RST_DONE=0

19. RESET--->IDLE===>RST_N=1 and RST_DONE=1.

20. IDLE--->RESET==>RST_N=0.

# 4.9. Implementation of the RAM Arbiter Module using Verilog

## 4.9.1. VHDL Code

```
library IEEE;
use IEEE.STD_LOGIC_1164.ALL;
use IEEE.STD_LOGIC_ARITH.ALL;
use IEEE.STD_LOGIC_UNSIGNED.ALL;
---------------------------------------------------------------------------------------------------------
-- Entity for ARBITER
---------------------------------------------------------------------------------------------------------
entity RAM_ARBITER_NEW is
generic
(
---------------------------------------------------------------------------------------------------------
-- Generics for scalability
---------------------------------------------------------------------------------------------------------
G_ADDR_WIDTH:          integer := 4;
G_DATA_WIDTH:          integer := 8;
G_REGISTERED_DATA:    integer :=0
-- G_ADDR_WIDTH = Number of bits required to address the ram
-- G_DATA_WIDTH = Number of bits in a data
-- G_REGISTERED_DATA =1 for registered data in output 0 for nonregistered
---------------------------------------------------------------------------------------------------------
);
port
(
```

```
------------------------------------------------------------------------------------
-- General Inputs & output
------------------------------------------------------------------------------------
RST_N:          in std_logic;
CLOCK:          in std_logic;
RST_DONE:   out std_logic;
------------------------------------------------------------------------------------
-- Inputs from --------client1--------------
------------------------------------------------------------------------------------
RD_EN_C1:       in  std_logic;                      --read enb--
WR_EN_C1:       in  std_logic;                      --write enb--
RDADDR_C1:   in  std_logic_vector(G_ADDR_WIDTH-1 downto 0);--read addr---
WRADDR_C1:   in  std_logic_vector(G_ADDR_WIDTH-1 downto 0);--write addr--
WRDATA_C1:   in  std_logic_vector(G_DATA_WIDTH-1 downto 0);--data in----
------------------------------------------------------------------------------------
-- Inputs from --------client2--------------
------------------------------------------------------------------------------------
DATAIN_C2:              in  std_logic_vector(G_DATA_WIDTH-1 downto 0);--input data--
REQUEST_C2:          in  std_logic;                      --request to access memory--
RD_NOT_WRITE_C2:   in  std_logic;                      --if '0' then write or read--
ADDR_C2:              in std_logic_vector(G_ADDR_WIDTH-1 downto 0);--addr for rd or wr--
------------------------------------------------------------------------------------
-- Output from --------client1--------------
------------------------------------------------------------------------------------
RDDATA_C1:           out std_logic_vector(G_DATA_WIDTH-1 downto 0);--data out--
------------------------------------------------------------------------------------
-- Output from --------client2--------------
------------------------------------------------------------------------------------
DATAOUT_C2:           out std_logic_vector(G_DATA_WIDTH-1 downto 0);--out data--
ACK_C2:           out std_logic);                      --acknowledgement--
end RAM_ARBITER_NEW;
------------------------------------------------------------------------------------
Architecture RTL of RAM_ARBITER_NEW is
signal WR_DATA: std_logic_vector(G_DATA_WIDTH-1  downto 0);-- temp WR data --
signal RD_DATA1: std_logic_vector(G_DATA_WIDTH-1  downto 0);-- temp RD data --
signal WR_ADDR:std_logic_vector(G_ADDR_WIDTH-1 downto 0); ---temp write address----
signal RD_ADDR:std_logic_vector(G_ADDR_WIDTH-1 downto 0); ---temp read address-----
signal RD_EN:std_logic;
signal WR_EN:std_logic;
component RAM is
generic
(
------------------------------------------------------------------------------------
-- Generics for scalability
------------------------------------------------------------------------------------
G_ADDR_WIDTH:              integer;
G_DATA_WIDTH:              integer
-- G_ADDR_WIDTH = Number of bits required to address the ram
-- G_DATA_WIDTH = Number of bits in a data
------------------------------------------------------------------------------------
);
```

```
port
(
-------------------------------------------------------------------------------------------------
-- RAM Inputs
-------------------------------------------------------------------------------------------------
CLOCK:      in std_logic;
RST_N:      in std_logic;
RD_EN:      in std_logic;                    --read enb--
WR_EN:      in std_logic;                    --write enb--
RD_ADDR:    in std_logic_vector(G_ADDR_WIDTH-1 downto 0);--read addr---
WR_ADDR:    in std_logic_vector(G_ADDR_WIDTH-1 downto 0);--write addr--
WR_DATA:    in std_logic_vector(G_DATA_WIDTH-1 downto 0);--data input----
RD_DATA:    out std_logic_vector(G_DATA_WIDTH-1 downto 0) --data output--
);
end component;
COMPONENT ARBITER_NEW is
generic
(
-------------------------------------------------------------------------------------------------
-- Generics for scalability
-------------------------------------------------------------------------------------------------
G_ADDR_WIDTH:          integer;
G_DATA_WIDTH:          integer;
G_REGISTERED_DATA:     integer
-- G_ADDR_WIDTH = Number of bits required to address the ram
-- G_DATA_WIDTH = Number of bits in a data
-- G_REGISTERED_DATA =1 for registered data in output 0 for nonregistered
-------------------------------------------------------------------------------------------------
);
port
(
-------------------------------------------------------------------------------------------------
-- General Inputs & output
-------------------------------------------------------------------------------------------------
RST_N:          in std_logic;
CLOCK:          in std_logic;
RST_DONE:       out std_logic;
-------------------------------------------------------------------------------------------------
-- Inputs from --------client1--------------
-------------------------------------------------------------------------------------------------
RD_EN_C1:       in std_logic;                    --read enb--
WR_EN_C1:       in std_logic;                    --write enb--
RDADDR_C1:      in std_logic_vector(G_ADDR_WIDTH-1 downto 0);--read addr---
WRADDR_C1:      in std_logic_vector(G_ADDR_WIDTH-1 downto 0);--write addr--
WRDATA_C1:      in std_logic_vector(G_DATA_WIDTH-1 downto 0);--data in----
-------------------------------------------------------------------------------------------------
-- Inputs from --------client2--------------
-------------------------------------------------------------------------------------------------
DATAIN_C2:      in std_logic_vector(G_DATA_WIDTH-1 downto 0);--input data--
REQUEST_C2:     in std_logic;                    --request to access memory--
RD_NOT_WRITE_C2:  in std_logic;                  --if '0' then write or read--
```

```
ADDR_C2:          in std_logic_vector(G_ADDR_WIDTH-1 downto 0);--addr for rd or wr--
----------------------------------------------------------------------------------------------------
-- Output from --------client1--------------
----------------------------------------------------------------------------------------------------
RDDATA_C1:        out std_logic_vector(G_DATA_WIDTH-1 downto 0);--data out--
----------------------------------------------------------------------------------------------------
-- Output from --------client2--------------
----------------------------------------------------------------------------------------------------
DATAOUT_C2:       out std_logic_vector(G_DATA_WIDTH-1 downto 0);--out data--
ACK_C2:       out std_logic;                    --acknowledgement--
RD_EN:        out std_logic;
WR_EN:        out std_logic;
WR_ADDR:      out std_logic_vector(G_ADDR_WIDTH-1 downto 0);
RD_ADDR:      out std_logic_vector(G_ADDR_WIDTH-1 downto 0);
WR_DATA:      out std_logic_vector(G_DATA_WIDTH-1 downto 0);
RD_DATA:      in std_logic_vector(G_DATA_WIDTH-1 downto 0));
end COMPONENT;
----------------------------------------------------------------------------------------------------
-------- RAM Arbiter Code --------------
----------------------------------------------------------------------------------------------------
begin
RAMCLIENT:RAM
GENERIC MAP(G_ADDR_WIDTH,G_DATA_WIDTH)
PORT MAP(CLOCK =>  CLOCK,
RST_N =>  RST_N,
RD_EN =>  RD_EN,
WR_EN =>  WR_EN,
RD_ADDR=> RD_ADDR,
WR_ADDR=> WR_ADDR,
WR_DATA=> WR_DATA,
RD_DATA=> RD_DATA1);
ARBITERCLIENT:ARBITER_NEW
GENERIC MAP(G_ADDR_WIDTH,G_DATA_WIDTH,G_REGISTERED_DATA)
PORT MAP(RST_N       =>  RST_N,
CLOCK        =>  CLOCK,
RST_DONE     =>  RST_DONE,
RD_EN_C1     =>  RD_EN_C1,
WR_EN_C1     =>  WR_EN_C1,
RDADDR_C1    =>  RDADDR_C1,
WRADDR_C1    =>  WRADDR_C1,
WRDATA_C1    =>  WRDATA_C1,
REQUEST_C2   =>  REQUEST_C2,
RD_NOT_WRITE_C2=> RD_NOT_WRITE_C2,
ADDR_C2      =>  ADDR_C2,
DATAIN_C2    =>  DATAIN_C2,
RD_EN        =>  RD_EN,
WR_EN        =>  WR_EN,
RD_ADDR      =>  RD_ADDR,
WR_ADDR      =>  WR_ADDR,
WR_DATA      =>  WR_DATA,
RD_DATA      =>  RD_DATA1,
```

```
DATAOUT_C2   => DATAOUT_C2,
ACK_C2       => ACK_C2,
RDDATA_C1    => RDDATA_C1);
end RTL;
```

---------------------------------------------------------------------------------------------------------------

## 4.9.2. Test Cases for RAM Arbiter

1. Only Client1 wants to write.

2. Only Client1 wants to read.

3. Only Client2 wants to write.

4. Only Client2 wants to read.

5. Client1 wants to read and write in different RAM location at same time.

6. Client1 wants to read and write in different RAM location at different time.

7. Client1 wants to read and write in same RAM location at same time.

8. Client1 wants to read and write in same RAM location at different time.

9. Client2 wants to read and write in different RAM location at same time.

10. Client2 wants to read and write in different RAM location at different time.

11. Client2 wants to read and write in same RAM location at same time.

12. Client2 wants to read and write in same RAM location at different time.

13. Client1 wants to write and Client2 wants to read in different RAM location at same time.

14. Client1 wants to write and client2 wants to read in different RAM location at different time.

15. Client1 wants to write and Client2 wants to read in same RAM location at same time.

16. Client1 wants to write and Client2 wants to read in same RAM location at different time.

17. Client1 wants to read and Client2 wants to write in same RAM location at same time.

18. Client1 wants to read and Client2 wants to write in same RAM location at different time.

19. Client1 wants to read and Client2 wants to write in different RAM location at same time.

20. Client1 wants to read and Client2 wants to write in different RAM location at different time.

21. Client1 wants to read and Client2 wants to read in same RAM location at different time.

22. Client1 wants to read and Client2 wants to read in same RAM location at same time.

23. Client1 wants to write and Client2 wants to write in different RAM location at different time.

24. Client1 wants to write and Client2 wants to write in different RAM location at same time.

25. Client1 wants to read and write in the same RAM location and Client2 also wants to read in the RAM location where Client1 has written at different time.

26. Client1 wants to read and write in the same RAM location and Client2 also wants to read in the RAM location where Client1 has written at same time.

27. Client1 wants to read and write in the same RAM location and Client2 also wants to write in the RAM location where Client1 has written at different time.

28. Client1 wants to read and write in the same RAM location and Client2 also wants to write in the RAM location where Client1 has written at same time.

29. Client2 wants to read and write in the same RAM location and Client1 also wants to write in the RAM location where Client2 has written at different time.

30. Client2 wants to read and write in the same RAM location and Client1 also wants to write in the RAM location where Client2 has written at same time.

31. Client2 wants to read and write in the same RAM location and Client1 also wants to read in the RAM location where Client2 has written at same time.

32. Client2 wants to read and write in the same RAM location and Client1 also wants to read in the RAM location where Client2 has written at different time.

33. If any client resets (RST_N=0) the system at any time.

34. If any client gives the inputs until RST_DONE is high.

### 4.9.3. VHDL Testbench

LIBRARY ieee;
USE ieee.std_logic_1164.ALL;
ENTITY RAM_ARBITER_TEST IS
END RAM_ARBITER_TEST;
ARCHITECTURE behavior OF RAM_ARBITER_TEST IS
COMPONENT RAM_ARBITER_NEW
PORT(
RST_N : IN  std_logic;
CLOCK : IN  std_logic;
RST_DONE : OUT  std_logic;
RD_EN_C1 : IN  std_logic;

```
WR_EN_C1 : IN  std_logic;
RDADDR_C1 : IN  std_logic_vector(3 downto 0);
WRADDR_C1 : IN  std_logic_vector(3 downto 0);
WRDATA_C1 : IN  std_logic_vector(7 downto 0);
DATAIN_C2 : IN  std_logic_vector(7 downto 0);
REQUEST_C2 : IN  std_logic;
RD_NOT_WRITE_C2 : IN  std_logic;
ADDR_C2 : IN  std_logic_vector(3 downto 0);
RDDATA_C1 : OUT  std_logic_vector(7 downto 0);
DATAOUT_C2 : OUT  std_logic_vector(7 downto 0);
ACK_C2 : OUT  std_logic
);
END COMPONENT;
--Inputs
signal RST_N : std_logic := '0';
signal CLOCK : std_logic := '0';
signal RD_EN_C1 : std_logic := '0';
signal WR_EN_C1 : std_logic := '0';
signal RDADDR_C1 : std_logic_vector(3 downto 0) := (others => '0');
signal WRADDR_C1 : std_logic_vector(3 downto 0) := (others => '0');
signal WRDATA_C1 : std_logic_vector(7 downto 0) := (others => '0');
signal DATAIN_C2 : std_logic_vector(7 downto 0) := (others => '0');
signal REQUEST_C2 : std_logic := '0';
signal RD_NOT_WRITE_C2 : std_logic := '0';
signal ADDR_C2 : std_logic_vector(3 downto 0) := (others => '0');
--Outputs
signal RST_DONE : std_logic;
signal RDDATA_C1 : std_logic_vector(7 downto 0);
signal DATAOUT_C2 : std_logic_vector(7 downto 0);
signal ACK_C2 : std_logic;
-- Clock period definitions
constant CLOCK_period : time := 50 ns;
BEGIN
-- Instantiate the Unit Under Test (UUT)
uut: RAM_ARBITER_NEW PORT MAP (
RST_N => RST_N,
CLOCK => CLOCK,
RST_DONE => RST_DONE,
RD_EN_C1 => RD_EN_C1,
WR_EN_C1 => WR_EN_C1,
RDADDR_C1 => RDADDR_C1,
WRADDR_C1 => WRADDR_C1,
WRDATA_C1 => WRDATA_C1,
DATAIN_C2 => DATAIN_C2,
REQUEST_C2 => REQUEST_C2,
RD_NOT_WRITE_C2 => RD_NOT_WRITE_C2,
ADDR_C2 => ADDR_C2,
RDDATA_C1 => RDDATA_C1,
DATAOUT_C2 => DATAOUT_C2,
ACK_C2 => ACK_C2
);
```

```
-- Clock process definitions
CLOCK_process :process
begin
CLOCK <= '0';
wait for CLOCK_period/2;
CLOCK <= '1';
wait for CLOCK_period/2;
end process;
-- Stimulus process
stim_proc: process
begin
-- hold reset state for 100 ns.
wait for 100 ns;
----------------------------------------------------------------------------------
----Test Case-1:  Only Client1 wants to write
----------------------------------------------------------------------------------
RST_N<='1';
wait for 500 ns;
WR_EN_C1<= '1';
WRADDR_C1 <="1010";
WRDATA_C1 <="10100011";
----------------------------------------------------------------------------------
----------------------------------------------------------------------------------
----Test Case-2:  Only Client1 wants to read
----------------------------------------------------------------------------------
RST_N<='1';
wait for 500 ns;
WR_EN_C1<= '1';
WRADDR_C1 <="1010";
WRDATA_C1 <="10100011";
wait for 1700 ns;
WR_EN_C1<= '0';
RD_EN_C1<= '1';
RDADDR_C1 <="1010";
----------------------------------------------------------------------------------
----------------------------------------------------------------------------------
----Test Case-3:  Only Client2 wants to write
----------------------------------------------------------------------------------
RST_N<='1';
WR_EN_C1<= '0';
REQUEST_C2<= '1';
RD_NOT_WRITE_C2<= '0';
ADDR_C2 <="1110";
DATAIN_C2 <="11100011";
----------------------------------------------------------------------------------
----------------------------------------------------------------------------------
----Test Case-4:  Only Client2 wants to read
----------------------------------------------------------------------------------
RST_N<='1';
WR_EN_C1<= '0';
REQUEST_C2<= '1';
```

```
RD_NOT_WRITE_C2<= '0';
ADDR_C2 <="1110";
DATAIN_C2 <="11100011";
wait for 1700 ns;
WR_EN_C1<= '0';
RD_NOT_WRITE_C2<= '1';
ADDR_C2 <="1110";
```
--------------------------------------------------------------------------------------
--------------------------------------------------------------------------------------
----Test Case-5:  Client1 wants to read and write in different RAM location at same time
--------------------------------------------------------------------------------------
```
RST_N<='1';
wait for 500 ns;
WR_EN_C1<= '1';
WRADDR_C1 <="1010";
WRDATA_C1 <="10100011";
wait for 1700 ns;
RD_EN_C1<= '1';
RDADDR_C1 <="1010";
WRADDR_C1 <="1110";
WRDATA_C1 <="10111011";
```
--------------------------------------------------------------------------------------
--------------------------------------------------------------------------------------
----Test Case-6:  Client1 wants to read and write in different RAM location at different time
--------------------------------------------------------------------------------------
```
RST_N<='1';
wait for 500 ns;
WR_EN_C1<= '1';
WRADDR_C1 <="1010";
WRDATA_C1 <="10100011";
wait for 1700 ns;
RD_EN_C1<= '1';
RDADDR_C1 <="1010";
wait for 500 ns;
WRADDR_C1 <="1110";
WRDATA_C1 <="10111011";
```
--------------------------------------------------------------------------------------
--------------------------------------------------------------------------------------
----Test Case-7:  Client1 wants to read and write in same RAM location at same time
--------------------------------------------------------------------------------------
```
RST_N<='1';
wait for 500 ns;
WR_EN_C1<= '1';
WRADDR_C1 <="1010";
WRDATA_C1 <="10100011";
wait for 1700 ns;
RD_EN_C1<= '1';
RDADDR_C1 <="1010";
WRADDR_C1 <="1010";
WRDATA_C1 <="10111011";
```
--------------------------------------------------------------------------------------

```
-------------------------------------------------------------------------------------------
----Test Case-8:  Client1 wants to read and write in same RAM location at different time
-------------------------------------------------------------------------------------------
RST_N<='1';
wait for 500 ns;
WR_EN_C1<= '1';
WRADDR_C1 <="1010";
WRDATA_C1 <="10100011";
wait for 1700 ns;
RD_EN_C1<= '1';
RDADDR_C1 <="1010";
wait for 500 ns;
WRADDR_C1 <="1010";
WRDATA_C1 <="10111011";
-------------------------------------------------------------------------------------------
-------------------------------------------------------------------------------------------
----Test Case-9:  Client2 wants to read and write in different RAM location at same time.
-------------------------------------------------------------------------------------------
RST_N<='1';
WR_EN_C1<= '0';
RD_EN_C1<= '0';
REQUEST_C2<= '1';
RD_NOT_WRITE_C2<= '0';
ADDR_C2 <="1010";
DATAIN_C2 <="11100011";
wait for 1700 ns;
RD_NOT_WRITE_C2<= '0';
ADDR_C2 <="1001";
DATAIN_C2 <="00100011";
RD_NOT_WRITE_C2<= '1';
ADDR_C2 <="1010";
-------------------------------------------------------------------------------------------
-------------------------------------------------------------------------------------------
----Test Case-10:  Client2 wants to read and write in different RAM location at different time.
-------------------------------------------------------------------------------------------
RST_N<='1';
WR_EN_C1<= '0';
RD_EN_C1<= '0';
REQUEST_C2<= '1';
RD_NOT_WRITE_C2<= '0';
ADDR_C2 <="1010";
DATAIN_C2 <="11100011";
wait for 1700 ns;
RD_NOT_WRITE_C2<= '0';
ADDR_C2 <="1001";
DATAIN_C2 <="00100011";
wait for 500 ns;
RD_NOT_WRITE_C2<= '1';
ADDR_C2 <="1010";
-------------------------------------------------------------------------------------------
```

```
-----------------------------------------------------------------------------------------------
----Test Case-11:  Client2 wants to read and write in same RAM location at same time.
-----------------------------------------------------------------------------------------------
RST_N<='1';
WR_EN_C1<= '0';
RD_EN_C1<= '0';
REQUEST_C2<= '1';
RD_NOT_WRITE_C2<= '0';
ADDR_C2 <="1010";
DATAIN_C2 <="11100011";
wait for 1700 ns;
RD_NOT_WRITE_C2<= '1';
ADDR_C2 <="1010";
RD_NOT_WRITE_C2<= '0';
ADDR_C2 <="1010";
DATAIN_C2 <="00100011";
-----------------------------------------------------------------------------------------------
-----------------------------------------------------------------------------------------------
----Test Case-12:  Client2 wants to read and write in same RAM location at different time.
-----------------------------------------------------------------------------------------------
RST_N<='1';
WR_EN_C1<= '0';
RD_EN_C1<= '0';
REQUEST_C2<= '1';
RD_NOT_WRITE_C2<= '0';
ADDR_C2 <="1010";
DATAIN_C2 <="11100011";
wait for 1700 ns;
RD_NOT_WRITE_C2<= '1';
ADDR_C2 <="1010";
wait for 500 ns;
RD_NOT_WRITE_C2<= '0';
ADDR_C2 <="1010";
DATAIN_C2 <="00100011";
-----------------------------------------------------------------------------------------------
-----------------------------------------------------------------------------------------------
----Test Case-13:  Client1 wants to write and client2 wants to read in different RAM location at
same time
-----------------------------------------------------------------------------------------------
RST_N<='1';
WR_EN_C1<= '0';
REQUEST_C2<= '1';
RD_NOT_WRITE_C2<= '0';
ADDR_C2 <="1110";
DATAIN_C2 <="11100011";
wait for 1700 ns;
WR_EN_C1<= '1';
RD_EN_C1<= '0';
WRADDR_C1 <="1001";
WRDATA_C1 <="10111011";
REQUEST_C2<= '1';
```

RD_NOT_WRITE_C2<= '1';
ADDR_C2 <="1110";
----------------------------------------------------------------------------------------------------
----------------------------------------------------------------------------------------------------
----Test Case-14: Client1 wants to write and client2 wants to read in different RAM location at different time
----------------------------------------------------------------------------------------------------
RST_N<='1';
WR_EN_C1<= '0';
REQUEST_C2<= '1';
RD_NOT_WRITE_C2<= '0';
ADDR_C2 <="1110";
DATAIN_C2 <="11100011";
wait for 1700 ns;
WR_EN_C1<= '1';
RD_EN_C1<= '0';
WRADDR_C1 <="1001";
WRDATA_C1 <="10111011";
wait for 500 ns;
REQUEST_C2<= '1';
RD_NOT_WRITE_C2<= '1';
ADDR_C2 <="1110";
----------------------------------------------------------------------------------------------------
----------------------------------------------------------------------------------------------------
----Test Case-15: Client1 wants to write and client2 wants to read in same RAM location at same time
----------------------------------------------------------------------------------------------------
RST_N<='1';
WR_EN_C1<= '0';
REQUEST_C2<= '1';
RD_NOT_WRITE_C2<= '0';
ADDR_C2 <="1110";
DATAIN_C2 <="11100011";
wait for 1700 ns;
WR_EN_C1<= '1';
RD_EN_C1<= '0';
WRADDR_C1 <="1110";
WRDATA_C1 <="10111011";
REQUEST_C2<= '1';
RD_NOT_WRITE_C2<= '1';
ADDR_C2 <="1110";
----------------------------------------------------------------------------------------------------
----------------------------------------------------------------------------------------------------
----Test Case-16: Client1 wants to write and client2 wants to read in same RAM location at different time
----------------------------------------------------------------------------------------------------
RST_N<='1';
WR_EN_C1<= '0';
REQUEST_C2<= '1';
RD_NOT_WRITE_C2<= '0';
ADDR_C2 <="1110";

```
DATAIN_C2 <="11100011";
wait for 1700 ns;
WR_EN_C1<= '1';
RD_EN_C1<= '0';
WRADDR_C1 <="1110";
WRDATA_C1 <="10111011";
wait for 500 ns;
REQUEST_C2<= '1';
RD_NOT_WRITE_C2<= '1';
ADDR_C2 <="1110";
```
-------------------------------------------------------------------------------------------------------
-------------------------------------------------------------------------------------------------------
----Test Case-17:  Client1 wants to read and Client2 wants to write in same RAM location at same time
-------------------------------------------------------------------------------------------------------
```
RST_N <='1';
WR_EN_C1<= '1';
RD_EN_C1<= '0';
WRADDR_C1<="1010";
WRDATA_C1<="10101111";
wait for 1700 ns;
RD_EN_C1<='1';
WR_EN_C1<='0';
RDADDR_C1<="1010";
REQUEST_C2<='1';
RD_NOT_WRITE_C2<= '0';
ADDR_C2<="1010";
DATAIN_C2<="10111011";
```
-------------------------------------------------------------------------------------------------------
-------------------------------------------------------------------------------------------------------
----Test Case-18:  Client1 wants to read and Client2 wants to write in same RAM location at different time
-------------------------------------------------------------------------------------------------------
```
RST_N <='1';
WR_EN_C1<= '1';
RD_EN_C1<= '0';
WRADDR_C1<="1010";
WRDATA_C1<="10101111";
wait for 1700 ns;
RD_EN_C1<='1';
WR_EN_C1<='0';
RDADDR_C1<="1010";
wait for 500 ns;
REQUEST_C2<='1';
RD_NOT_WRITE_C2<= '0';
ADDR_C2<="1010";
DATAIN_C2<="10111011";
```
-------------------------------------------------------------------------------------------------------

-------------------------------------------------------------------------------------------------------
----Test Case-19:  Client1 wants to read and Client2 wants to write in different RAM location at same time
-------------------------------------------------------------------------------------------------------
RST_N <='1';
WR_EN_C1<= '1';
RD_EN_C1<= '0';
WRADDR_C1<="1000";
WRDATA_C1<="10101111";
wait for 1700 ns;
RD_EN_C1<='1';
WR_EN_C1<='0';
RDADDR_C1<="1000";
REQUEST_C2<='1';
RD_NOT_WRITE_C2<= '0';
ADDR_C2<="1010";
DATAIN_C2<="10100011";
-------------------------------------------------------------------------------------------------------
-------------------------------------------------------------------------------------------------------
----Test Case-20:  Client1 wants to read and Client2 wants to write in different RAM location at different time
-------------------------------------------------------------------------------------------------------
RST_N <='1';
WR_EN_C1<= '1';
RD_EN_C1<= '0';
WRADDR_C1<="1000";
WRDATA_C1<="10101111";
wait for 1700 ns;
RD_EN_C1<='1';
WR_EN_C1<='0';
RDADDR_C1<="1000";
wait for 500 ns;
REQUEST_C2<='1';
RD_NOT_WRITE_C2<= '0';
ADDR_C2<="1010";
DATAIN_C2<="10100011";
-------------------------------------------------------------------------------------------------------
-------------------------------------------------------------------------------------------------------
----Test Case-21:  Client1 wants to read and Client2 wants to read in same RAM location at different time.
-------------------------------------------------------------------------------------------------------
RST_N <='1';
WR_EN_C1<= '1';
RD_EN_C1<= '0';
WRADDR_C1<="1010";
WRDATA_C1<="10101111";
wait for 1700 ns;
RD_EN_C1<='1';
WR_EN_C1<='0';
RDADDR_C1<="1010";

```
wait for 300 ns;
REQUEST_C2<='1';
RD_NOT_WRITE_C2<='1';
ADDR_C2<="1010";
wait for 200 ns;
RD_EN_C1<='0';
```
-----------------------------------------------------------------------------------------------------
-----------------------------------------------------------------------------------------------------

----Test Case-22:  Client1 wants to read and Client2 wants to read in same RAM location at same time.
-----------------------------------------------------------------------------------------------------
```
RST_N <='1';
WR_EN_C1<= '1';
RD_EN_C1<= '0';
WRADDR_C1<="1010";
WRDATA_C1<="10101111";
wait for 1700 ns;
RD_EN_C1<='1';
WR_EN_C1<='0';
RDADDR_C1<="1010";
REQUEST_C2<='1';
RD_NOT_WRITE_C2<='1';
ADDR_C2<="1010";
```
-----------------------------------------------------------------------------------------------------
-----------------------------------------------------------------------------------------------------

----Test Case-23:  Client1 wants to read and Client2 wants to read in different RAM location at different time.
-----------------------------------------------------------------------------------------------------
```
RST_N <='1';
WR_EN_C1<= '1';
RD_EN_C1<= '0';
WRADDR_C1<="1001";
WRDATA_C1<="10101111";
wait for 1700 ns;
RD_EN_C1<='1';
WR_EN_C1<='0';
RDADDR_C1<="1001";
wait for 300 ns;
REQUEST_C2<='1';
RD_NOT_WRITE_C2<='1';
ADDR_C2<="1010";
wait for 200 ns;
RD_EN_C1<='0';
```
-----------------------------------------------------------------------------------------------------
-----------------------------------------------------------------------------------------------------

----Test Case-24:  Client1 wants to read and Client2 wants to read in different RAM location at same time.
-----------------------------------------------------------------------------------------------------
```
RST_N <='1';
WR_EN_C1<= '1';
RD_EN_C1<= '0';
```

```
WRADDR_C1<="1001";
WRDATA_C1<="10101111";
wait for 1700 ns;
RD_EN_C1<='1';
WR_EN_C1<='0';
RDADDR_C1<="1001";
REQUEST_C2<='1';
RD_NOT_WRITE_C2<='1';
ADDR_C2<="1010";
```
---------------------------------------------------------------------------------------------------------------
---------------------------------------------------------------------------------------------------------------
----Test Case-25:  Client1 wants to read and write in the same RAM location and Client2 also
wants to read in the RAM location where Client1 has written at different time.
---------------------------------------------------------------------------------------------------------------
```
RST_N <='1';
WR_EN_C1<= '1';
RD_EN_C1<= '0';
WRADDR_C1<="1001";
WRDATA_C1<="10101111";
wait for 1700 ns;
RD_EN_C1<='1';
RDADDR_C1<="1001";
WRADDR_C1<="1001";
WRDATA_C1<="10100011";
wait for 300 ns;
REQUEST_C2<='1';
RD_NOT_WRITE_C2<='1';
ADDR_C2<="1001";
wait for 200 ns;
RD_EN_C1<='0';
```
---------------------------------------------------------------------------------------------------------------
---------------------------------------------------------------------------------------------------------------
----Test Case-26:  Client1 wants to read and write in the same RAM location and Client2 also
wants to read in the RAM location where Client1 has written at same time.
---------------------------------------------------------------------------------------------------------------
```
RST_N <='1';
WR_EN_C1<= '1';
RD_EN_C1<= '0';
WRADDR_C1<="1001";
WRDATA_C1<="10101111";
wait for 1700 ns;
RD_EN_C1<='1';
RDADDR_C1<="1001";
WRADDR_C1<="1001";
WRDATA_C1<="10100011";
REQUEST_C2<='1';
RD_NOT_WRITE_C2<='1';
ADDR_C2<="1001";
```
---------------------------------------------------------------------------------------------------------------

```
-------------------------------------------------------------------------------------
----Test Case-27:  Client1 wants to read and write in the same RAM location and Client2 also
wants to write in the RAM location where Client1 has written at different time.
-------------------------------------------------------------------------------------
RST_N <='1';
WR_EN_C1<= '1';
RD_EN_C1<= '0';
WRADDR_C1<="1001";
WRDATA_C1<="10101111";
wait for 1700 ns;
RD_EN_C1<='1';
RDADDR_C1<="1001";
WRADDR_C1<="1001";
WRDATA_C1<="10100011";
wait for 300 ns;
REQUEST_C2<='1';
RD_NOT_WRITE_C2<='0';
ADDR_C2 <="1001";
DATAIN_C2 <="11100011";
wait for 200 ns;
WR_EN_C1<='0';
-------------------------------------------------------------------------------------
-------------------------------------------------------------------------------------
----Test Case-28:  Client1 wants to read and write in the same RAM location and Client2 also
wants to write in the RAM location where Client1 has written at same time.
-------------------------------------------------------------------------------------
RST_N <='1';
WR_EN_C1<= '1';
RD_EN_C1<= '0';
WRADDR_C1<="1001";
WRDATA_C1<="10101111";
wait for 1700 ns;
RD_EN_C1<='1';
RDADDR_C1<="1001";
WRADDR_C1<="1001";
WRDATA_C1<="10100011";
REQUEST_C2<='1';
RD_NOT_WRITE_C2<='0';
ADDR_C2 <="1001";
DATAIN_C2 <="11100011";
-------------------------------------------------------------------------------------
-------------------------------------------------------------------------------------
----Test Case-29:  Client2 wants to read and write in the same RAM location and Client1 also
wants to write in the RAM location where Client2 has written at different time.
-------------------------------------------------------------------------------------
RST_N <='1';
WR_EN_C1<= '0';
RD_EN_C1<= '0';
REQUEST_C2<='1';
RD_NOT_WRITE_C2<='0';
```

ADDR_C2 <="1001";
DATAIN_C2 <="11100011";
wait for 1700 ns;
RD_NOT_WRITE_C2<='1';
ADDR_C2 <="1001";
wait for 300 ns;
WRADDR_C1<="1001";
WRDATA_C1<="10101111";
wait for 200 ns;
WR_EN_C1<= '1';
-------------------------------------------------------------------------------------------------------------
-------------------------------------------------------------------------------------------------------------
----Test Case-30:  Client2 wants to read and write in the same RAM location and Client1 also
wants to write in the RAM location where Client2 has written at same time.
-------------------------------------------------------------------------------------------------------------
RST_N <='1';
WR_EN_C1<= '0';
RD_EN_C1<= '0';
REQUEST_C2<='1';
RD_NOT_WRITE_C2<='0';
ADDR_C2 <="1001";
DATAIN_C2 <="11100011";
wait for 1700 ns;
WR_EN_C1<= '1';
RD_NOT_WRITE_C2<='1';
ADDR_C2 <="1001";
WRADDR_C1<="1001";
WRDATA_C1<="10101111";
-------------------------------------------------------------------------------------------------------------
-------------------------------------------------------------------------------------------------------------
----Test Case-31:  Client2 wants to read and write in the same RAM location and Client1 also
wants to read in RAM location where Client2 has written at same time.
-------------------------------------------------------------------------------------------------------------
RST_N <='1';
WR_EN_C1<= '0';
RD_EN_C1<= '0';
REQUEST_C2<='1';
RD_NOT_WRITE_C2<='0';
ADDR_C2 <="1001";
DATAIN_C2 <="11100011";
wait for 1700 ns;
RD_EN_C1<= '1';
RD_NOT_WRITE_C2<='1';
ADDR_C2 <="1001";
RDADDR_C1 <="1001";
-------------------------------------------------------------------------------------------------------------
-------------------------------------------------------------------------------------------------------------
----Test Case-32:  Client2 wants to read and write in the same RAM location and Client1 also
wants to read in the RAM location where Client2 has written at different time.
-------------------------------------------------------------------------------------------------------------
RST_N <='1';

```
WR_EN_C1<= '0';
RD_EN_C1<= '0';
REQUEST_C2<='1';
RD_NOT_WRITE_C2<='0';
ADDR_C2 <="1001";
DATAIN_C2 <="11100011";
wait for 1700 ns;
RD_NOT_WRITE_C2<='1';
ADDR_C2 <="1001";
wait for 300 ns;
RD_EN_C1<= '1';
RDADDR_C1 <="1001";
```
-------------------------------------------------------------------------------------------------------------
-------------------------------------------------------------------------------------------------------------
----Test Case-33:  If any client resets (RST_N=0) the system at any time.
-------------------------------------------------------------------------------------------------------------
```
RST_N<='1';
WR_EN_C1<= '1';
RD_EN_C1<= '0';
WRADDR_C1<="1010";
WRDATA_C1<="10101111";
wait for 1700 ns;
RST_N<='0';
RD_EN_C1<= '1';
WR_EN_C1<='0';
RDADDR_C1<="1010";
REQUEST_C2<='1';
RD_NOT_WRITE_C2<='0';
ADDR_C2<="0110";
DATAIN_C2<="10111011";
wait for 500 ns;
RST_N<='1';
RDADDR_C1<="1010";
wait for 300 ns;
RDADDR_C1<="0110";
```
-------------------------------------------------------------------------------------------------------------
-------------------------------------------------------------------------------------------------------------
----Test Case-34:  If any client gives the inputs until RST_DONE is high.
-------------------------------------------------------------------------------------------------------------
```
RST_N<='1';
wait for 200 ns;
WR_EN_C1<= '1';
RD_EN_C1<= '0';
WRADDR_C1<="1010";
WRDATA_C1<="10101111";
RST_N<='0';
wait for 1700 ns;
RD_EN_C1<= '1';
WR_EN_C1<='0';
RDADDR_C1<="1010";
REQUEST_C2<='1';
```

```
RD_NOT_WRITE_C2<='0';
ADDR_C2<="0110";
DATAIN_C2<="10111011";
wait for 800 ns;
RST_N<='1';
RDADDR_C1<="1010";
-----------------------------------------------------------------------------------------------------------
wait;
end process;
END;

-----------------------------------------------------------------------------------------------------------
```

## 4.9.4. Waveforms

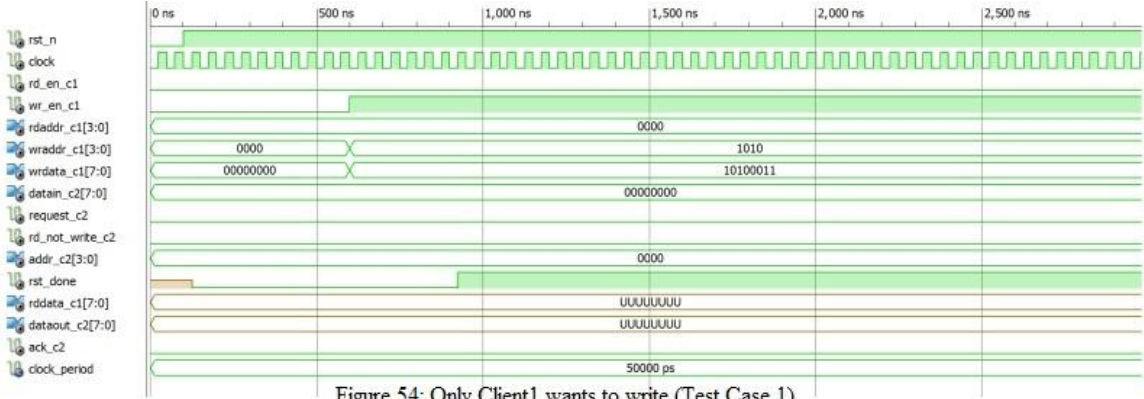

Figure 54: Only Client1 wants to write (Test Case 1)

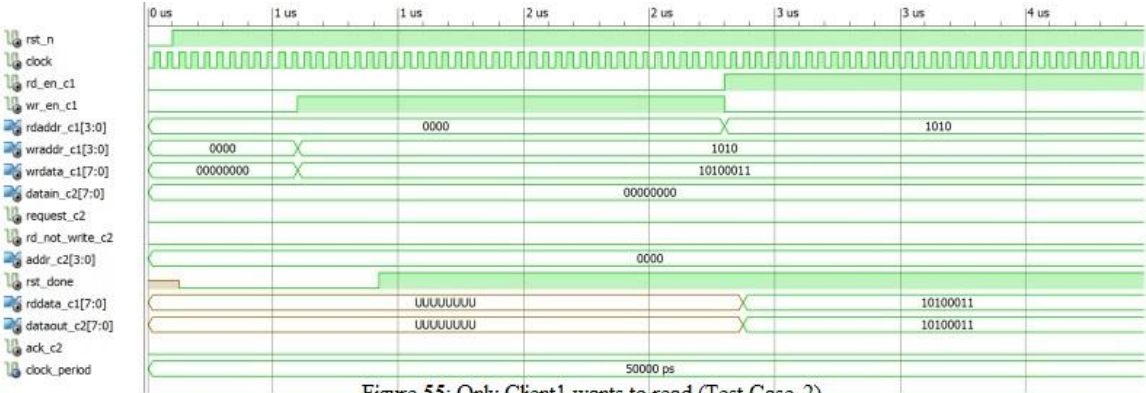

Figure 55: Only Client1 wants to read (Test Case-2)

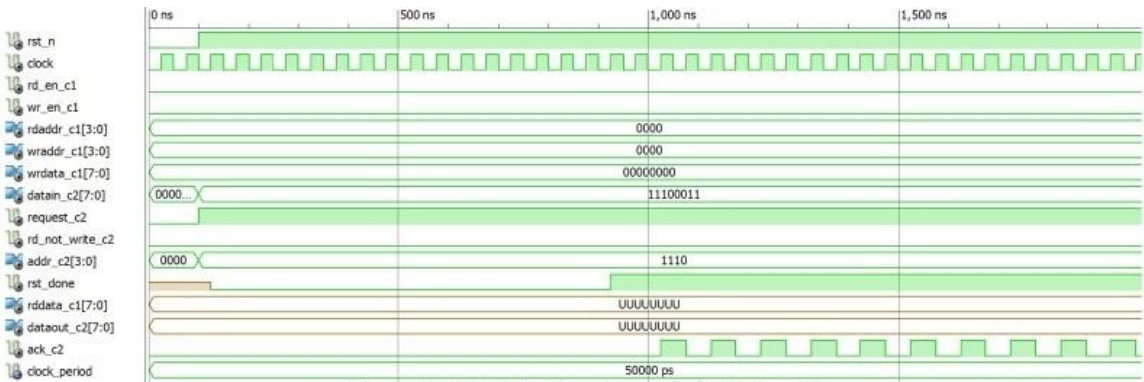

Figure 56 : Only Client2 wants to write (Test Case 3)

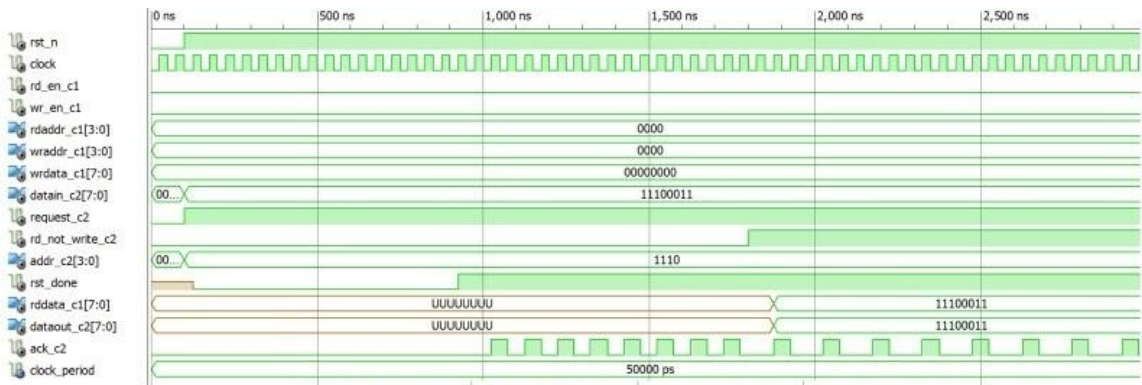

Figure 57: Only Client2 wants to read (Test Case 4)

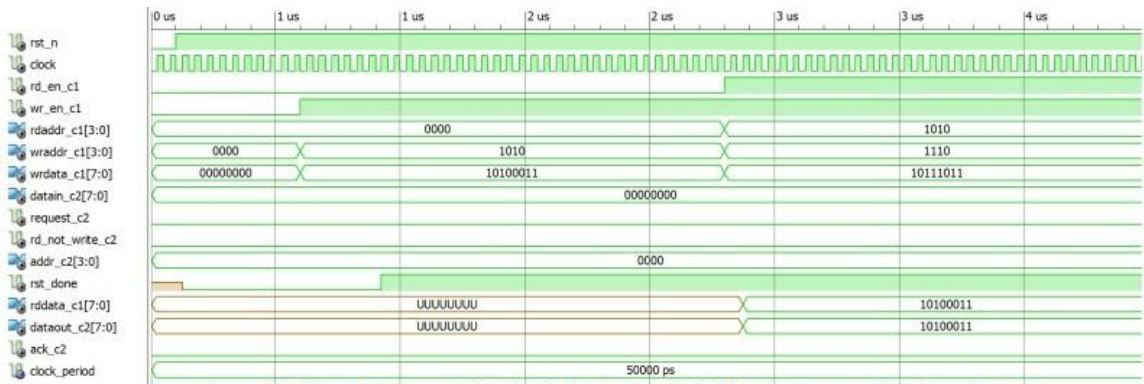

Figure 58: Client1 wants to read and write in different RAM location at same time (Test Case 5)

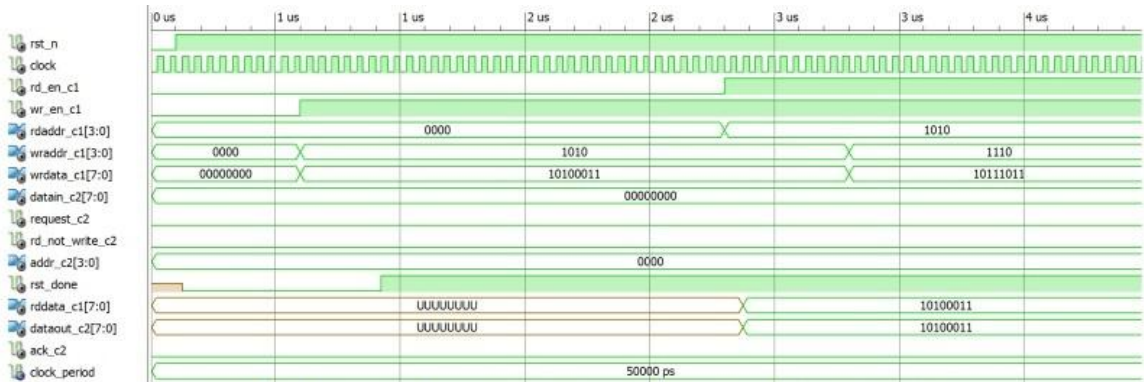

Figure 59: Client1 wants to read and write in different RAM location at different time (Test Case 6)

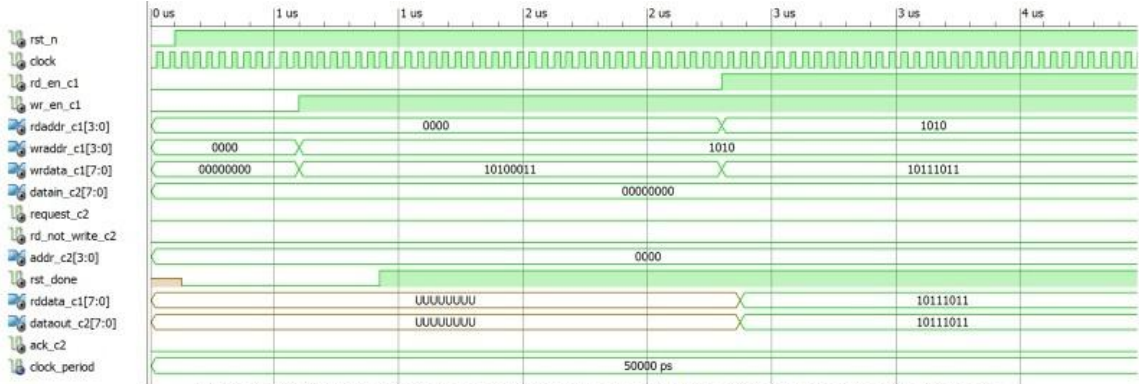

Figure 60: Client1 wants to read and write in same RAM location at same time (Test Case 7)

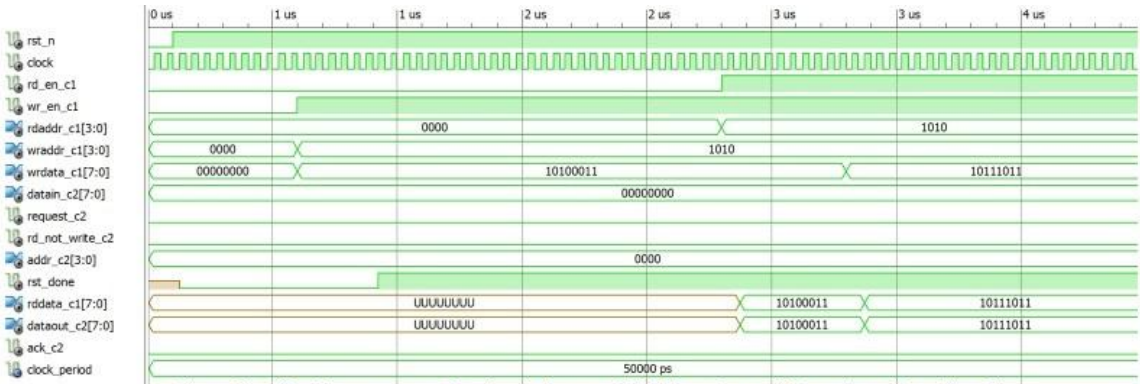

Figure 61: Client1 wants to read and write in same RAM location at different time (Test Case 8)

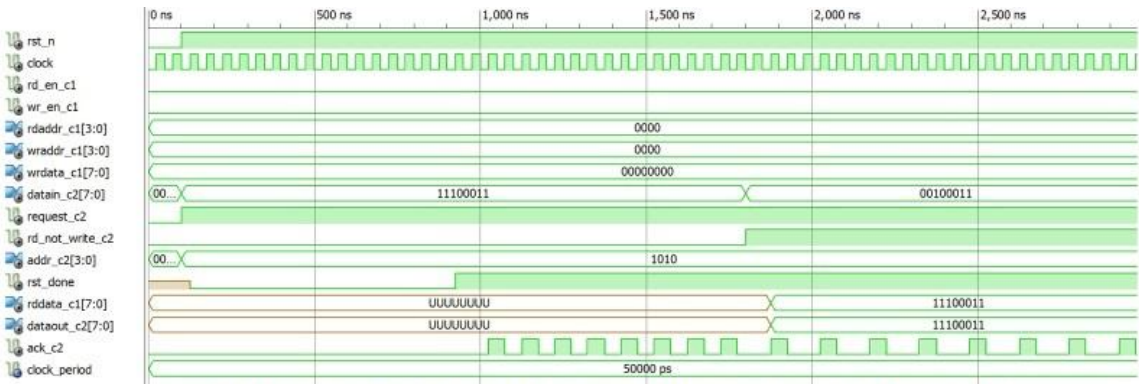

Figure 62: Client2 wants to read and write in different RAM location at same time (Test Case 9)

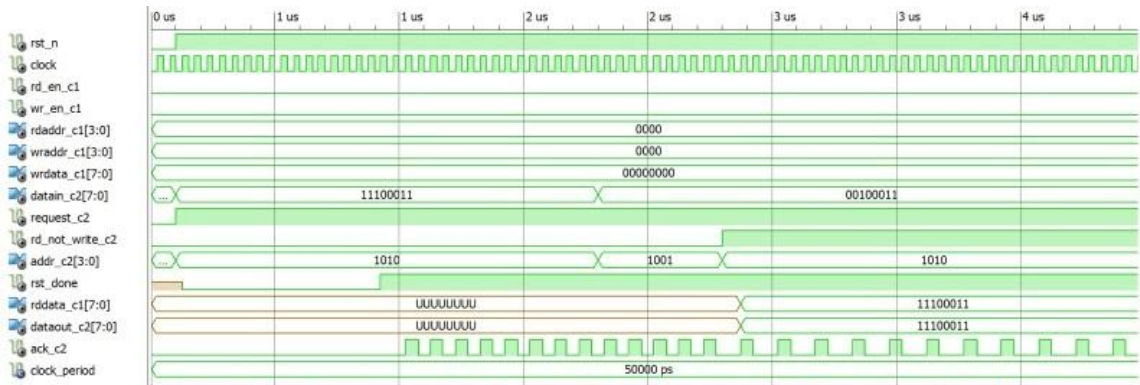

Figure 63: Client2 wants to read and write in different RAM location at different time (Test Case 10)

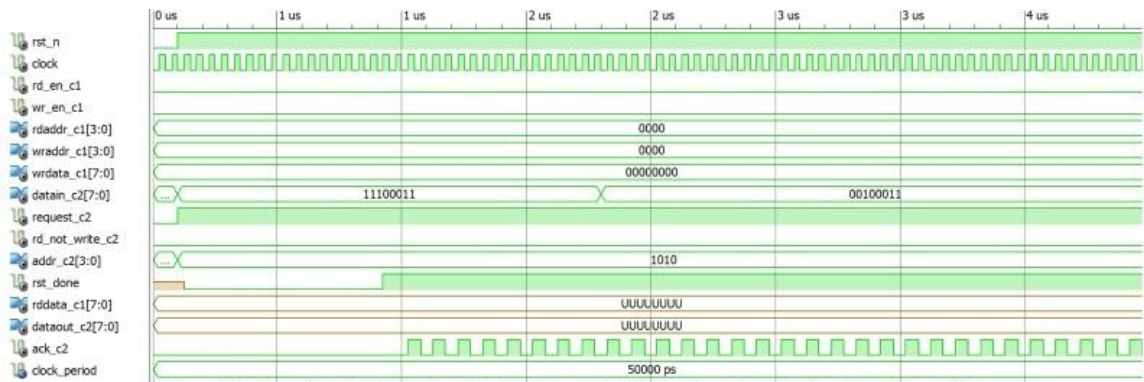

Figure 64: Client2 wants to read and write in same RAM location at same time (Test Case 11)

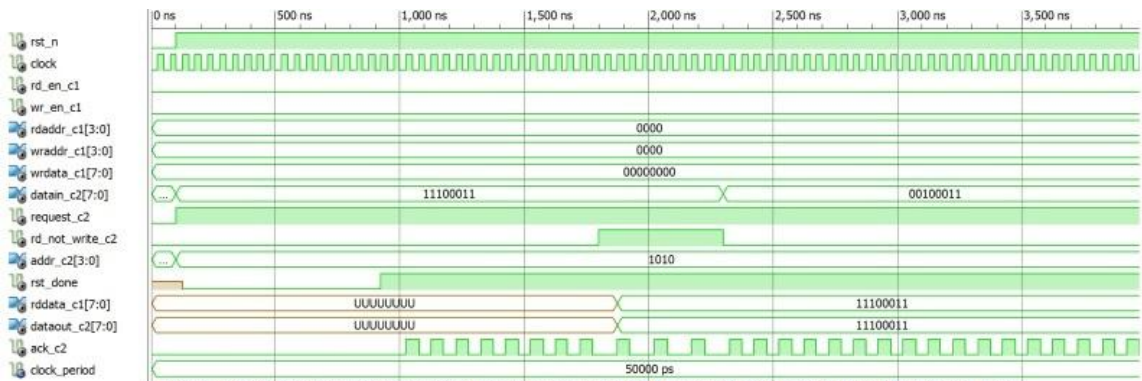

Figure 65: Client2 wants to read and write in same RAM location at different time (Test Case 12)

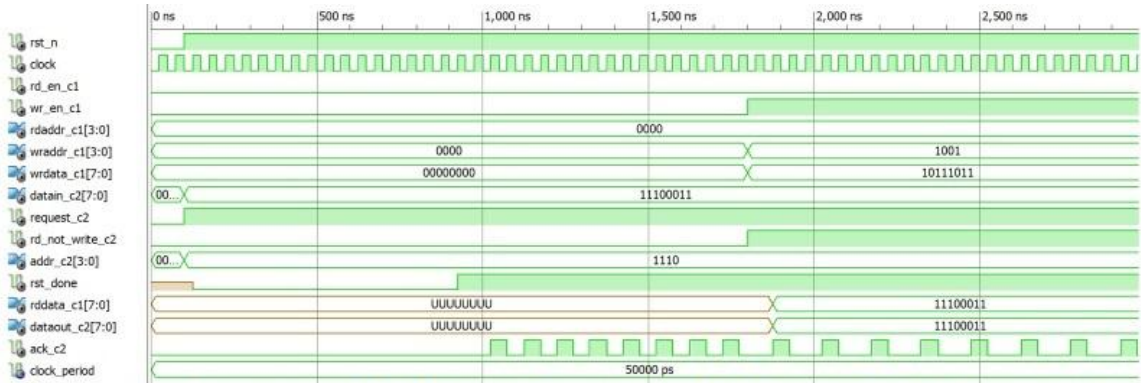

Figure 66: Client1 wants to write and Client2 wants to read in different RAM location at same time (Test Case 13)

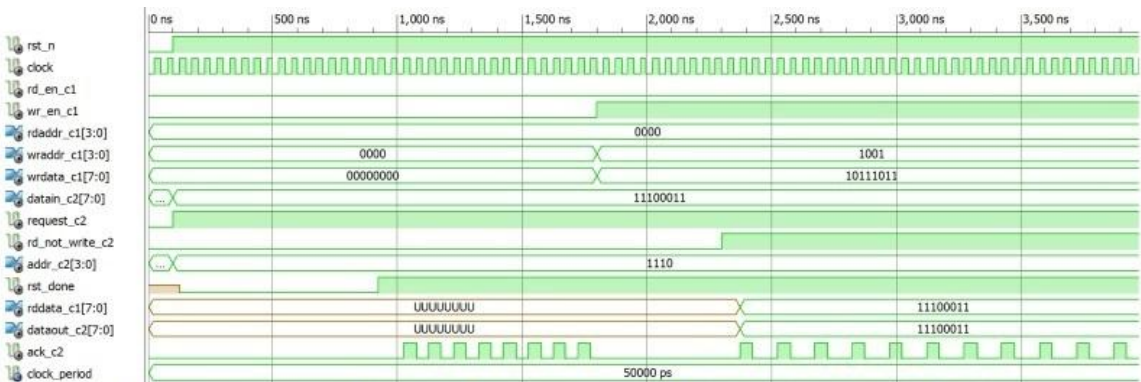

Figure 67: Client1 wants to write and client2 wants to read in different RAM location at different time (Test Case 14)

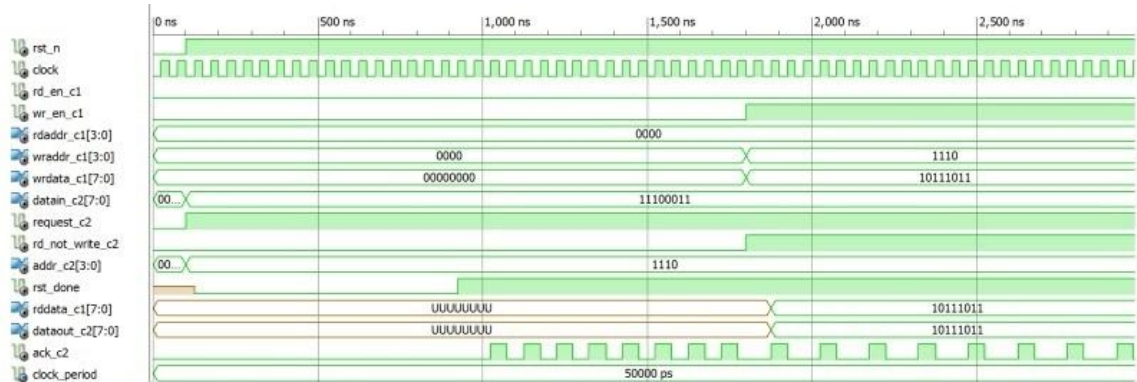

Figure 68: Client1 wants to write and Client2 wants to read in same RAM location at same time. (Test Case 15)

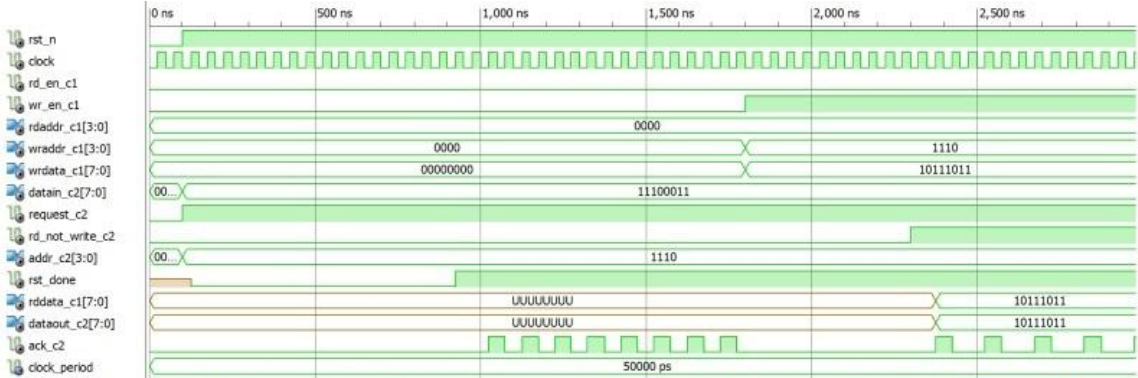

Figure 69: Client1 wants to write and Client2 wants to read in same RAM location at different time (Test Case 16)

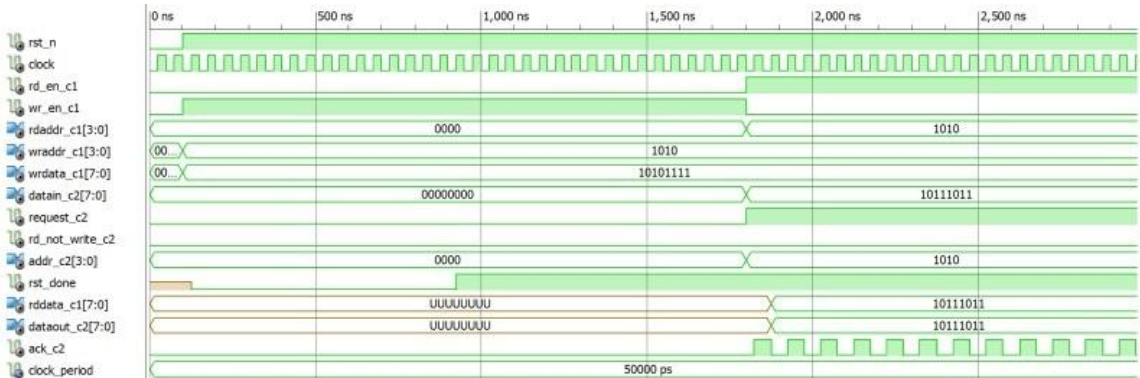

Figure 70: Client1 wants to read and Client2 wants to write in same RAM location at same time (Test Case 17)

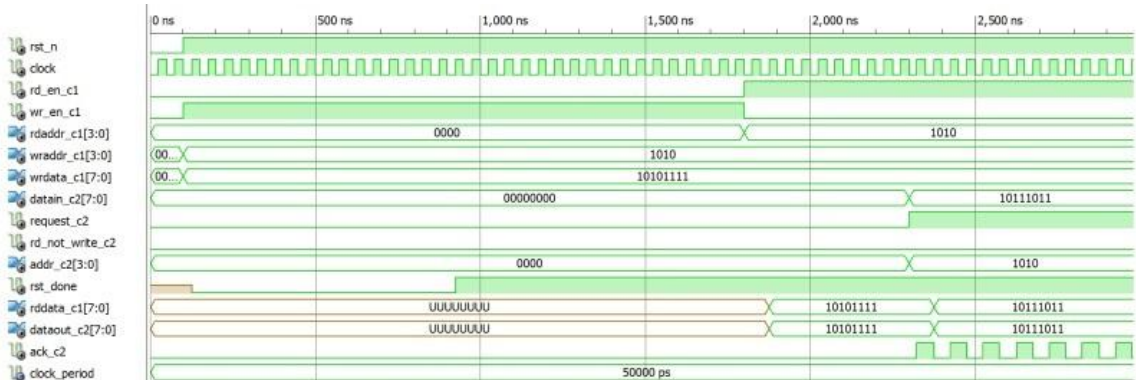

Figure 71: Client1 wants to read and Client2 wants to write in same RAM location at different time (Test Case 18)

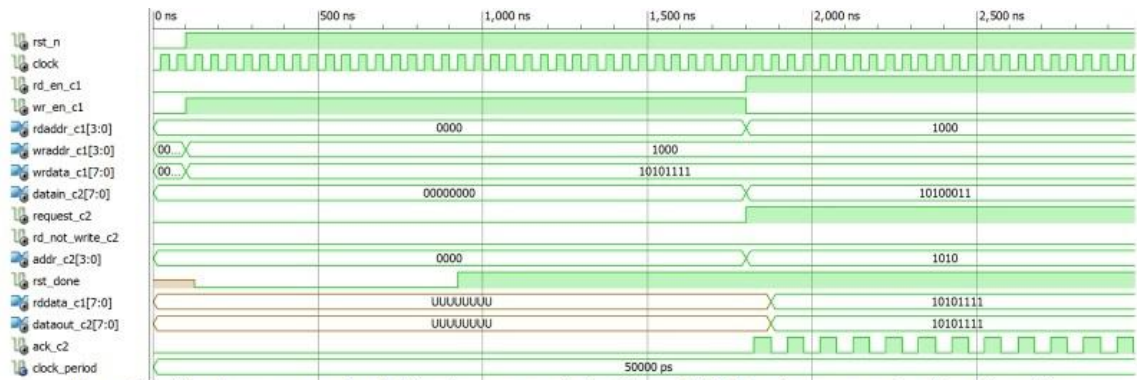

Figure 72: Client1 wants to read and Client2 wants to write in different RAM location at same time (Test Case 19)

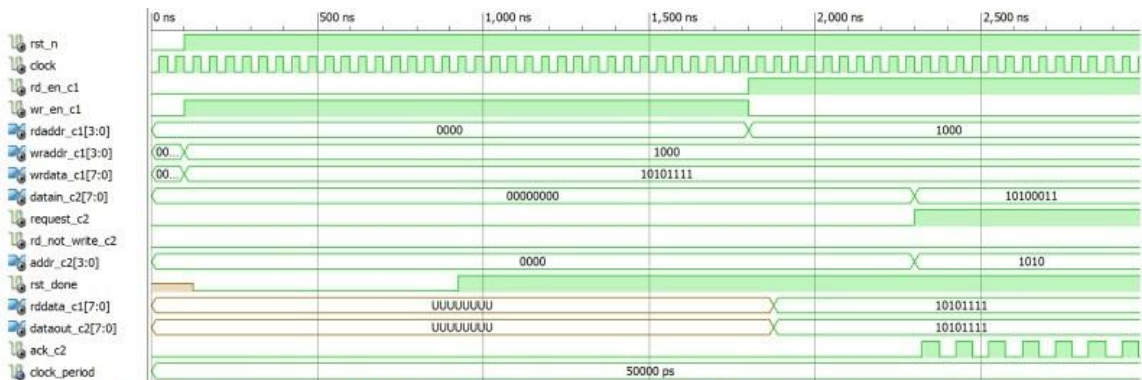

Figure 73: Client1 wants to read and Client2 wants to write in different RAM location at different time (Test Case 20)

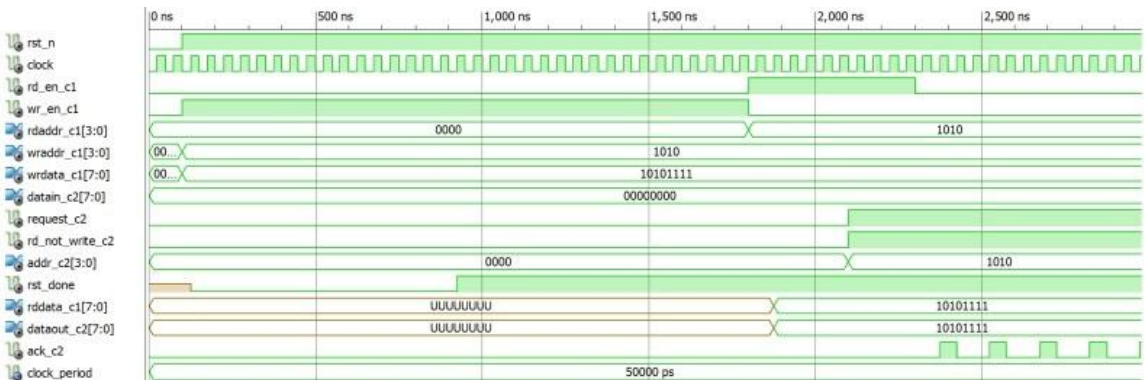

Figure 74: Client1 wants to read and Client2 wants to read in same RAM location at different time (Test Case 21)

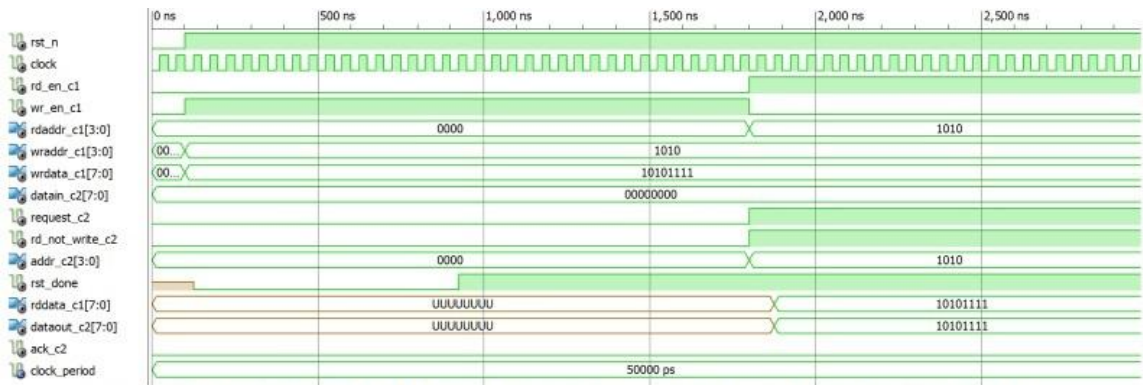

Figure 75: Client1 wants to read and Client2 wants to read in same RAM location at same time (Test Case 22)

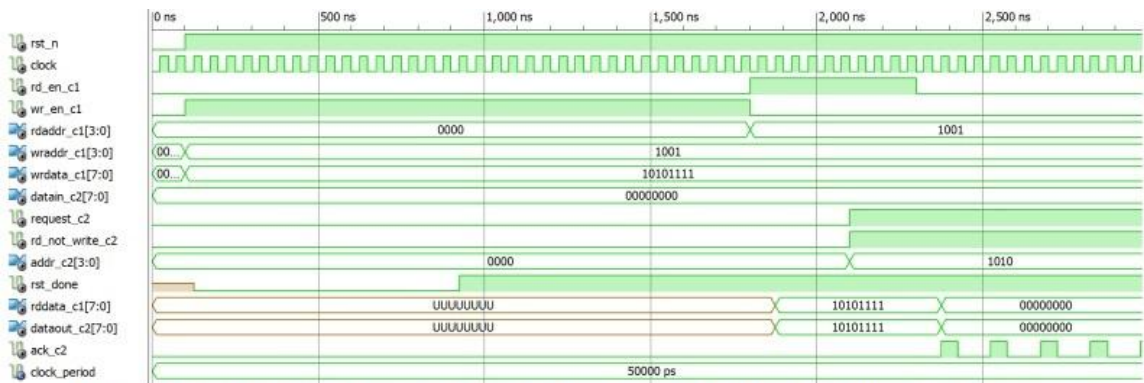

Figure 76: Client1 wants to write and Client2 wants to write in different RAM location at different time (Test Case 23)

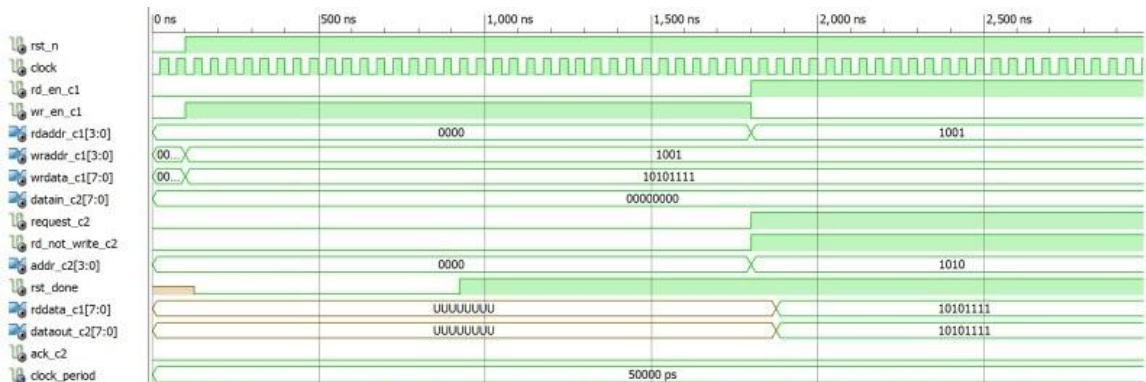

Figure 77: Client1 wants to write and Client2 wants to write in different RAM location at same time (Test Case 24)

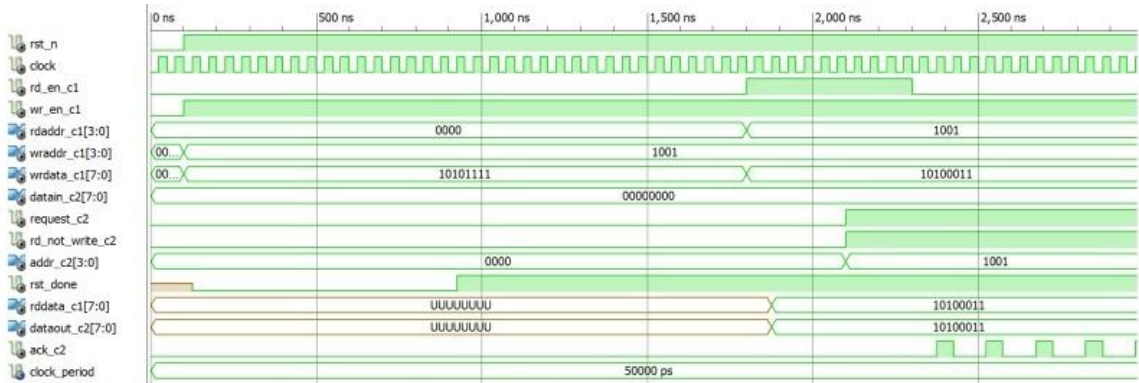

Figure 78: Client1 wants to read and write in the same RAM location and Client2 also wants to read in the RAM location where Client1 has written at different time (Test Case 25)

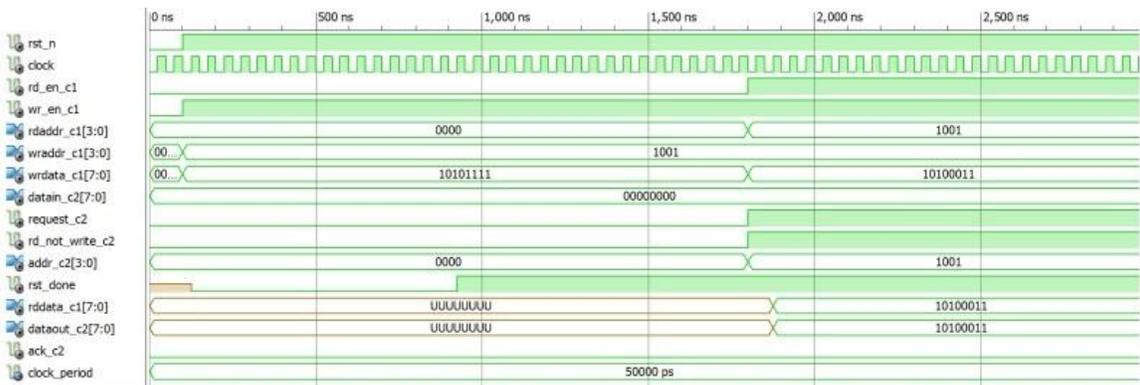

Figure 79: Client1 wants to read and write in the same RAM location and Client2 also wants to read in the RAM location where Client1 has written at same time (Test Case 26)

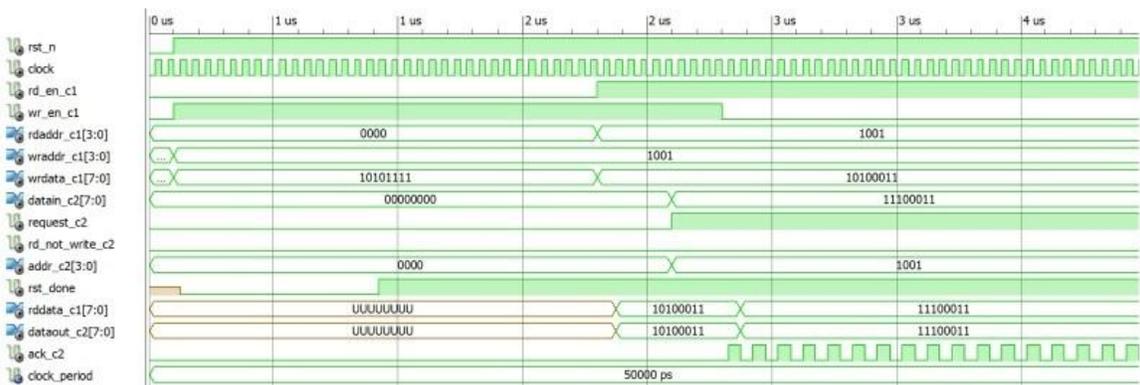

Figure 80: Client1 wants to read and write in the same RAM location and Client2 also wants to write in the RAM location where Client1 has written at different time (Test Case 27)

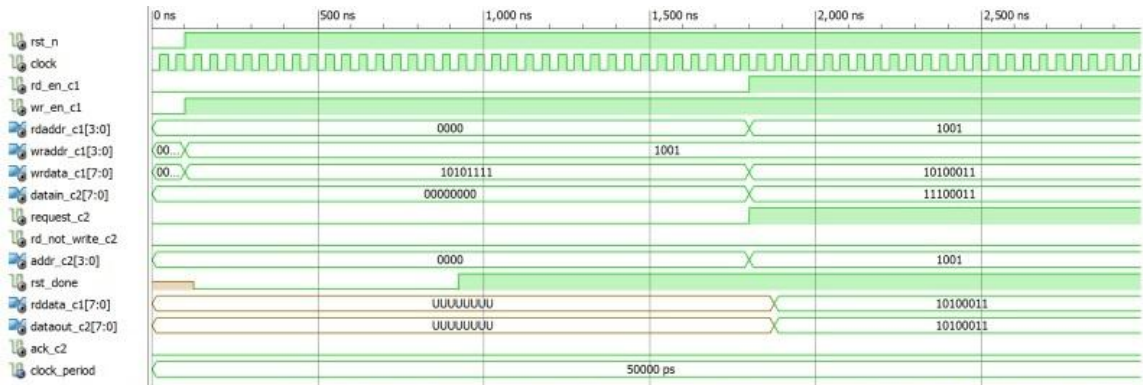

Figure 81: Client1 wants to read and write in the same RAM location and Client2 also wants to write in the RAM location where Client1 has written at same time (Test Case 28)

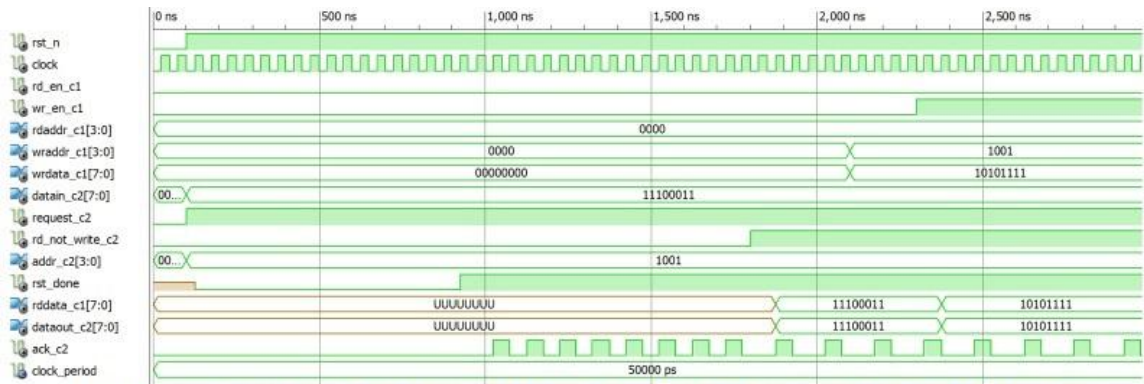

Figure 82: Client2 wants to read and write in the same RAM location and Client1 also wants to write in the RAM location where Client2 has written at different time (Test Case 29)

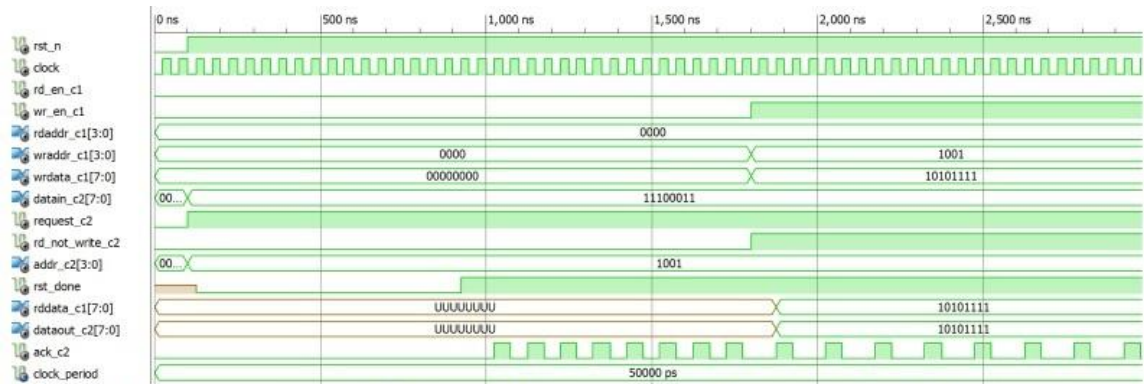

Figure 83: Client2 wants to read and write in the same RAM location and Client1 also wants to write in the RAM location where Client2 has written at same time (Test Case 30)

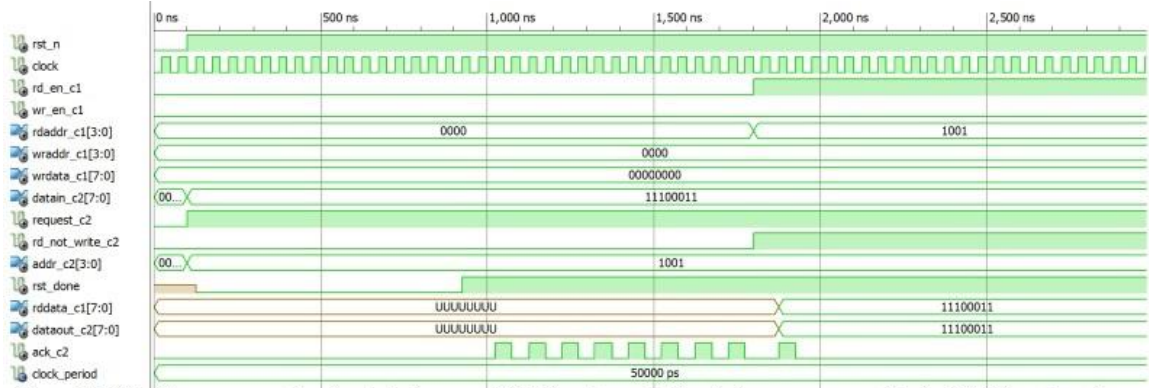

Figure 84: Client2 wants to read and write in the same RAM location and Client1 also wants to read in the RAM location where Client2 has written at same time (Test Case 31)

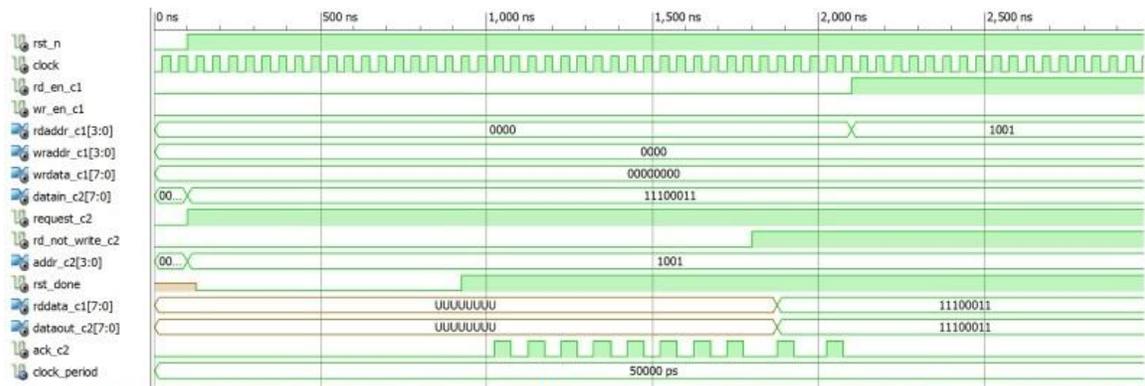

Figure 85: Client2 wants to read and write in the same RAM location and Client1 also wants to read in the RAM location where Client2 has written at different time (Test Case 32)

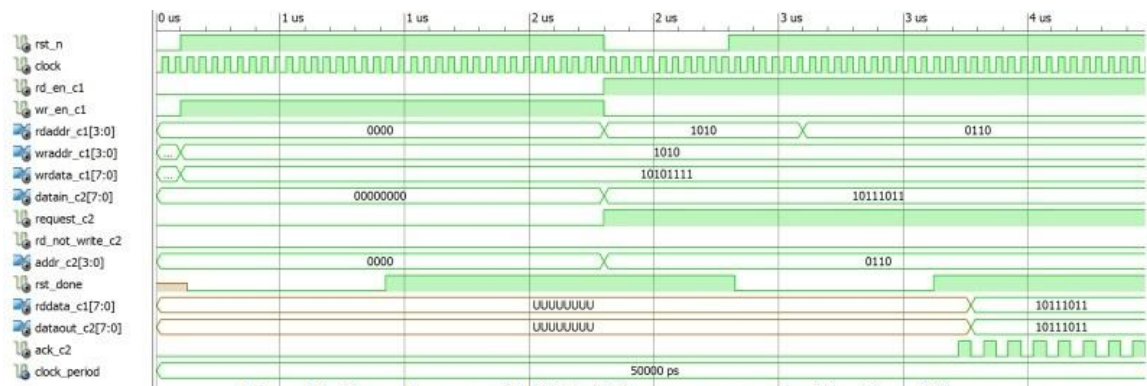

Figure 86 : If any client resets (RST_N=0) the system at any time (Test Case 33)

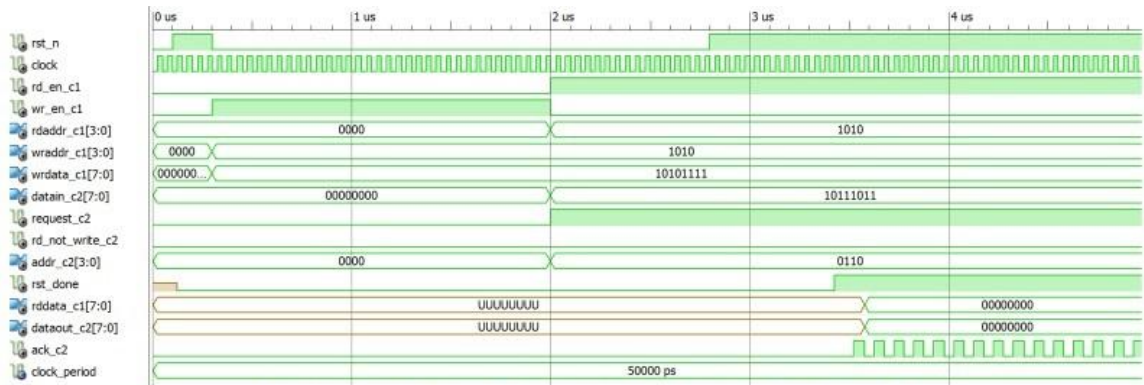

Figure 87: If any client gives the inputs until RST_DONE is high (Test Case 34)

## 4.9.5. Analysis of the Test Cases

**1. Only Client1 wants to write.**
In this case, when the RST_N pin is enabled, the arbiter starts its operation. Within the first RAM_DEPTH cycle, Client1 enables its WR_EN_C1 pin and writes the data "1010011" at the address location "1010".From the point the RAM Arbiter operation starts to take place within 500 ns the RST_DONE pin goes high. A typical RAM_DEPTH cycle duration is of 1700 ns.

**2. Only Client1 wants to read.**
The Arbiter starts working as soon as the RST_N pin is enabled. The clock period is set to 50ns. Within the first RAM_DEPTH cycle, a data "10100011" is written at the location "1010". The WR_EN_C1 pin is enabled to facilitate the operation. During the next RAM_DEPTH cycle, Client1 wishes to read at the RAM location "1010" by enabling the RD_EN_C1 pin. Immediately, the information is sent to the RAM. The RDDATA_C1 pin gives the output "10100011".

**3. Only Client2 wants to write.**
Here in this case, the Arbiter starts working as soon as the RST_N pin is set high. The clock period is set to 50 ns. As the arbitration scheme used is a priority based one, so Client2 needs to issue a signal on the REQUEST_C2 pin in order to access the RAM. Here the priority is given to Client1.Within the first RAM_DEPTH cycle, REQUEST_C2 pin is enabled to allow Client2 to access the RAM. Client2 uses two states of a single pin RD_NOT_WRITE, to perform both of its read and write operation. This pin only works if REQUEST_C2 pin is enabled. Upon enabling REQUEST_C2, RD_NOT_WRITE is set to low to allow for the write operation. Data "11100011" is written at the location "1110". Contrary to the previous case, here the RST_DONE goes high after a relative long time as since in this priority based scheme, Client1 enjoys a higher priority compared to Client2; hence it takes a relatively long time to set up the system for Client2. The arbiter acknowledges the write operation by setting up periodic pulses at the ACK_C2 output pin.

**4. Only Client2 wants to read.**

In this case, within the first RAM_DEPTH cycle, data "11100011" is written in the address location "1110" by enabling the REQUEST_C2 pin and setting RD_NOT_WRITE pin to an active low state. The ACK_C2 acknowledges the write request with normal periodic pulses. However in the second RAM_DEPTH cycle, Client2 activates the high signal on RD_NOT_WRITE pin and issues a signal to the arbiter to read from location "1110". The read operation is acknowledged by the Arbiter by issuing clock pulses at the ACK_C2 pin whose clock period is two times that of the former one. The output RDDATA_C1 pin gives "11100011".

**5. Client1 wants to read and write in different RAM location at same time.**

Within the first RAM_DEPTH cycle data "10100011" is written at the address location "1010" by Client1 by enabling the WR_EN_C1 pin. In the successive RAM_DEPTH cycle, Client1 wishes to perform the simultaneous operation of reading and writing at different RAM location by enabling both the RD_EN_C1 and WR_EN_C1 pins. This is handled successfully by the Arbiter. Reading was done at the address location "1010" and writing of the data "10111011" was done at the location "1110".

**6. Client1 wants to read and write in different RAM location at different time.**

Within the first RAM_DEPTH cycle data "10100011" is written at the address location "1010" by Client1 by enabling the WR_EN_C1 pin. In the successive RAM_DEPTH cycle, Client1 wishes to perform the operation of reading and writing at different RAM location by enabling both the RD_EN_C1 and WR_EN_C1 pins. However in this case the client chooses to read and write at different time within the same RAM_DEPTH cycle. As shown there is a time delay of 500 ns between the successive operations. Reading was done at the address location "1010" and writing of the data "10111011" was done at the location "1110".

**7. Client1 wants to read and write in same RAM location at same time.**

In this case, within the first RAM_DEPTH cycle data "10100011" is written at the address location "1010" by Client1, enabling the WR_EN_C1 pin. In the successive RAM_DEPTH cycle, Client1 wishes to perform the operation of reading and writing at the same RAM location by enabling both the RD_EN_C1 and WR_EN_C1 pins. Reading was done at the address location "1010" and writing of the data "10111011" was done at the same location. In such cases we deploy the strategy of making the client read the updated value compared to the old one. One who has gone through the code would know that we made use of a temporary register in the Arbiter itself which holds the data for the current RAM_DEPTH cycle. Since such is the case, one could see that we get the output "101110011" on the output pin RDDATA_C1, which is nothing but the temporary data disguised to the client who perceives it to be coming from the RAM.

**8. Client1 wants to read and write in same RAM location at different time.**

Within the first RAM_DEPTH cycle data "10100011" is written at the address location "1010" by Client1, enabling the WR_EN_C1 pin. In the successive RAM_DEPTH cycle, Client1 wishes to perform the operation of reading and writing at the same RAM location by enabling both the RD_EN_C1 and WR_EN_C1 pins. Reading was done at the address location "1010" and writing of the data "10111011" was done at the same location. In such a case we deploy the strategy of making the client read the updated value compared to the old one. However in this case the client chooses to read and write at different time within the same RAM_DEPTH cycle. As shown there is a time delay of 500ns between the successive operations. We made use of a temporary register in the Arbiter itself which holds the data for the current clock cycle. Since such is the case, one could see that we get the output "101110011" on the output pin RDDATA_C1, which is nothing but the temporary data disguised to the client who perceives it to be coming from the RAM.

**9. Client2 wants to read and write in different RAM location at same time.**

Client2 initially performs the write operation in the first RAM_DEPTH cycle at the address location "1010". The data is "11100011". This is achieved by setting the REQUEST_C2 and the RD_NOT_WRITE pin to active high and low respectively. During the next RAM_DEPTH cycle, Client2 performs the read and write operation at the locations "1010" and "1001" respectively. But if we look into the graph closely, one would find no trace of the data being written at the address location "1001". This is because in the third RAM_DEPTH cycle, since Client2 uses a single pin for read & write operation so it can only perform a single operation at a given time contrary to Client1. In the third RAM_DEPTH cycle, RD_NOT_WRITE pin is set high, and the read operation is performed. We could see that data "00100011" is given at the pin DATAIN_C2 but the data is waiting to get recognized at the arbiter, which waits for the RD_NOT_WRITE signal to go low. However in this case since the RAM does not read & write at the same RAM location hence, there is no address clash problem.

**10. Client2 wants to read and write in different RAM location at different time.**

Client2 initially performs the write operation in the first RAM_DEPTH cycle at the address location "1010". The data is "11100011". This is achieved by setting the REQUEST_C2 and the RD_NOT_WRITE pin to active high and low respectively. During the next RAM_DEPTH cycle, Client2 performs the read and write operation at the locations "1010" and "1001" respectively at a different time. Contrary to the previous case, here we could see that Client2 could perform both the operations very smoothly without hiccups. During the second RAM_DEPTH cycle, write operation is performed at the location "1001", which is immediately sent to the input pins of the RAM. In the successive clock cycle, the reading operation is performed at the address location "1010". The interval between the two operations is 500 ns. The ACK_C2 signal gives us the correct signals verifying our claims.

**11. Client2 wants to read and write in same RAM location at same time.**

This case is similar in working to Test Case 9; the only difference is that the RAM locations are the same for both read and write operation. Since Client2 can perform only a single operation at any given time, hence one of the operations seems to be omitted as seen from the waveforms.

**12. Client2 wants to read and write in same RAM location at different time.**

This case is similar in working to Test Case 10; the only difference is that the RAM locations are the same for both read and write operation. There is a time delay of 500 ns between the read and write operation.

**13. Client1 wants to write and Client2 wants to read in different RAM location at same time.**

In this case Client1 performs the write operation in the second RAM_DEPTH cycle. In the first RAM_DEPTH cycle, the Arbiter is enabled by the RST_N pin. Along with which Client2 sends a REQUEST_C2 signal for sharing the RAM. We observe that when Client1 is not using the read as well as write function, the arbiter grants access to Client2 to read or write from the RAM. Similar is the case for the write operation also. In the first RAM_DEPTH cycle, Client2 write data "11100011" at the address location "1110'. During the second RAM_DEPTH cycle, both the client performs their respective operations smoothly.

**14. Client1 wants to write and client2 wants to read in different RAM location at different time.**

Client1 performs the write operation in the second RAM_DEPTH cycle. In the first RAM_DEPTH cycle, the Arbiter is enabled by the RST_N pin. Along with which Client2 sends a REQUEST_C2 signal for sharing the RAM. Client2 write data "11100011" at the address location "1110'. During the second RAM_DEPTH cycle, both the client performs their respective operations at different times i.e. there is a time delay of 500 ns. Client1 performs the write operation at the address location "1001". After about 500 ns, Client2 performs the read operation at the address location "1110".

**15. Client1 wants to write and Client2 wants to read in same RAM location at same time.**

In the first RAM_DEPTH cycle, Client2 performs the write operation by enabling the requisite pins. During the second RAM_DEPTH cycle, since Client1 is not using the read function, the arbiter grants the read operation to Client2 without any hassle and both the clients perform the operation smoothly. The RD_ADDR and the WR_DATA pins supply the data to the RAM. One thing to notice here is that the output which is seen at the output pins RDDATA_C1 and DATAOUT_C2 does not itself come from the RAM. This is because since the both the clients performs read and write at the same address location, hence when client2 reads from the location where client1 has written it gets the updated value. This updated value comes from the Arbiter's temporary register rather than the RAM itself.

### 16. Client1 wants to write and Client2 wants to read in same RAM location at different time.

In the first RAM_DEPTH cycle, Client2 performs the write operation by enabling the requisite pins. During the second RAM_DEPTH cycle, since Client1 is not using the read function, the arbiter grants the read operation to Client2 without any hassle and both the clients perform the operation smoothly as they operate at different times. The RD_ADDR and the WR_DATA pins supply the data to the RAM. One thing to notice here is that the output which is seen at the output pins RDDATA_C1 and DATAOUT_C2 does not itself come from the RAM. This is because since the both the clients performs read and write at the same address location, hence when client2 reads from the location where client1 has written it gets the updated value. This updated value comes from the Arbiter's temporary register rather than the RAM itself. Our claims are testified by the appropriate pulses we get from the ACK_C2 pin output. In contrast to the previous test case, the only thing that is different here is that the operations are performed at a time interval of 500 ns.

### 17. Client1 wants to read and Client2 wants to write in same RAM location at same time for Arbiter.

This case is similar to test case 15 which had been discussed earlier. The only difference is that here Client1 performs a write operation in the first RAM_DEPTH cycle. Since there is no conflict of interest among the clients, the system allows both the clients to access the system simultaneously.

### 18. Client1 wants to read and Client2 wants to write in same RAM location at different time.

This case is similar to test cased 16.

### 19. Client1 wants to read and Client2 wants to write in different RAM location at same time.

Client1 during the first RAM_DEPTH cycle performs the write operation at the address location "1000". The data written is "10101111". During the second RAM_DEPTH cycle, since the address location for both the clients are different hence, the arbiter grants them access to the RAM simultaneously. Client1 reads from address location "1000" and Client2 writes to the address location "1010".

### 20. Client1 wants to read and Client2 wants to write in different RAM location at different time.

This case is very similar in operation the previous test case. The only difference is that during the second RAM_DEPTH cycle the operations of both the clients differ by 500 ns.

### 21. Client1 wants to read and Client2 wants to read in same RAM location at different time.

During the first RAM_DEPTH cycle, Client1 perform write operation at the address location "1010". Subsequently in the next RAM_DEPTH cycle, both Client1 and Client2 try to access the RAM to read the same RAM address location. Since it a priority based arbitration system, Client1 gets access to the RAM compared to Client2. It is only when Client1 does not require the read operation any longer that the system grants access to Client2 which reads from the same location. There is a time delay of 500 ns after which Client1 sets its RD_EN_C1 to active low signal. Since both access the same RAM location at different time, this does not affect the clients on the whole.

### 22. Client1 wants to read and Client2 wants to read in same RAM location at same time.

The present case is similar to the previous test case differing only in the fact that here the scheme of priority based arbitration is clearly visible. During the second RAM_DEPTH cycle, both the clients try to access the same RAM location at the same time but only Client1 gets access to the RAM. Since there is no output at ACK_C2 pin it bears testimony to our claims.

### 23. Client1 wants to write and Client2 wants to write in different RAM location at different time.

It is similar in behavior to test case 21. The only difference is that the following case deals with the "write" operation compared to the "read" operation in the former one.

### 24. Client1 wants to write and Client2 wants to write in different RAM location at same time.

It is similar in behavior to test case 22. The only difference is that the following case deals with the "write" operation compared to the "read" operation in the former one. One thing to note is that the case of whether both the clients access the same or different RAM location is insignificant here since it is a priority based arbitration scheme.

### 25. Client1 wants to read and write in the same RAM location and Client2 also wants to read in the RAM location where Client1 has written at different time.

In this test case Client1 performs the write operation at the address location "1001". In the second RAM_DEPTH cycle, Client1 as it enjoys the higher priority among the two, can simultaneously perform the read and write operation. It writes the data "10100011" at the previous address location However as previously also discussed, the output at the RDDATA_C1 pin is the apparent output which comes from the temporary register located in the Arbiter itself which provides this output when client1 performs the read operation. Since Client2 access the system at a different time (i.e. after 500 ns), the arbiter grants permission to Client2 to read from the same location as Client1 is not using the read operation. This can be observed by the setting low of the RD_EN_C1 signal near 2500 ns mark in the waveform graph.

**26. Client1 wants to read and write in the same RAM location and Client2 also wants to read in the RAM location where Client1 has written at same time.**

Client1 performs the write operation at the address location "1001". In the second RAM_DEPTH cycle, Client1 as it enjoys the higher priority among the two, can simultaneously perform the read and write operation. It writes the data "10100011" at the previous address location. However as previously also discussed, the output at the RDDATA_C1 pin is the apparent output which comes from the temporary register located in the Arbiter itself which provides this output when client1 performs the read operation. Since Client2 access the system at the same time, hence Client2 is unable to access the RAM at all.

**27. Client1 wants to read and write in the same RAM location and Client2 also wants to write in the RAM location where Client1 has written at different time.**

In this test case Client1 performs the write operation at the address location "1001". In the second RAM_DEPTH cycle, Client1 as it enjoys the higher priority among the two, can simultaneously perform the read and write operation. It writes the data "10100011" at the previous address location. However as previously also discussed, the output at the RDDATA_C1 pin comes from the temporary register located in the Arbiter itself which provides this output when client1 performs the read operation. Since Client2 access the system at a different time (i.e. after 500 ns), the arbiter grants permission to Client2 to write at same location as Client1 is not using the write operation. One thing to observe here is that when Client2 starts to write the new data at the same address location, the output changes at the output pin instantaneously since Client1 continues to read. This clearly shows that instead of the data coming from the RAM itself, the output comes from the temporary register located in the Arbiter.

**28. Client1 wants to read and write in the same RAM location and Client2 also wants to write in the RAM location where Client1 has written at same time.**

Client1 performs the write operation at the address location "1001". In the second RAM_DEPTH cycle, Client1 as it enjoys the higher priority among the two, can simultaneously perform the read and write operation. It writes the data "10100011" at the previous address location. However as previously also discussed, the output at the RDDATA_C1 pin comes from the temporary register located in the Arbiter itself which provides this output when client1 performs the read operation. Since Client2 tries access the system at a same time, it has to wait until Client1 has finished its write operation.

**29. Client2 wants to read and write in the same RAM location and Client1 also wants to write in the RAM location where Client2 has written at different time.**

This test case is similar to the previous test case. The thing to observe here is that during the second RAM_DEPTH cycle, compared to the former case, here Client1 performs the write operation at the time when Client2 performs the read operation at the same time. There would be no basic difference between the previous waveform graph and the current one since the read and write operation is being performed by both the clients and not one, hence it does not really affect us whether the operations are being performed instantaneously or at different time.

**30. Client2 wants to read and write in the same RAM location and Client1 also wants to write in the RAM location where Client2 has written at same time.**

During the first RAM_DEPTH cycle, Client2 performs the write operation at the address location "1001". The data written is "11100011". However in the second RAM_DEPTH cycle, since Client2 has a lower priority in the system, it can only read or write at any given time. The Arbiter only allows Client2 to read from the address location "1001". This can verified from the pulses at the output pins of ACK_C2.Moreover, since Client2 reads, the arbiter at a later time allows Client1 to write to the RAM at the same address location. Almost instantaneously, the data given is sent to the input pins of the RAM and is also reflected at the output pins of DATAOUT_C2 and RDDATA_C1, the reason for which had already been discussed. We observe that the write operation of Client2 during the second RAM_DEPTH cycle is omitted.

**31. Client2 wants to read and write in the same RAM location and Client1 also wants to read in the RAM location where Client2 has written at same time.**

In this case, Client2 performs a write operation during the first RAM_DEPTH cycle. In the next cycle, one would observe that around 1700-1800 ns that Client2 performs the read operation on the address location "1010".This is more clear from the type of the output pulses at the output pin of ACK_C2. But almost within 25-50 ns, we see that the Arbiter grants the read operation to Client1 who enjoys a higher priority. The ACK_C2 signals stops giving any output. Since both Client1 and Client2 try to perform the read operation at the same time, it may be somewhat difficult to comprehend the change with the naked eye.

**32. Client2 wants to read and write in the same RAM location and Client1 also wants to read in the RAM location where Client2 has written at different time for Arbiter.**

Client2 performs a write operation during the first RAM_DEPTH cycle. In the next cycle, one would observe that around 1700-1800 ns that Client2 performs the read operation on the address location "1010".This is more clear from the type of the output pulses at the output pin of ACK_C2. But almost within 200-300 ns, we see that the Arbiter grants the read operation to Client1 who enjoys a higher priority. The ACK_C2 signals stops giving any output. Since both Client1 and Client2 try to perform the read operation at different time, the change is over a larger period of time as compared to the previous case and one could see the change quite easily.

### 33. If any client resets (RST_N=0) the system at any time.

In this we find out that whenever the RST_N goes low the systems resets and initializes all the data at all the address locations to 0. However one thing to observe here is that any data which is written after the RST_N goes low is not erased or resetted if the RST_DONE pin is still enabled. From the waveform graph we see that when RST_N and RST_DONE both goes high, data output "10111011" is given at the output pins which indicates our previously made statements. The reason for this is quite obvious as the write operation done by Client2 was after the system was reset.

### 34. If any client gives the inputs until RST_DONE is high.

In this case, we observe that any operation from any of the clients is nullified by the RAM Arbiter if RST_DONE is set to active low even when the RST_N pin is high.

# CONCLUSION

In this project we have successfully laid out the architectural design of a RAM Arbiter. We have used Xilinx as the platform to perform the design throughout. With the help of VHDL we first designed a RAM followed it up with our Arbiter module and lastly port mapped the two to design the RAM Arbiter.

Here, the completion of each objective of the project led to the design of an efficient RAM arbiter. Firstly, the single system interface with memory was essential in utilizing the memory and carrying our system accesses. Second, the requirements that were set at the beginning of the project for the arbiter were important in meeting the goals. Thirdly, the validation process was necessary to debug and verify all key features of the design. With the success of each goal, the arbiter can be considered validated and successful. All the blocks as well as the design considerations which had been taken into account have been implemented through the platform. To design the FSM block, a sum total of thirty four test cases have been considered and the output waveforms have been shown along with the analysis done for each of the test cases. One of the most primary problems associated with arbiters i.e. the address clash problem has been taken care of very smoothly. The procedures for handling such problems have been discussed in detail in the earlier parts of the project.

With regards to the arbiter itself, it could be further expanded to three or four systems. If that were the case, adding additional arbitration schemes would be ideal since for more than two or three systems the fixed priority arbiter would need some design changes, whereas a round robin arbiter would be simple to implement for high number of systems. For a three system fixed priority arbiter it would be simple to add an extra priority level for the third system and also add the logic to handle that. Extra timeout counters would also be necessary in order to track the timeouts for multiple systems. For n systems, there would be n-1 timeout counters for each of the systems that have priority lower than the highest. This would allow the timeouts for each of the systems to be independent of other systems. This is the limiting factor in adding additional systems to the fixed priority arbiter, as it gets increasingly more complex for each addition of a system. This is why implementation of multiple arbitration schemes is also an important addition to the arbiter.

If improvements are to be made, the modularity of the design should be upheld. This project not only provides a useful flexible arbiter design as set out to do, but also provides a basis for future work in arbiter design. Thus the architectural design of the RAM Arbiter has been executed through Xilinx by VHDL Coding effectively and efficiently.